\documentclass[12pt,twoside,eqsecnum]{article}
\usepackage{cite}
\usepackage{a4wide}
\usepackage[T1]{fontenc}
\usepackage{pstricks,pst-all,pst-coil,pst-circ,pst-plot,multido,pstricks-add,fp,
pst-3d,pst-3dplot,pst-vue3d,pst-eucl,amsbsy}
\def\vec#1{\boldsymbol{#1}}
\newrgbcolor{hotpink}{0.9 0 0.5}
\newrgbcolor{hotgreen}{0.1 0.8 0.3}
\usepackage{graphicx}
\usepackage{palatino}
\newcommand{\mathsym}[1]{{}}
\newcommand{\unicode}[1]{{}}
\usepackage{wrapfig}
\usepackage{subfigure}
\usepackage{amsfonts}
\usepackage{amsmath,amssymb,amsbsy}
\usepackage{aas_macros}
\usepackage{slashbox}
\usepackage{slashed}
\def\vec#1{\boldsymbol{#1}}
\def\slj#1#2#3{{}^{#1}\text{#2}_{#3}}
\def\dd{\mathrm{d}}
\def\RE{\mathop{\Re{\rm e}}\nolimits}
\def\IM{\mathop{\Im{\rm m}}\nolimits}
\usepackage{lscape}
\begin{document}
 \title{An introduction to the quark model}
\author{Jean-Marc Richard\\
{\small Universit\'e de Lyon \& Institut de Physique Nucl\'eaire de Lyon}\\
{\small IN2P3-CNRS \& UCB, 4 rue Enrico Fermi, 69622 Villeurbanne, France}\\
{\small\sf j-m.richard@ipnl.in2p3.fr}}
\date{\small \today}
\maketitle
\begin{abstract}\noindent
This document contains a review on the quark model, prepared for lectures at
the
Niccol\`o Cabeo School at Ferrara in May 2012. It includes  some historical
aspects, the spectral properties of the 2-body
and 3-body Schr\"odinger operators applied to mesons and baryons, the link
between meson and baryon spectra, the role
of flavour independence, and the speculations about stable or metastable multiquarks.
The analogies between few-charge systems and few-quark bound
states will be underlined.
\end{abstract}
%\newpage
\tableofcontents
\clearpage
\markboth{\sc An introduction to the quark model\hspace*{1cm}
}{\sc Prelude}
\section{Prelude: few charge systems in atomic physics}
\label{se:at}
\vskip .01cm
\begin{flushright}
 \sl Si  parva licet componere magnis
\footnote{if it is allowed to compare small things with great} \\
\rm Virgil
\end{flushright}
\vspace{-1cm}
\subsection{Introduction}\label{at:sub:intro}
The spectrum of few-charge systems was among the very first applications of
quantum
physics. The Bohr--Sommerfeld rules explain the
energy levels of the hydrogen atom $(p,e^-)$, and can be easily extended to any
$(m_1^+,m_2^-)$ pair with arbitrary masses. The solution of the three-body
problem for ($m_1^+,m_2^-,m_3^-)$
turned out less obvious, and required the Schr\"odinger equation, and the
associated variational
methods. It also revealed some surprises, with the stability imposing 
constraints on the masses $m_i$.

The stability of the positronium molecule $(e^+,e^+,e^-,e^-)$, or any similar
system with equal
masses (in the limit where annihilation is neglected) was suggested by Wheeler
in 1945 \cite{Wheeler:1945} and
proved in 1947 \cite{PhysRev.71.493},
but the (indirect) experimental evidence was published only in 2007
\cite{2007Natur.449..195C}. This indicates
how patient one should be when predicting novel structures.

The quantum mechanics of few-charge systems was a great source of inspiration
for building the quark model, in its minimal non-relativistic version.
Amazingly, some techniques developed
to extrapolate into flavour space the few-quark spectra with
fla\-vour-independent
forces turned out useful to understand the stability patterns in atomic
physics. I refer to \cite{2005PhR...413....1A} for a review about few-charge
systems. Here, I will just stress a few points that are connected to  the quark
model.
\subsection{The atomic two-body problem}\label{at:sub:2b}
\subsubsection{Central potential}
Out of the Hamiltonian
\begin{equation}\label{at:eq:H}
\frac{\vec p_1^2}{2\,m_1}+\frac{\vec p_1^2}{2\,m_1}-\frac{e^2}{r_{12}}~,
\end{equation}
the centre of mass can be removed and one is left with the intrinsic Hamiltonian
describing the relative motion, which reads
\begin{equation}\label{at:eq:Hi}
H=\frac{\vec p^2}{2\,\mu}-\frac{e^2}{r}~,
\end{equation}
where $\mu$ is the reduced mass, given by $\mu^{-1}=m_1^{-1}+m_2^{-1}$, and
$\vec r=\vec r_2-\vec r_1$ and $\vec p=
(\vec p_2-\vec p_1)/2$ are conjugate variables for the relative motion.

The Hamiltonian $H$ has seemingly two parameters, the reduced mass $\mu$ and the
strength $e^2$, but it can be rescaled. It is sufficient to solve once for ever
the universal spectral problem
\begin{equation}\label{at:eq:Hu}
h=-\Delta-r^{-1}~,
\end{equation}
and to apply a simple factor $2\,\mu\,e^4$ to the eigenvalues and
$(2\,\mu\,e^2)^{-1}$ to the distances
to recover the actual energies and wave-functions of \eqref{at:eq:Hi}.

There is a similar universality for all harmonic oscillators $\vec p^2/m+K\,\vec
r^2$, and more generally
for all power-law interactions \cite{Quigg:1977xe}
\begin{equation}\label{at:eq:pw}
H=\frac{\vec p^2}{2\,\mu}+g\,\epsilon(\alpha)\,r^\alpha~,
\end{equation}
where $g>0$, $\epsilon$ is the sign function,  to ensure that the force is
attractive,
and $\alpha$ is not  too large if negative, otherwise the Hamiltonian would not
be bounded from below.

A common feature of the Coulomb and Hookes potentials in three dimensions, is
that they
support an infinite number of bound states, however weak is the strength of the
potential. This contrasts with short-range potentials such as the Yukawa
interaction
$-g\,\exp(- b\,r)/r$ of nuclear physics, which requires a minimal strength $g$
to achieve binding. Similarly, the effective potential between neutral atoms or
the effective potential between two hadrons are of short-range nature, and thus
do not necessarily support bound states even when they are attractive. 

The Coulomb Hamiltonian \eqref{at:eq:Hu} has a well-known spectrum, with a
discrete part $\epsilon_{n,l}=-1/[4\,(n+\ell)^2]$, where $n=1,\,2,\ldots$ is
the radial number, and $\ell=0,\,1\,\ldots$ the orbital momentum ($n+\ell$ is
the principal quantum number). There is also a continuum for~$\epsilon\ge0$. 
\subsubsection{Spin-dependent corrections}
The Coulomb interaction can be derived from Quantum ElectroDynamics (QED) in
the non-relativistic limit, and in the approximation of the lowest order in the
coupling constant, which corresponds to one-photon exchange. 

There are several interesting corrections, which have been probed successfully.
In particular, the vector nature of the photon gives very characteristic spin
corrections. 

For $\ell\ge1$, there are spin-orbit and tensor terms which contribute to the
fine structure: levels with the same orbital momentum $\ell$ but
different coupling of $\ell$ to the spins have slightly different energies.

For $\ell=0$, there is spin--spin term, which splits the spin-singlet form the
spin triplet states. It reads
\begin{equation}\label{at:eq:ss}
V_{ss}=\frac{e^2}{m_1\,m_2}\,\frac{2\,\pi}{3}\,\delta^{(3)}(\vec r)\,
\vec \sigma_1.\vec\sigma_2~,
\end{equation}
in its most simplified form, to be treated at first order in perturbation
theory, using the wave functions generated by the central potential $-e^2/r$.
The short-range character, and the dependence on the masses should be kept in
mind when an analogue will be proposed for the quark--antiquark interaction.
This interaction is responsible for the shift between the ortho- and
para-hydrogen states and the transition between them produces the famous 21\,cm
line, whose gravitational red-shift and Doppler shift gives valuable
information in astrophysics. The analogue will be measured at CERN for
antihydrogen, to probe the matter--antimatter symmetry.

Note that in the case of the positronium atom, the hyperfine splitting
also receives some contribution from the annihilation diagrams. 
 
\subsection{Three-unit-charge ions}\label{at:sub:3b}
The best known 3-body system is atomic physics is the neutral Helium atom,
$(\alpha,e^-,e^-)$, but its
stability is obvious since once the first electron is attached to the $\alpha$
particle, there
remains some long-range attraction to bind the second electron.

The binding of $\mathrm{H}^-=(p,e^-,e^-)$ is less obvious, as the second
electron feels a neutral system
at large distances. Unlike the case of the Helium atom, neither an
effective-charge ansatz
$\psi(r_2,r_3)=\exp(-a (r_2+r_3))$, nor a more general Hartree wave function
$f(r_2)\,f(r_3)$
suffices to demonstrate the binding variationally. But a more elaborate wave
function does. 
For a review on two-electron systems, see, e.g., \cite{2010AmJPh..78...86H}.

More interesting, perhaps, is the dependence upon the masses
\cite{2005PhR...413....1A}: the molecular hydrogen ion
$\mathrm{H}_2{}^+=(p,p,e^-),$ is very stable, with a variety of 
excitations, the hydrogen ion 
$\mathrm{H}^-=(p,e^-,e^-)$ and the positronium ion $\mathrm{Ps}^-=(e^+,e^-,e^-)$
are bound by a small margin, but the ``protonium ion'' $(p,\bar
p,e^\pm)$ is not stable. (Here $p$
and $\bar p$ simply denotes the mass and the Coulomb
charge; any hadronic interaction is neglected.)

There is an increase of stability near the symmetry line $m_2=m_3$ of
$(m_1^\pm,m_2^\mp,m_3^\mp)$. 
Schematically, the system
has there two threshold configurations $(m_1^+,m_2^-)+m_{3}^-$, and
$(m_1^+,m_3^-)+m_{2}^-$,
which are degenerate, and thus interfere optimally to achieve the best binding.

This mass dependence is sometimes acute. For instance, a fictitious H${}^-$ 
with one electron heavier by about 10\% would not be be stable. 
This is the case, e.g., for the (very) exotic $(p,\mu^-,\pi^-)$ ion. 
\subsection{Three-body exotic ions}\label{at:sub:3bexo}
It is sometimes believed that $\mathrm{H}^-$ has no excited bound state.
This is both true and false.  This is true if you define a bound state
as lying below the lowest threshold, which is made of $\mathrm{H}=(p,e^-)$ in
the 1S
state and an
isolated electron. However, there is a level with total orbital momentum $L=1$
and parity $+1$, that cannot break into $\mathrm{H}(1S)+e^-$ without involving
spin forces and radiative effects. For this state, the lowest  threshold for
spontaneous
dissociation consists of 
$\mathrm{H}=(p,e^-)$ in the 2P state and an electron. A bound state 
lying below this threshold exists for
$\mathrm{H}^-=(p,e^-,e^-)$, but it exhibits even a more striking mass dependence
when the
constituents are modified \cite{PhysRevA.80.054502}. Allowing a very small mass
difference
between the two electrons (less than 1\%) or replacing the proton by a positron
spoils the binding. On the other hand, if the masses are inverted, i.e., for
the unnatural-parity state of $\mathrm{H}_2{}^+=(e^-,p,p)$, a comfortable
binding is observed. 
\subsection{Molecules with four unit charges}
The best known case corresponds to the hydrogen molecule
$\mathrm{H}_2=(p,p,e^-,e^-)$,
bound well below the threshold for dissociation into two hydrogen atoms, with a
variety of excitations. There are also variants with one or two protons
replaced by an isotope. The most suited framework for studying H$_2$ is the
Born--Oppenheimer
approximation. For a given position of the protons, the electronic energy is
computed, and interpreted as an effective proton--proton interaction
supplementing their direct electrostatic repulsion. The Schr\"odinger equation
is then solved for the two-body problem of the protons and generates the ground
state and its radial and orbital excitations. Another set of levels correspond
to the electron cloud being in an excited state. (Later the analogue will be
the hybrid hadrons, with the gluon field linking the quarks being excited).
Note that one hardly treats the hydrogen molecule as a ``di-proton'' bound to a
``di-electron''.

In atomic physics and ab-initio chemistry, one usually starts from the limit
where the proton and other nuclei are infinitely heavy, and then their finite
mass can be treated as a correction, through the Hughes--Eckart terms or
similar. In 1945, Wheeler \cite{Wheeler:1945} addressed the somewhat opposite 
issue of a proton with the same mass as the electron, i.e., the problem of the
stability of the positronium molecule, $(e^+,e^+,e^-,e^-)$ (in the limit where
annihilation is neglected). In 1946, Ore performed an elaborate 4-body
calculation, and concluded that such a configuration is hardly stable
\cite{PhysRev.70.90}. But the following year, Hylleraas and the very same Ore
published an elegant analytic proof of the stability of the positronium
molecule \cite{PhysRev.71.493}. It took 60 years to obtain at last an indirect
experimental evidence
of the existence of this molecule \cite{2007Natur.449..195C}.

Meanwhile the dependence upon the masses was studied in some details.
In
Refs.~\cite{1971SSCom...9.2037A,1972PMag...26..143A,1993PhRvL..71.1332R,
1994PhRvA..49.3573R}, it was shown how
$(M^+,M^+,m^-,m^-)$ gets more binding, as compared to its threshold, when the
mass ratio departs from $M/m=1$. On the other hand, it was observed that
when allowing for two different masses in another way, $(M^+,m^+,M^-,m^-)$,
stability is lost for $M/m\gtrsim 2.2$ (or, $M/m\lesssim 1/2.2$), because the
molecule cannot compete any more with the lowest
threshold $(M^+,M^-)+(m^+,m^-)$ \cite{PhysRevA.55.200,Varga99}. 

Many less symmetric configurations are bound, such as the positronium hydride
$\mathrm{PsH}=(p,e^+,e^-,e^-)$, or any configurations in which two particles
have the same charge and the same mass, that is to say, $(a^+,b^+,c^-,c^-)$,
whatever are the masses of $a$ and $b$~\cite{1997EL.....37..183V}.

The lessons, to be kept in mind when arriving at the section on multiquarks,
are:
\begin{itemize}
 \item the role of the masses is important. Look at $(a^+,b^+,c^-,d^-)$ vs.\
its lowest threshold supposed to be $(a^+,c^-)+(b^+,d^-)$. For both the 4-body
system and its threshold, the potential energy $\sum g_{ij}\,r_{ij}^{-1}$ has a
cumulated strength $\sum g_{ij}=-2$. So, there is no obvious reason why a
molecule should lie lower that two atoms. Hence any favourable breaking of
symmetry in the kinetic energy is welcome. Conversely, any unfavourable breaking
can spoil the binding. 
\item the four-body problem is delicate, even when the interaction is known
perfectly,
\item one should be patient to see the experimental confirmation of theoretical
predictions of exotic states.
\end{itemize}
%
%\clearpage
\markboth{\sc An introduction to the quark model\hspace*{1cm}
}{\sc History}
\section{A brief historical survey}\label{se:hi}
\begin{flushright}
 \sl History is Philosophy teaching by examples\footnote{According to Michel
Casevitz, the sentence is not by Thucydid, but a British commentator}\\
\rm Thucydid
\end{flushright}
\vspace{-1cm}
\subsection{Prehistory}\label{hi:sub:pre}
There are several books relating the birth of particle physics, with
entertaining anecdotes. Segr\`e, for instance, was an acute observer
\cite{segre2007x}.
A very comprehensive review is given by Pais \cite{pais1988inward}. For the
nuclear forces, see \cite{brown1996origin}. Books and reviews devoted to the
quark model include
\cite{kokkedee1969quark,morpurgo1999relation,lipkin1973quarks,
close1979introduction,
flamm1982introduction,LeYaouanc:111431,Grosse:847188}.
There are also reviews covering baryons
\cite{Korner:1994nh,Capstick:2000qj,Klempt:2009pi} or mesons
\cite{Novikov:1977dq,Klempt:2007cp,Voloshin:2007dx}, in particular the synthesis
by the ``quarkonium working
group'' \cite{Brambilla:2004wf,Brambilla:2010cs}.

After the Rutherford experiment, indicating how compact is the atomic nucleus,
and the discovery of the neutron by Chadwick, it became necessary to understand
how the nucleus is built out of protons and neutrons. The mechanism of Yukawa:
the exchange of a massive boson, turned out the be successful, with the 
discovery of the pion at Bristol in 1947.

Among the theoretical activity stimulated by the Yukawa model, two points at
least deserve attention. 
\begin{enumerate}
 \item The mass of the Yukawa boson is constrained by the ratio of the 3-body to
the 2-body binding energy, as stressed by Thomas in a celebrated paper
\cite{PhysRev.47.903}, who anticipated what is known today as ``Borromean
binding'', and, more generally, ``Efimov physics''. 
\item At first sight, three potentials are needed to
build the nucleus: proton--proton, proton--neutron and neutron--neutron, and it
is natural to seek for some simplification. A
tempting scenario is that where solely the proton--neutron interaction exists,
but this is
clearly contradicted by the data on proton  scattering. What eventually
prevailed
is isospin symmetry. With the proton and the neutron forming a
doublet, there are only two potentials in the limit where isospin is conserved,
one for total isospin $I=0$, and another one for $I=1$.
\end{enumerate}
\subsection{Early hadrons}
The pion was discovered in 1947 and seen in three charge states,
$\pi^+,\,\pi^0$, and $\pi^-$ which form
an
isospin triplet. Hence in 1947, we had 5 hadrons: 2 nucleons and 3 pions. 
But already two more were expected, as the existence of antimatter was (not so
easily) predicted as a consequence of the Dirac equation, and the positron was
discovered in cosmic rays by Anderson. The antiproton and the antineutron were
anticipated as well.
%\footnote{From its magnetic moment, it should be noted,
%however, that the proton does not obey exactly the Dirac equation.}\@
A dedicated accelerator was built at Berkeley, the Bevatron, and the antiproton
was, indeed, discovered in 1955, and the antineutron shortly after.

With 7 hadrons, 2 nucleons, 2 antinucleons, and 3 pions, the world of hadronic
physics would be reasonably sized, and one could envisage to work on the
interaction among these few hadrons. However, several complications occurred
almost simultaneously.

First, the Yukawa picture of nuclear forces, though very efficient for the
long-range part, faced difficulties at shorter distances. More attraction was
needed, and also some spin-orbit component, that neither pion-exchange or
iterated pion-exchange were able to provide. The work by Breit, among others,
was remarkable \cite{PhysRev.51.248}. For further details, see, e.g., the review
\cite{brown1996origin}. An explicit
scalar exchange (call it $\sigma$ or $\epsilon$) and an explicit vector
exchange were needed. While the former was about isospin independent, and thus
provided by the exchange of an isoscalar scalar meson, the latter was sought
to be different in $np$ vs. $pp$ scattering, and thus called for both an
isoscalar and an isovector vector meson. The $\omega$ and $\rho$ were thus
predicted!

Second, the interaction of pions with nucleons was shown to produce new 
particles, nucleon resonances, in particular the $\Delta(1232)$, which has
isospin 3/2, i.e., exists in four possible electric charges. Similarly,
proton--nucleon or proton--nucleus scattering, or proton--antiproton
annihilation were able to produce several new mesons, the ones desired to
improve the theory of nuclear forces, and others. These hadrons are not stable,
with for instance $\Delta\to N+\pi$ or $\rho\to\pi+\pi$, but were named hadrons
as well, baryons or mesons. 

The $\Delta$, at first sight, appears as a consequence of the interaction
between $\pi$ and $N$, and the $\rho$ as a resonance of the $\pi\pi$
interaction. Chew and his collaborators generalised this scenario, and
suggested the concept of
``bootstrap'' or  ``nuclear democracy'' \cite{chew1961s,jacob1964strong}:
everything is made of
everything, and any hadron is both a building block and  the result of the
interaction of the other hadrons.
For instance, $\Delta$ is made of $N\pi$, $N\pi\pi$, etc., and, as well $N$ is
made of $\Delta\pi$, etc. This gives an infinite set of coupled equations, of
which it was hoped one could extract a finite set as a first
tractable approximation to
the spectrum and to the dynamics. The success was, however, extremely limited.
Ball, Scotti and Wong, for instance, stressed that describing
mesons as resulting of the nucleon--antinucleon interaction hardly gives the
observed ``exchange degeneracy'' (named after the phenomenology of high-energy
scattering), the property that an isoscalar and an isovector mesons with the
same quantum numbers have very often the same mass~\cite{PhysRev.142.1000}. In
the baryon sector, the next state after the $\Delta$ with isospin $I=3/2$ and
spin $J=3/2$ was sometimes predicted to have $I=5/2$ and $J=5/2$!

Third came strangeness. New particles were observed in the 50s and 60s,
decaying weakly though massive enough to decay to existing particles, and
produced
by pairs with strict rules: $K^+$ together with $\Lambda$ for instance, but
never $K^-$ together $\Lambda$. A new quantum number, strangeness $S$, was
introduced to
summarise the properties of these new particles: strangeness is conserved in
production processes by strong interaction, and thus $\Lambda (S=-1)$  can be
produced in association with $K^+(S=1)$, but not $K^-$ which has $S=-1$.
On the other hand, strangeness is not conserved in the decay by weak
interaction, as $\Lambda (S=-1)\to N+\pi$, or $\Lambda\to p+e^-+\bar\nu$. The
weak interaction of strange particle was beautifully linked to that involved in
ordinary beta decay. 
\subsection{Generalised isospin}
There is of course the exception of the $\pi$ meson, with a mass of about
0.14\,GeV,  significantly lighter than the mass of the $K$, about 0.49\,GeV,
and the exception of light scalar mesons with long-standing questions about
their structure. 
Otherwise, one observes that strange particles do not differ much form the
non-strange ones. For instance, 
the mass of the $\Lambda$ baryon is about 1.1\,GeV, just slightly above that
of the nucleon at 0.94\,GeV, and the mass of the $K^*$, 0.89\,GeV, is close to
that of the vector mesons $\rho$ and $\omega$, about 0.78\,GeV. It was thus
tempting to put strange and non-strange particles in multiplets generalising
isospin.
Since isospin is built on the SU(2) group, the minimal extension is SU(3).
Later, it was renamed SU(3)$_\text{F}$, to differentiate it
from the SU(3) group associated with colour. 

A specific model was proposed by Sakata \cite{Sakata:1956hs}, with
$(n,p,\Lambda)$ as the
building blocks of
matter, and mesons as baryon--antibaryon pairs. We already mentioned that this
picture of mesons is difficult to accommodate with the long-range
baryon--baryon interaction as given in the Yukawa picture. But the Sakata model
had also problems with baryons. There are too many baryons with low mass. Take
for instance the lowest baryons with spin $J=1/2$. Besides $p$, $n$, and
$\Lambda$, there is a triplet de singly-strange ($S=-1$) baryons
$(\Sigma^+,\Sigma^0,\Sigma^-)$ of mass about 1.3\,GeV, and a pair of doubly
strange ($S=-2$) baryons $(\Xi^0,\Xi^-)$ with mass about 1.5\,GeV. In the Sakata
model,
they should belong to higher representation, and, in a dynamical picture,
contain an additional baryon--antibaryon pair. This is hard
to believe. 

To take care of this problem, Ne'emann and Gell-Mann suggested to keep the SU(3)
group
as the basic symmetry, but to put the known $J=1/2$ baryons in an octet
representation. This is the famous ``eightfold way'' \cite{gell2000eightfold}.
The group SU(3) has eight
generators, instead of three for SU(2) ($I_\pm$ and $I_3$). Each  multiplet can
be
characterised by the dimension of the representation (this generalises the
$2\,I+1$ multiplicity for SU(2)), and two generators that commute, which are
usually taken as $I_3$ and strangeness $S$, or equivalently the
\emph{hypercharge} defined as $Y= b+S$, where $b$ is the baryon number.
\begin{figure}[!tch]
 \centering
%
% octet des baryons
%
\includegraphics{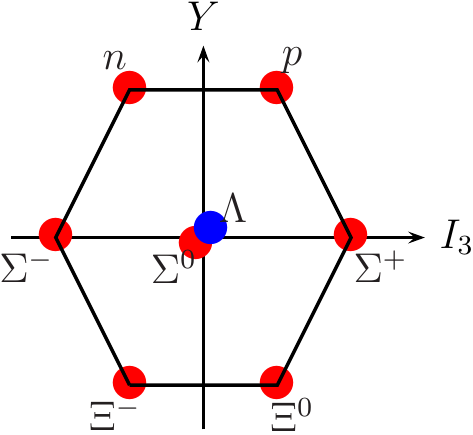}
 \caption{\label{hi:fig:octet} Octet of the lowest spin $J=1/2$ baryons}
\end{figure}
SU(3) is rather good symmetry. Its breaking can be described by simple terms,
treated at first order, which are proportional to some  generators of the
group. More elaborate mechanisms were proposed for breaking SU(3).
One example is the famous Gell-Mann--Okubo formula for the baryon masses
\begin{equation}\label{hi:eq:GMO1}
M=M_0+ a\,Y+ b(I(I+1)-Y^2/4)~,
\end{equation}
from which one gets
\begin{equation}\label{hi:eq:GMO2}
2(N+\Xi)=3\,\Lambda+\Sigma~,
\end{equation}
where each particle stands for its mass, in surprisingly good agreement with
experiment. 
\subsection{The success of the eightfold way}
When the eightfold way was proposed, only 9 baryons with spin $J=3/2$ were
known, with a low mass: four $\Delta$ of charge ranging from $-1$ to $+2$, thre
$\Sigma^*$ with strangeness $S=-1$ and two $\Xi^*$ with strangeness $S=-2$. At
the 1962 Rochester Conference held in Geneva, 
Gell-Mann pointed out they would fit very well a decuplet representation of
SU(3), provided the last member does also exist. He named it $\Omega^-$, as a
kind of ultimate achievement, in the biblical sense. The masses of the 9
existing members being observed to grow linearly with strangeness, i.e.,
following the pattern%
\begin{equation}\label{hi:eq:esd}
\Sigma^*-\Delta=\Xi^*-\Sigma^*~,
\end{equation}
known as the ``equal spacing rule of the decuplet'',  it was tempting to
extrapolate to include $\Omega^--\Xi^*$ in the equality, thus predicting the
mass of the $\Omega^-$ at about 1.67\,GeV. Some physicists were sceptical about
the  possibility of producing and detecting easily the $\Omega^-$.  It was
nevertheless seen in an experiment at Brookhaven led by Nick Samios, with
exactly the mass predicted by Gell-Mann. This was at the end of 1963, and
published in  1964~\cite{Barnes:1964pd}, which otherwise was
one of the best vintages ever in Burgundy!
\begin{figure}[!tch]
 \centering
\includegraphics{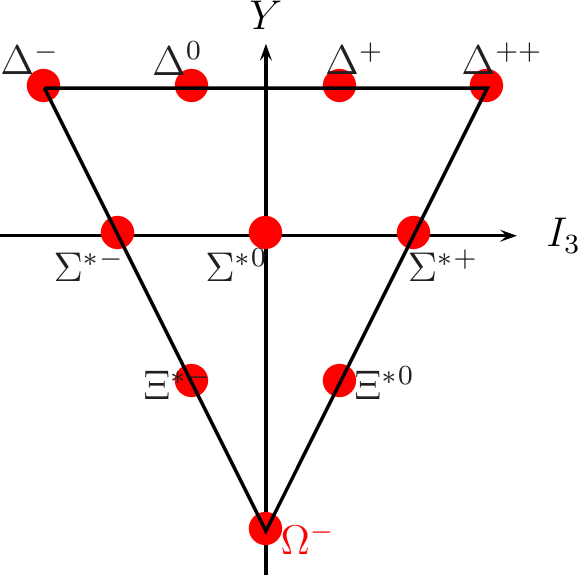}
 \caption{\label{hi:fig:decuplet} Decuplet of the lowest spin $J=3/2$ baryons.
The
$\Omega^-$ was missing till its discovery in 1964.}
\end{figure}
\subsection{The fundamental representation: quarks}\label{hi:sub:quarks}
\begin{figure}[!hbc]
 \begin{center}
\includegraphics{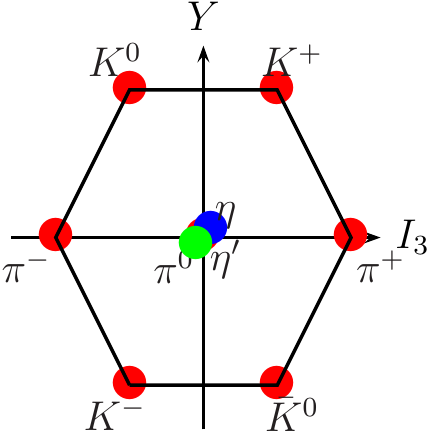}
\hspace*{1cm}
\includegraphics{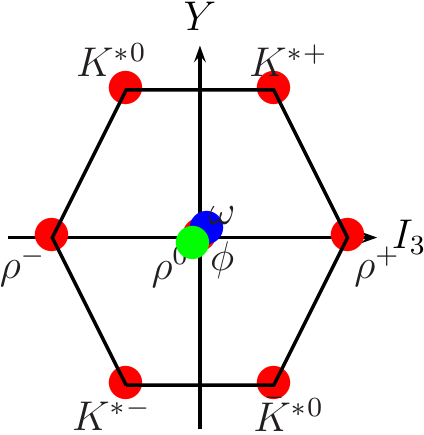}
\end{center}
 \caption{Octet and singlet of pseudoscalar (left) and vector (right) mesons.}
 \label{hi:fig:mesons}
\end{figure}

The pseudoscalar mesons ($\pi$, $\eta$, $K$, $\bar K$) 
were accommodated in an octet and a singlet. As there is presumably mixing
between the singlet and the isoscalar member of the octet, it became customary
to talk about ``nonet''. The same holds for the vector mesons. See
Fig.~\ref{hi:fig:mesons}.

We thus had already in the hadronic world octets, singlets and decuplets, but no
triplet, which corresponds to the fundamental representation! Gell-Mann
proposed that the fundamental representation is populated by three yet not
discovered -- or hypothetical -- particles. He was of course fully aware that
any representation can be built by combining the fundamental representation,
$3$, and its conjugate, $\bar 3$, in the same way that any value of the spin
can be built by adding elementary spins 1/2.

The word ``quark'' was taken from the sentence ``Three Quarks for Muster Mark''
in Joyce's Finnegans Wake. In another context (see next subsection), the word
``ace'' was suggested, but did not prevail. The individual quarks were
sometimes named $p$, $n$ and $\lambda$, in reference to the Sakata model, but
quickly the naming scheme $u$, $d$, $s$ was adopted by the community. 

\begin{figure}[htp]
 \centering
\includegraphics{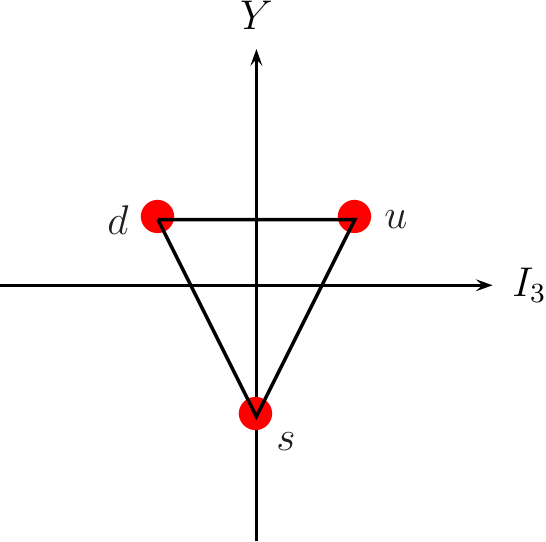}
\hspace*{1cm}
\includegraphics{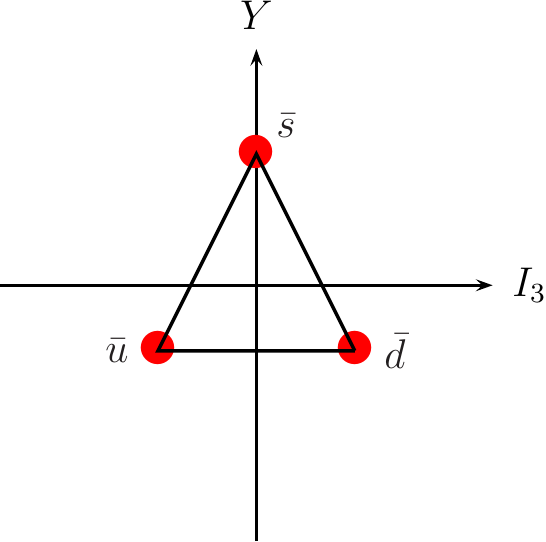}
 \caption{\label{hi:fig:quarks} Triplet of quarks and antitriplet of antiquarks}
\end{figure}

The properties of the quarks are summarized in Table \ref{hi:tab:quarks} and
Fig.~\ref{hi:fig:quarks}
\begin{table}[!!htp]
\caption{\label{hi:tab:quarks}Properties of the quarks}
\vskip -.5cm
 \centering
$$
 \begin{array}{cccrrrr}
q &b & I & I_3 & Y & S& Q\\
\hline\\[-4pt]
u &\frac13&\frac12 &\frac12& \frac13& 0 &\frac23 \\[5pt]
d & \frac13&\frac12 &-\frac12& \frac13& 0 &-\frac13 \\[5pt]
s & \frac13&0&0& -\frac23& -1 &\frac23 \\[2pt]
\hline
 \end{array} 
$$
\end{table}

The basic reduction of products of representations that are important for the
quark model are
\begin{equation}\label{hi:eq:prosu3}
 3\times3\times3=1+8+8+10~,\quad
3\times\bar 3=1+8~.
\end{equation}

Historians could debate endlessly whether the pioneers considered
quarks as a handy mathematical tool to build the representation of SU(3), or
had already in mind a physical interpretation of the quarks as the constituents
of the hadrons. Anyhow, this was one of the major breakthroughs
in physics.

\subsection{The OZI rule}\label{hi:sub:ozi}
Another approach was followed by Zweig. See, e.g., his recollection at
the Conference Baryon80  \cite{Zweig:1980nu} or at the celebration
of Gell-Mann's 80th birthday \cite{Zweig:2010jf}.

The $\phi$ meson, of mass about 1.02\,GeV, was discovered in 1962
\cite{Bertanza:1962zz,Schlein:1963zz}. It is an
isoscalar, vector meson, like the $\omega$, but with peculiar decay patterns.
While it could easily decay into pions, it prefers the $K\bar K$ channels, for
which the phase-space is meagre (the $K$ has a mass of about 0.49\,GeV). 

\begin{figure}[!htp]
 \centering
\includegraphics{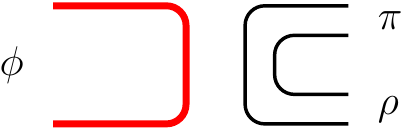}
\hspace*{.5cm}
%Doubly OZI suppressed
\includegraphics{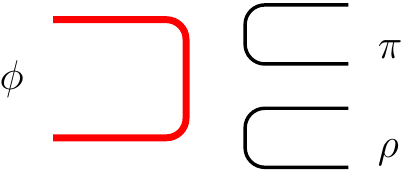}
\hspace*{.5cm}
 % Desintegration du phi en K K
\includegraphics{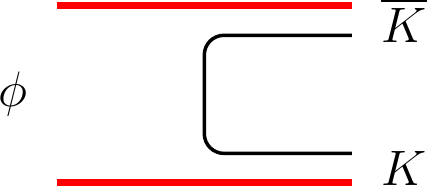}
 \caption{\label{hi:fig:phi} Decay of the $\phi$ meson: Zweig forbidden (left,
centre) or allowed (right)}
\end{figure}

Zweig's explanation is that this favoured decay is dictated  by $\phi$ meson
content. He named the constituent ``aces'', but we will call them quarks to
conform with the current usage. 
The idea is that the decay preferentially keeps the existing content. In the
modern language, the $\phi$ decay is described by the diagrams of
Fig.~\ref{hi:fig:phi}.

He thereby invented the  ``Zweig rule'', also called Okubo--Iizuka--Zweig
 (OZI), or (A--Z) rule since many authors contributed, from Alexander to Zweig.
For a review, see, e.g., \cite{Nomokonov:2002jb}.
As we shall see shortly, a nice surprise was that the rule works even better
for heavier quarks.

The rule governing the $\phi$ decay was extended to other decays and to
reaction mechanisms. Quark lines should not start from  and end in the same
hadron,
i.e., disconnected diagrams are suppressed. Instead quark lines should better
link two hadrons in the initial or final state. 

\begin{figure}[!htp]
 \centering
% \includegraphics[width=.3\textwidth]{FigsQuark/FigsQ-fig6.pdf}
 % Rearrangemeent
\includegraphics{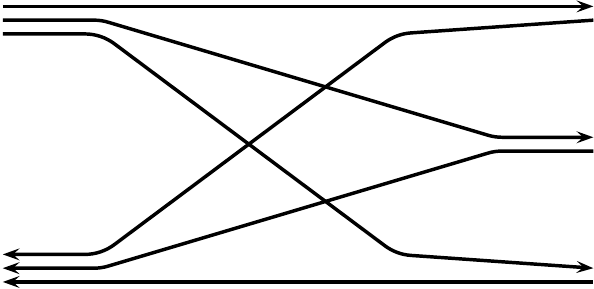}
\hspace*{1cm}
% planar
\includegraphics{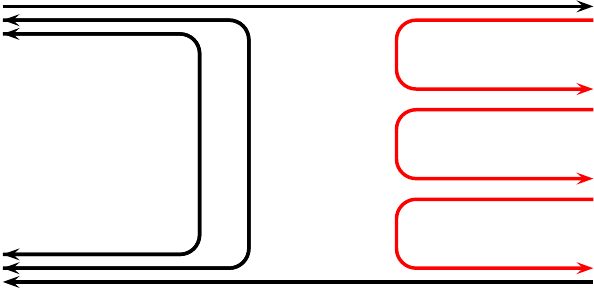}
 \caption{\label{hi:fig:annih} Possible diagrams contributing to
baryon--antibaryon annihilation. The hierarchy of these diagrams requires an
extension of the Zweig rule that remains a little controversial.}
\end{figure}

The OZI rule then got variants. It is sometimes argued that the dominant
processes correspond to planar diagrams, while non-planar (but still connected)
diagrams are suppressed. For instance, there are dozens of measured branching
ratios for antinucleon--nucleon annihilation at rest or in flight. In
Fig.~\ref{hi:fig:annih}, the rearrangement diagram (left) clearly keeps the
initial constituents,  while the diagram on the right is more planar. Clearly,
the former does not produce enough kaons, while the second predicts much more
kaons than observed.

Note that antiproton annihilation is an indirect evidence for quarks, but this
was understood much after the first measurements. If guidance is sought from
QED,
annihilation has a rather small cross section, because it is a very short-range
process. The results obtained at Berkeley for  the elastic and annihilation
cross-sections of antiproton scattering on nucleons indicated that
the latter is larger. To reproduce these results with an empirical (complex)
potential, one needs a large size for the annihilation part, about 0.8\,fm or
more. This was a puzzle. One now understands that the proton and the antiproton
are composite, with a size of the order of 0.5--1\,fm. Hence, when they
overlap, they can rearrange their constituents into quark--antiquark pairs. This
is similar to the rearrangement collisions in molecular physics, but has little
to
do with $e^+e^-$ annihilation in QED.
\subsection{First quark models}\label{hi:sub:first}
Greenberg, in a celebrated paper \cite{Greenberg:1964pe}, tried to understand
the structure of baryons
made of quarks. In this paper, he suggested a kind a harmonic oscillator as a
first approximation to describe the quark motion. He addressed the problem of
the statistics and suggested a kind of para-statistics, that eventually became 
the colour degree of freedom.

The work of Dalitz was done almost simultaneously. In the summer of 1965, and
in particular in his lectures at the School of Les Houches
\cite{Dalitz:1965fb}, he constructed
his first version of the harmonic oscillator model of baryons. As Greenberg, he
faced the problem of the statistics of the quark.

Dalitz's work was the starting point of a series of studies about baryons in the
harmonic oscillator model with contributions by Horgan, Hey, Kelly, Reinders,
Gromes, Stamatescu, Stancu, Cutosky, etc., culminating with Isgur and  Karl.
Potential models not based on harmonic confinement were proposed
somewhat later. The references will be given in the section on baryons.

Note also the contribution by Becchi and Morpurgo about the possibility of
describing hadrons made of quarks in a non-relativisitic approximation
\cite{kokkedee1969quark,morpurgo1999relation,Becchi:1965zz,Morpurgo:1965xy}%
.\footnote{I thank Pr.~Morpurgo for a correspondence related to this subject}
\subsection{Heavy quarks}\label{hi:sub:heavy}
The physics of kaons has been very stimulating along the years: strangeness led
to the quarks, the $\theta-\tau$ puzzle led to parity violation and to $K^0\bar
K^0$ mixing, whose detailed scrutiny revealed  $CP$ violation
\cite{pais1988inward}. Another problem
in the weak decay of kaons inspired Glashow, Illiopoulos and Maiani
\cite{Glashow:1970gm}, who
predicted a fourth quark, named ``charmed quark'' and abbreviated as $c$, whose
mass should not be too high.
In some processes, diagrams involving a $u$ and diagrams involving a $c$ cancel
out. This is the GIM mechanism. 

A new spectroscopy was thus predicted, with charmed mesons such as
$(c\bar u)$,
or charmed baryons such as $(csu)$, double-charm baryons, etc. See, e.g.,
Gaillard, Lee and Rosner \cite{Gaillard:1974mw}. Of course, $(c\bar c)$ mesons
were predicted as well. 

However, when in November 1974 (this was sometimes called the October
revolution), the $J/\psi$ was discovered simultaneously at SLAC and Brookhaven,
and the $\psi'$ shortly after at SLAC, they were not immediately recognised as
$(c\bar c)$, i.e., charmonium. The surprise was that they are extremely narrow
resonances. 
We
now understand that the Zweig rule works better and better when the quark
become heavier. The spectrum of charmonium was completed with various P state,
the $\psi''=\psi(3770)$ which is a D state, and after some time, with the
spin-singlet states.

Charmed mesons \cite{Goldhaber:1976xn} and baryons
\cite{Cazzoli:1975et} were discovered as well, and
this sector is now rather rich, tough the double- and triple-charm baryons still
await discovery. 

The charmonium gave a decisive impulse to the quark model in the meson sector. 
Thanks to Regge and others, we had already an idea about sequences of mesons
with increasing spin~$J$. In the quark model, this corresponds to orbital
excitations of the quark--antiquark motion. With charmonium, the new feature is
that \textsl{i)} the levels are better seen, since the lowest states are narrow,
\textsl{ii)} there is a clear evidence for the radial degree of freedom.
Explicit quark models were developed to describe the $(c\bar c)$ spectrum.
They will be reviewed in Sec.~\ref{se:mes}.

At the time of the discovery of charm, in the 70s, there were already
speculations about a symmetry between quarks and leptons. The light quarks
$(u,d)$ are the partners of $(e^-,\nu_e)$. The strange and charmed quarks
belong in the family of $(\mu^-,\nu_\mu$). 

Note that the leptons are ahead. The $\mu^-$ was discovered in
1936,\footnote{In 1947, the discovery of the pion was confused by the decay
of  pions into muons} and  was completely unexpected (``who ordered the
muon? asked Rabbi). The $\tau$ lepton was discovered at SLAC in 1977. The
partners of the $(\tau^-,\nu_\tau)$ pairs were thus anticipated and named
$(t,b)$, as ``top'' and ``bottom''. And a spectroscopy of hadrons containing
$b$ or $t$, or both, was predicted.
However, at a Conference in Hamburg, a German minister who had some
knowledge of English, suggested to replace bottom by ``beauty''. And $t$ was
renamed ``truth quark'', but this is not very often used.

In 1977, Leon Lederman, who missed the charmonium by a small margin, did an
experiment similar to Ting's, but with a more powerful beam and an improved
detector, and announced in 1977 the discovery of the $\Upsilon$ and $\Upsilon'$
particles, immediately interpreted as $(b\bar b)$ bound states
\cite{Herb:1977ek,Innes:1977ae}.
We shall come back in the next sections about the role of flavour independence.
Just a word here. Lederman noticed that, within the precision of his
measurement, $\Upsilon'-\Upsilon\simeq \psi'-J/\psi$, and submitted to
local theorists the question whether there exists a potential such that changing
the reduced mass
keep the spacings $\Delta E$ unchanged (remember that $\Delta E\propto m$ for
the Coulomb potential and
$\Delta E\propto m^{-1/2}$ for the harmonic oscillator). Quigg and Rosner found
that the logarithmic potential has this property for all
spacings~\cite{1977PhLB...71..153Q}. In fact, the
logarithmic potential was already used in empirical speculations about the
not-yet-discovered bottomonium, but this property was not stressed clearly
enough~\cite{Machacek:1976ws}.

Predictions for $(t\bar t)$ were  revised continuously, as the mass of the top
was pushed higher and higher by the negative results of the experimental
searches.
It was then stressed (see, e.g., \cite{Kuhn:1982ua} for an early review) that
if the top quark is very heavy, it will decay before hadronising.
\subsection{Confirmation}\label{hi:sub:conf}
On the way from the early days of the quark model to the recent state of art,
there are many beautiful and decisive contributions that unfortunately, I cannot
review here, due  to the lack of time. I will just list of few of them:
deep inelastic scattering,
the parton model,
the MIT bag model, and its many variants, 
QCD,
its property of asymptotic freedom,
lattice QCD,
QCD sum rules, etc., etc.
\clearpage
%\goodbreak\goodbreak\goodbreak\goodbreak\goodbreak\goodbreak
\markboth{\sc An introduction to the quark model\hspace*{1cm}
}{\sc Mesons}
\section{The quark--antiquark model of mesons}\label{se:mes}
\begin{flushright}
\sl I married them \\
\rm Friar Laurence, Romeo and Juliet
\end{flushright}
\vspace{-1cm}
\subsection{Introduction}\label{mes:sub:intro} 
The quark model of mesons has not been developed ``from the bottom
up'',
but mainly  ``from the charm down to light sector, and from the charm to
the beauty quark''.  

At the end of 1974, when the new particles seen at Brookhaven and SLAC were
identified as $(c\bar c)$ bound states, explicit models were proposed to
calculate the spectrum and the radiative transitions
\cite{PhysRevLett.34.43,PhysRevLett.34.46,PhysRevLett.34.365,PhysRevLett.34.369}
. The potential proposed in \cite{PhysRevLett.34.369} is know as the
funnel potential or Cornell potential, and reads,
\begin{equation}
 \label{mes:eq:funnel} V(r)=-\frac{a}{r}+b\,r+c
\end{equation}
A new era of meson spectroscopy was opened, with explicit calculations of the
meson properties in models which were first guessed empirically and are now
more seriously inspired by QCD.
\subsection{Quantum numbers}
The total angular momentum $J$ results from the coupling of the spin
$\mathsf{S}$ of the
quarks  and their orbital momentum $\ell$,
from which one also gets the parity $(-1)^{\ell+1}$ and charge conjugation
$C=(-1)^{\ell+\mathsf{S}}$. The
lowest states are given in Table~\ref{mes:tab:qn}.
\begin{table}[htp]
 \caption{Quantum numbers of the lowest quarkonium states}
 \label{mes:tab:qn}
\vspace*{.1cm}
 \centering
$
\begin{array}{lllllllllll}
{}^{2\,\mathsf{S}+1}\ell_J & \slj1S0 & \slj3S1 & \slj1P1 & \slj3P0 & \slj3P1 &
\slj3P2 %
&\slj1D2  &\slj3D1 & \slj3D2 & \slj3D3 \\
J^{PC} & 0^{-+} & 1^{--} & 1^{+-} & 0^{++} & 1^{++} & 2^{++} %
&2^{-+} & 1^{--} & 2^{--} & 3^{--}
 \end{array}
$
 \end{table}

Comments are in order
\begin{itemize}
 \item some sets of quantum numbers are absent. For instance, a state with
$J^{PC}=1^{-+}$ would be exotic,
\item some $J^{PC}$ occur twice. There is the possibility of, e.g.,
$\slj3S1-\slj3D1$ mixing, as for the deuteron,
\item the above quantum numbers are repeated for the radial excitations,
labelled with the radial number $n$. In the literature, one can find either
$n=0,1\,\ldots$ as in the familiar one-dimensional oscillator, or
$n=\ell+1,2,\ldots$ as in the Coulomb problem. In the latter case, the first P
state is labelled 2P, the first D state 3D. We shall adopt
$n=1,2,\ldots$,
i.e., a counting where the radial wave function has $(n-1)$ nodes.
\end{itemize}
\subsection{Spin averaged spectrum}\label{mes:sub:central}
With a potential such as \eqref{mes:eq:funnel} which is central, without spin
dependence, the energy depends only on $\ell$ and $n$. The radial equation reads
\begin{equation} \label{mes:eq:radial}
 -u''(r)+\frac{\ell(\ell+1)}{r^2}\,u(r)+ m\,V(r)\,u(r)=m\,E\,u(r)~,
\end{equation}
where $m=m_c$ is the mass of the charmed quark in the case of quarkonium, and
otherwise $m$ is twice the reduced mass.  For a pure Coulomb interaction,
this equation can be rescaled to
$-v''(r)+(\ell(\ell+1)/r^2-1/r-\epsilon)v(r)=0$,
with eigenenergies $\epsilon=-1/(4\,n^2)$, with $n=\ell+1,\ell+2,\ldots$. For a
purely linear interaction, another rescaling leads to
$-v''(r)+(\ell(\ell+1)/r^2+r-\epsilon)v(r)=0$, and in the case where $\ell=0$,
it
reduces to a shifted Airy equation: the eigenenergies are the negative of
the zeros $a_i$ of the Airy function $\mathop{\rm Ai}(x)$, and the
$n^\text{th}$ eigenfunction is just the very same Airy function shifted at
$a_n$, $v_n(r)\propto \mathop{\rm Ai}(r+a_n)$, the normalised version being
\begin{equation} \label{mes:eq:linear-s}
 v_n(r)=\frac{\mathop{\rm Ai}(r+a_n)}{\mathop{\rm Ai}'(a_n)}~.
\end{equation}
If both the Coulomb and the linear terms are present, one can rescale to a
one-parameter problem, which can be chosen as 
\begin{equation} \label{mes:eq:cb+linear}
 \left[-\Delta -\frac{\lambda}{r}+r-\epsilon\right]\psi(\vec r)=0~,
\end{equation}
where $\lambda$ can be expressed in terms of the quark mass $m$ and the
strength parameters $a$ and $b$ of the Cornell potential.

If the two-body problem is solved, one can tune the parameters to reproduce the
low
levels of charmonium. This was done by several groups in the late 70s, and the
authors were able to predict the missing states.

Note that this exercise did not provide with a sharp determination of the
parameters, e.g., $a$, $b$ and $c$ in \eqref{mes:eq:funnel}, and the mass $m_c$
of the charmed quark.  Additional constraints were used, such as the leptonic
widths, and the rates of the $\gamma$-transitions. Still some flexibility was
allowed, and, when the first bottomonium levels were found, the game became more
challenging: to reproduce simultaneously the $(c\bar c)$
and the $(b\bar b)$ spectra.

Indeed, even so the interquark potential was not derived from QCD in early
quarkonium phenomenology,  it was
assumed that it is universal, or ``flavour independent''. In QCD the gluons
couple
to the colour. Hence it is reasonable to assume that the potential is flavour
independent. 

We already mentioned the logarithmic solution for
$V(r)$ \cite{1977PhLB...71..153Q}. Then the radius scales as $m^{-1/2}$, and
the square of the wave function at the origin, $|\phi(0)|^2$, that enter
several decay widths, scale as $m^{-3/2}$. As $\ln r=\lim (r^\alpha-1)/\alpha$
as $\alpha\to 0$, a generalisation consists of a power-law interaction
$V(r)=A\,r^\alpha + B$. Martin used this functional form to fit the quarkonium
levels, and obtained a rather good fit \cite{Martin:1980rm} with a small value
of $\alpha$,
i.e., a potential close to the logarithmic one. See, also, \cite{Muraki:1978yn}.

\begin{figure}[htp]
 \centering
 \includegraphics[width=.5\textwidth]{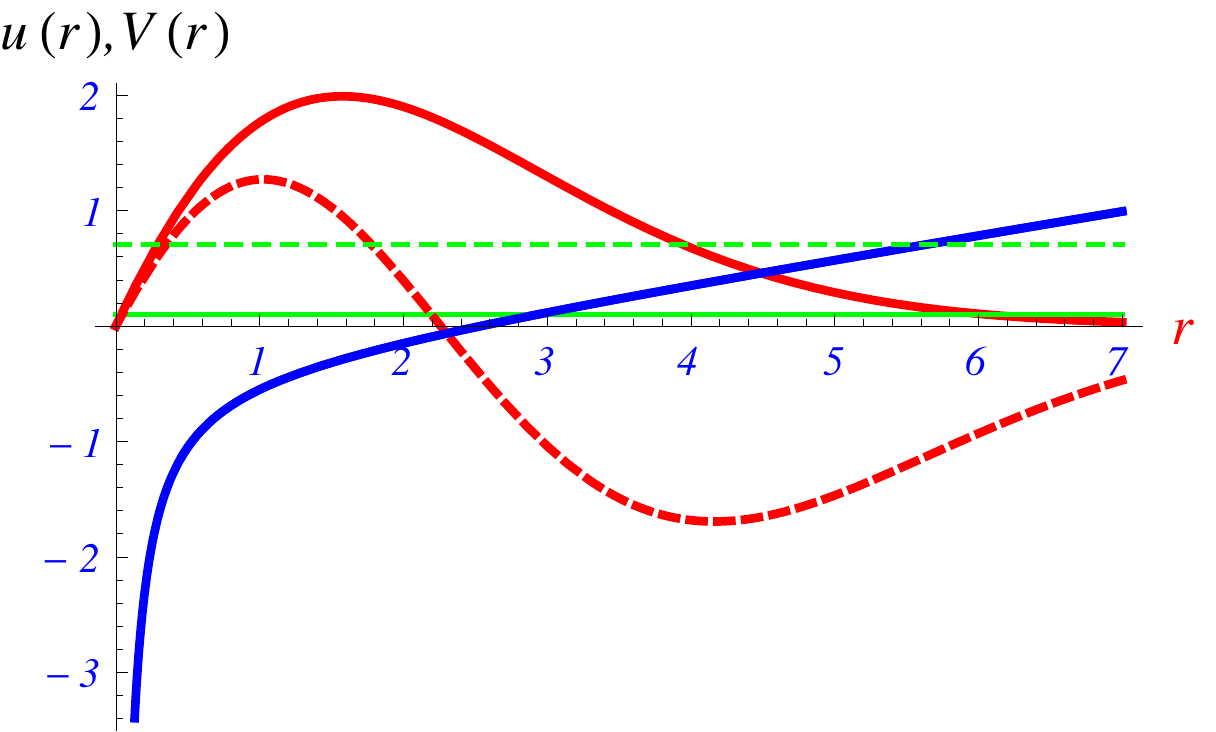}
%{./FigsMath/P12V.pdf}
% PotCornell.pdf: 350x210 pixel, 72dpi, 12.35x7.41 cm, bb=0 0 350 210
\caption{Simple central potential $-0.4/r+0.2\,r-0.35$ and
two first S-wave levels for a quark mass $m_c=1.5$. Units are GeV for $V$,
GeV$^{-1}$ for $r$, and arbitrary for the reduced radial functions $u(r)$.}
 \label{mes:fig:pot}
\end{figure}

A typical charmonium potential is shown in Fig.~\ref{mes:fig:pot}, as well as
the reduced radial wave function $u(r)$ for the 1S and 2S states.  In units of
GeV for $V$ and GeV$^{-1}$ for $r$, it reads $V(r)=-.4/r+0.2\,r-0.35$. It is
just for illustration purpose, without any  attempt to achieve the best fit.

The corresponding results are listed in Table~\ref{mes:tab:QQbar}. This should
be considered as a kind of $0$th order starting point which could be
refined by tuning the parameters, fitting some higher levels, and including
some decay properties in the constraints. 
\begin{table}[htp]
\caption{Rough fit to the spin-averaged levels of quarkonium (for 1D, the
experimental value corresponds to $\psi''$ which is $\slj3D1$, and the state
seen by CLEO \cite{Bonvicini:2004yj} and BABAR \cite{delAmoSanchez:2010kz},
which is presumably $\slj3D2$}
 \label{mes:tab:QQbar}
 $$\begin{array}{lllllllllll}
&\multicolumn{4}{c}{c\bar c}&\multicolumn{5}{c}{b\bar b}\\
&1\mathrm{S} & 2\mathrm{S} & 1\mathrm{P} & 1\mathrm{D} &     1\mathrm{S} &
2\mathrm{S} & 1\mathrm{P} & 1\mathrm{D}& 2\mathrm{P}\\
\mathrm{Model}&3.07 & 3.68 & 3.48 & 3.78 &9.47 &\phantom{1}9.99 & 9.87 &
10.11& 10.23\\
\mathrm{exp.}&3.07 & 3.67 & 3.52& 3.77 & 9.44 &           10.01           & 9.89
& 10.16&10.26
 \end{array}
$$
\end{table}

\begin{figure}[htp!!!]
\begin{center}
\includegraphics{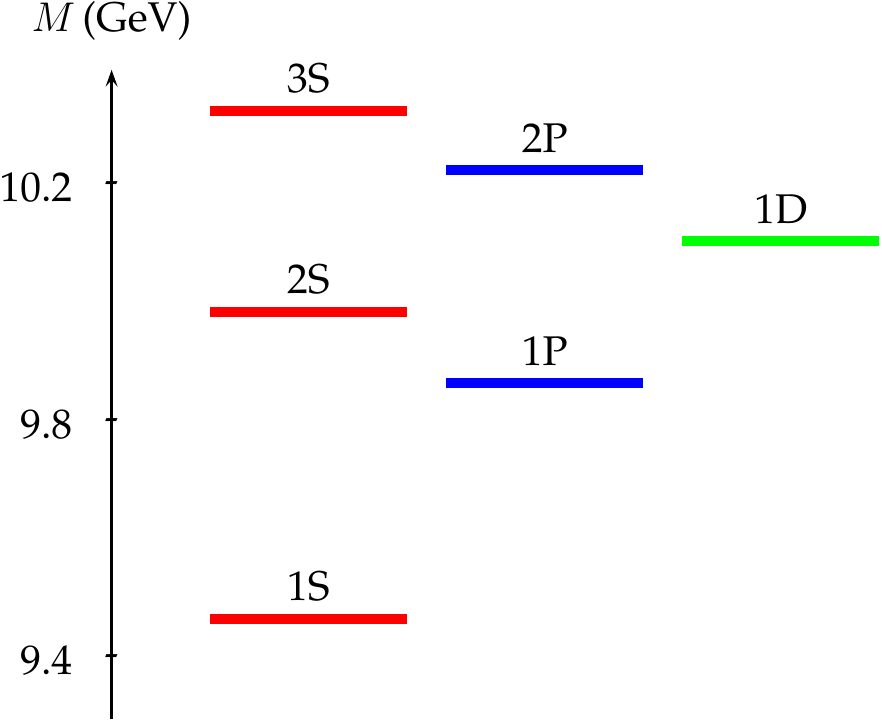}
\end{center}
 \caption{Predictions for the $b\bar b$ spectrum with a very simple potential.}
 \label{mes:fig:bbar}
\end{figure}

In Fig.\ \ref{mes:fig:bbar} are shown the levels of bottomonium using the
simple potential $-0.4/r+0.2\,r-0.35$  and a mass $m_b=4.5\,$GeV. For the very
beginners, its a good exercise to reproduce these numbers. The hierarchy of the
excitations corresponds to the observation.

\begin{figure}[htp]
\begin{center}
\includegraphics{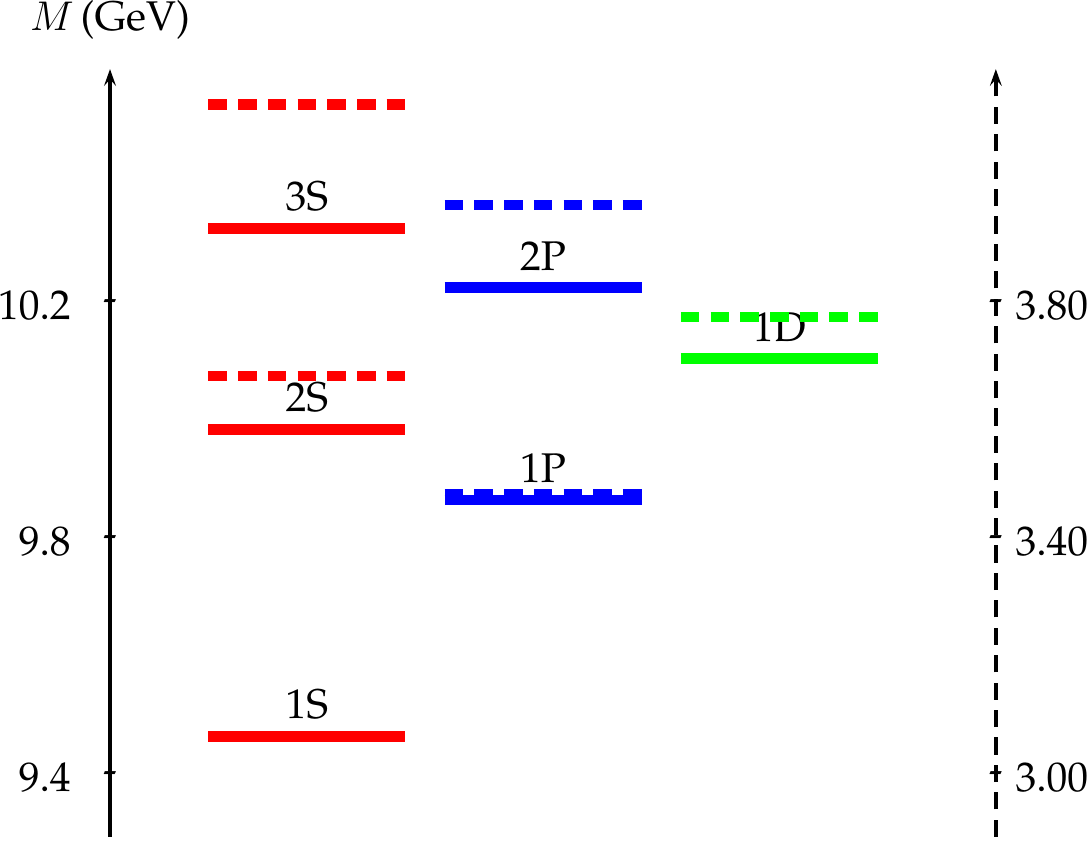}
\end{center}
\caption{Comparison of the predictions for the $b\bar b$ (solid lines) and
$(c\bar c)$ (dotted lines) levels with a very simple potential.}
 \label{mes:fig:bbar-ccbar}
\end{figure}

 In Fig.\ \ref{mes:fig:bbar-ccbar}, we supplement the previous figure by the
charmonium levels, computed with the same simple potential. The 1S levels are
arranged to coincide. The potential is tuned to produce about the same spacing
for the lowest states. But for the higher states, the spectrum 
becomes dominated by the linear part of the interaction, and the spacing is
significantly higher for $(c\bar c)$ than for $(b\bar b)$.

\subsection{Improvements to the potential}\label{mes:sub:imp}
Perhaps, one should not try to improve too much the simple potential model of
mesons. Take for instance proton--nucleus scattering in nuclear physics, where
the so-called
Glauber approximation works very well. For years, many corrections have been
estimated, and each correction gave a large effect. At the end, one understood
that the most important corrections cancel each other. There might be something
similar in the quark model. Anyhow, let us list a few possible improvements.
\subsubsection{More elaborate potentials}
So far, we mentioned the simplest choices for the interquark potential, such as
power-law, logarithm, or Coulomb-plus-linear. 

The success of these empirical potentials, with a short-range part finite or
less singular than $1/r$, and a confining part less sharp than $r$ can be
understood as follows: the simplest interaction corresponds to a linear
confinement, with a cigar-shaped gluon flux linking the two gluons, and a
Coulomb-part to which one-gluon exchange contributes. But the linear part is
smoothed by pair-creation effects; if one tries to increase the separation
between $c$ and $\bar c$, a pair of light quark is created, and this results in
a softening of the interaction \cite{Poggio:1978xg,Gonzalez:2003gx}. At
short-distance, there is the asymptotic freedom, which weakens the Coulomb term,
see, for instance, the potential by Buchmuller et al.~\cite{Buchmuller:1980su}
or Richardson~\cite{Richardson:1978bt}.
\subsubsection{Relativistic corrections}
The simplest potential models use the Schr\"odinger equation. However, the
kinetic energy turns out not to be very small as compared to the rest mass of
the quarks. Models have been devised with a relativistic form of the kinetic
energy
\begin{equation}\label{mes:eq:rel-kin}
\frac{\vec p^2}{2\,m}\to \sqrt{m^2+\vec p^2}-m~,
\end{equation}
See, e.g., Basdevant and Boukraa \cite{Basdevant:1984rk}, Godfrey and Isgur
\cite{Godfrey:1985xj}, etc. 
This is of course just one step towards a fully relativistic and covariant
picture, with retardation effects, etc., one of the most serious attempts being
the work of the Bonn group \cite{Koll:2000ke,Ricken:2000kf}, which later also
described the baryons.
\subsubsection{Strong decay of quarkonia}\label{mes:sub:strong}
As already stressed, the miracle of charmonium and bottomonium is the
suppression of the decay into light hadrons, which proceeds through internal
$Q\bar Q$ annihilation into gluons and subsequent hadronisation of the emitted
gluons. Hence the states below the relevant $(Q\bar q)+(\bar Q q)$ threshold
are narrow, and the states lying above easily decay into a pair of flavoured
mesons, by a Zweig-allowed mechanism.

The most widely used model for describing  $(Q\bar Q)\to (Q\bar q)+(\bar Q q)$
is based on quark pair creation. For, say
\begin{equation}\label{mes:eq:qpc1}
A(q_1,\bar q_2)\to B(q_1,\bar q_3)+C(q_3,\bar q_2)~,
\end{equation}
a prescription is given for the creation of the extra pair $(q_3,\bar q_3)$,
and the amplitude includes the coefficients for the spin--isospin recoupling
and the overlap of the wave-functions, say
\begin{equation}\label{mes:eq:qpc2}
\mathcal{M}=\int\dd\tau \Psi(1,\bar2)
\,\mathcal{O}(3,\bar3)\,\Psi^*(1,\bar3)\,\Psi(3,\bar2)~,
\end{equation}
where $\dd\tau$ is meant for all relative variables. The most popular is the
co-called $\slj3P0$ model proposed by Micu \cite{Micu:1968mk} and extensively
used and developed by the Orsay group \cite{yaouanc1988hadron} and many others.
The Cornell model of charmonium \cite{Eichten:1978tg} is very similar.

The early applications were mostly for nodeless states, for which 
\eqref{mes:eq:qpc2} gives some slight enhancement or suppression of
$|\mathcal{M}|^2$ multiplying pure phase effects. The application to high
charmonium states $\psi^{(n)}$ revealed more dramatic variations. In particular,
if you consider the three OZI-allowed decays
\begin{equation}\label{mes:eq:qpc3}
\psi^{(n)}\to D\bar D,\ D^*\bar D+\mathrm{c.c.},\ D^*\bar D{}^*~,
\end{equation}
each, schematically, call for a certain momentum $p$ in the initial-state wave
function. If $p$ lies at a node or a bump, the transition is suppressed or
enhanced. On the basis of simple spin counting, one expects for
\eqref{mes:eq:qpc3} that the rates, divided by
phase-space, are proportional to 
\begin{equation}\label{mes:eq:qpc4}
R[D\bar D,\ D^*\bar D+\mathrm{c.c.},\ D^*\bar D{}^*]\propto 1: 4 : 7~,
\end{equation}
while the experimental results for $\psi(4.04)$ show a clear dominance of $
D^*\bar D{}^*$. This led to suggest a molecular structure for this
$\psi(4.04)$ \cite{DeRujula:1976qd} (see the section on multiquarks). In fact,
the formalism of
quark-pair creation shows that the decays into $D\bar D$ or $D^*\bar
D+\mathrm{c.c.}$ are
suppressed by the node structure of the initial state
\cite{LeYaouanc:1977ux}.  See also, \cite{LeYaouanc:1977gm,Eichten:1979ms}.
In a more detailed analysis, Eichten et al.~\cite{Eichten:2005ga} and Fernandez
et al.~\cite{Ortega:2012rs} concluded that the situation is neither as simple a
pure
$(c\bar c)$ with nodes or a pure molecule, namely that this state is a
charmonium with an abundant molecular
component.\footnote{I thank F.~Fernandez for an interesting correspondence on
this point}
% Ironically, Entem et al.\ 
% posted very recently a paper on ArXiv \cite{Segovia:2012aa}, concluding that
% this state is more likely a molecule.
% ``Le flux les apporta, le reflux les remporte'' (The incoming tide
% brought them here, the outgoing tide carries them away), Corneille, Le Cid.
%
%
\subsubsection{Coupling to decay channel}
\begin{figure}[htp]
\begin{center}
\includegraphics{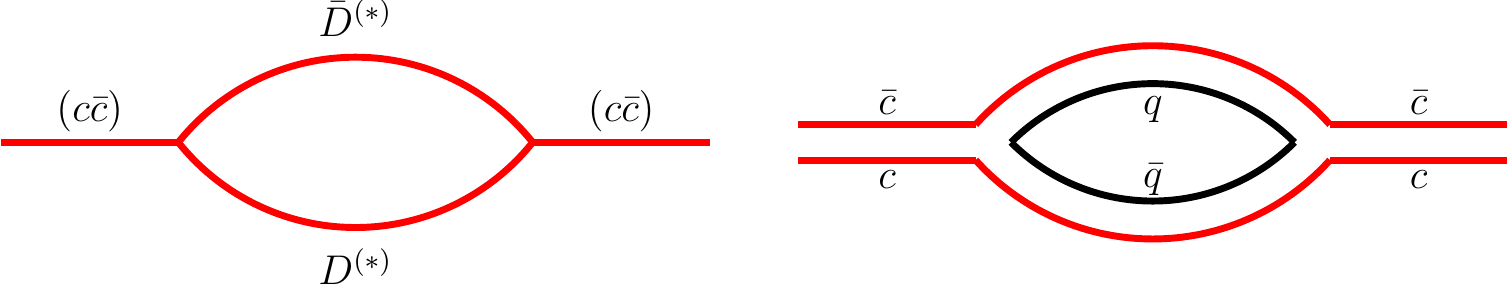}
\end{center}
\caption{Mesonic loop correction to the charmonium, seen at the hadron level
(left) or quark level (right)}
 \label{mes:fig:loop}
\end{figure}

The coupling to $(Q\bar q)+(\bar Q q)$  also influence the energy and the
internal structure of the quarkonia. 
The quark model is just a first approximation, with the minimal component in
Fock space, i.e., the first term in
\begin{equation}\label{mes:eq:fock}
|\psi\rangle=a\,|c\bar c\rangle+\sum_i b_i \,|c\bar cq_i\bar q_i\rangle+\cdots
\end{equation}
with the second term tentatively saturated by the lowest $D^{(*)}\bar D^{(*)}$
component and the $D_s$ analogues. In the model by Eichten et
al.~\cite{Eichten:1978tg,Eichten:1979ms}, these
contributions are generated from the main $c\bar c$ component  by an explicit
operator
which creates a pair of light or strange quarks out of the vacuum. This provides
the states above the threshold with an explicit decay width.  The branching
ratios are found to be in good agreement with the observed ones. There is also a
dispersive
part, i.e., a shift of the energy due to meson loops (see Fig.\
\ref{mes:fig:loop}).  If one
introduces an explicit mass difference between $D$ and $D^*$, these meson loops
contribute to the fine and hyperfine splittings. See also
\cite{Heikkila:1983wd}, and  the recent
contributions by Barnes et al. \cite{Barnes:2007xu}, T. Burns
\cite{Burns:2011fu}, etc., where it is stressed that the loops with $D$, $D^*$,
$D_s$, etc., tend to cancel one another.

However, very close to the $D\bar D$ threshold, there is an effect on the
$\psi'$, which can couple to $D\bar D$, while the $\eta_c'$ only couples to
$D\bar D{}^*+\mathrm{c.c.}$ and $D^*\bar D{}^*$. Hence the $\psi'-\eta_c'$ is
sensitive to the coupling to virtual decay channels
\cite{Martin:1982nw,Eichten:2004uh}.

We shall come back on these meson--meson configurations  in the section on
multiquarks.
\subsubsection{Fine structure}
Three of the four members of the 1P level of $(c\bar c)$ were discovered rather
early, thanks to the transitions
\begin{equation}\label{mes:eq:chi}
\psi'\to\chi_J+\gamma~,\quad \chi_J\to J/\psi+\gamma~,
 \end{equation}
with $J=0,1,2$. The masses have later been measured with a very high precision
using an antiproton beam \cite{Patrignani:2004nf}. 

In the quark model, they are the spin triplet
states $\slj3P0$, $\slj3P1$ and $\slj3P2$ (with a small admixture of $\slj3F2$).
The formalism for the splittings and for the electromagnetic transitions is
adapted from nuclear and atomic physics, and is reviewed, e.g., in Jackson's
lectures at SLAC in 1976 \cite{Jackson:1976mt}.

The potential is written as 
\begin{equation}\label{mes:eq:potspin}
V(r)+V_{ss}(r)\,\vec\sigma_1.\vec\sigma_2+V_{ls}(r)\,\vec\ell.\vec
s+V_t(r)\,S_{12}~,
\end{equation}
where $\vec\ell$ is the orbital momentum, $\vec
s=(\vec\sigma_1+\vec\sigma_2)/2$ the total spin of the quarks, and the tensor
operator is $S_{12}=3\,\vec\sigma_1.\hat r\,\vec\sigma_1.\hat
r-\vec\sigma_1.\vec\sigma_2$. 
If the two last terms in \eqref{mes:eq:potspin} are treated at first order, one
gets the masses
\begin{equation}\label{mes:eq:fine}
\begin{aligned}
 M(\slj3P0)&=M_t- 2\,\langle V_{ls}\rangle -4\, \langle V_t\rangle~,\\[3pt]
 M(\slj3P1)&=M_t- \phantom{1}\, \langle V_{ls}\rangle +2\,\langle
V_t\rangle~,\\[-1pt]
 M(\slj3P2)&=M_t+ \phantom{1}\, \langle V_{ls}\rangle -\frac25\,\langle
V_t\rangle~,
\end{aligned}
\end{equation}
in terms of a an average  triplet mass $M_t$ and two matrix elements. This can
be inverted as
\begin{equation}\label{mes:eq:fine-i}
\begin{aligned}
 M_t&=\frac19\left[M(\slj3P0)+3\,M(\slj3P1)+5\,M(\slj3P2)\right]~,\\
\langle V_{ls}\rangle&=\frac{1}{12}\left[-2\,M(\slj3P0)-3\,M(\slj3P1)+5\,
M(\slj3P1)\right]~,\\
\langle V_{t}\rangle&=\frac{5}{72}\left[2\,M(\slj3P0)-3\,M(\slj3P1)+
M(\slj3P1)\right]~.
\end{aligned}
\end{equation}
Note that if the spin-orbit and tensor terms are treated beyond perturbation
theory, leading to the computed masses $M(\slj3PJ)$, and if $M_t$ still denotes
the computed result with only $V(r)+V_{ss}(r)$, 
then one gets
\begin{equation}\label{mes:eq:fine-ii}
 M_t\ge \frac19\left[M(\slj3P0)+3\,M(\slj3P1)+5\,M(\slj3P2)\right]~,
\end{equation}
i.e., 
$M_t$ is \emph{above} the naive weighted average. This should be kept in mind,
when estimating the sign of $V_{ss}$ by comparing the spin-triplet states with
the spin-singlet state governed by $V(r)-3\,V_{ss}(r)$ (see below the
paragraph on $\slj1P1$).

From the date of the Particle Data Group \cite{Nakamura:2010zzi} on the 1P
levels, of charmonium, \eqref{mes:eq:fine-i} gives
\begin{equation}\label{mes:eq:1Pcc}
M_t=3.525~,\quad 
 \langle V_{ls}\rangle=0.035~,\quad\text{and}\quad 
 \langle V_t\rangle=0.01\,\text{GeV}~.
\end{equation}

For the 1P level of $(b\bar b)$, the result is 
\begin{equation}\label{mes:eq:1Pbb}
M_t=9.900~,\quad 
 \langle V_{ls}\rangle=0.014~,\quad\text{and}\quad 
 \langle V_t\rangle=0.003\,\text{GeV}~,
\end{equation}
and for 2P
\begin{equation}\label{mes:eq:1P}
M_t=10.260~,\quad 
 \langle V_{ls}\rangle=0.009~,\quad\text{and}\quad 
 \langle V_t\rangle=0.002\,\text{GeV}~.
\end{equation}

\subsubsection{Hyperfine splittings}
This is a more delicate issue on the experimental side. The simple mechanisms
explaining the fine structure (see next section) also predicted some hyperfine
splitting, with 
a pseudoscalar state named $\eta_c$ about 0.1\, GeV lighter than its
vector partner  $J/\psi$,.
The $\eta_c$  is the $\slj1S0$ state of $(c\bar c)$, with $n=1$. It was thus
anticipated
that the M1 transition from $J/\psi$ to $\eta_c$ would not have a very large
probability \cite{Feinberg:1975hk,Kang:1978yw}.

A first candidate for $\eta_c$ was claimed in Germany
\cite{Braunschweig:1977dc}, with a surprising
shift $\delta=m(J/\psi)-m(\eta_c)\sim0.3\,$GeV. Leutwyler and Stern
\cite{Leutwyler:1977pv} claimed
that such a large shift was an unavoidable consequence of the relativistic
character of the dynamics.  In contrast, most potential builders predicted a
smaller $\delta$, once the first $\eta_c$ was announced, 
recognized to face difficulties to accommodate such a large value of $\delta$
\cite{Schnitzer:1975ux,Schnitzer:1976td,Pham:1977px,%
Khare:1978hh,Krammer:1979gr}.

The German $\eta_c$ was not confirmed \cite{Partridge:1979tn}, and instead, a
more plausible one was
seen at 2.980\,GeV \cite{Himel:1980dj,Partridge:1980vk}, and confirmed in other
places, including LEP and
antiproton--proton collisions. The corresponding splitting is $\delta\simeq
0.116\,$GeV. 

The $\eta_c(2S)$, sometimes named $\eta_c'$, was then predicted with a
splitting $\delta'=\psi'-\eta'_c$ of about 80\,MeV. The ratio $\delta'/\delta$
is mainly governed by the ratio $R=|\phi_2(0)|^2/|\phi_1(0)|^2$ of the wave
function at the origin. This ratio can be estimated in potential models (for
instance, this ratio is exactly $R=1$ for a purely linear potential). It can
also be deduced from the ratio of the leptonic widths. The $\eta_c(2S)$  was
unfortunately forgotten in the elaborate Cornell model with explicit account for
the coupling to the virtual decay channels \cite{Eichten:1979ms}. This omission
was repaired in
\cite{Martin:1982nw,Eichten:2004uh}, and the effect of meson loops found to be
sizeable. The $\eta_c(2S)$ does not
couple to the nearby $D\bar D$, but the $\psi'$ does, and is pushed down. This
reduces $\delta'$. The $\eta_c(2S)$ was eventually found at BELLE, BABAR and
CLEO,  in
$B$ decay, in $\gamma\gamma$ spectra and double-charmonium production
\cite{Choi:2002na,Aubert:2003pt,Asner:2003wv}. 
It means that if one looks at 
$e^+e^-\to J/\psi+X$ and believes  the Zweig rule, $X$ is dominated by
$(c\bar c)$, and, indeed, already known states have been seen in $X$ as
striking peaks. The $\eta_c(2S)$ was confirmed in some other experiments. The
splitting is $\psi'-\eta_c'=0.049\,$GeV.

From the very beginning, it was suspected that the $h_c$, corresponding to the
$\slj1P1$ state of $(c\bar c)$, will be very difficult to produce. See, e.g.,
Renard \cite{Renard:1976up}. In most models, $h_c$ is predicted to lie very
close to the centre of
gravity of the spin-triplet, $M_t$. A first indication was provided by the R704
experiment at CERN, the last experiment at the ISR ring \cite{Baglin:1986yd}.
The $h_c$ was later found at
CLEO (Cornell) \cite{Rubin:2005px} and E835 (Fermilab) \cite{Andreotti:2005vu}.
The mass of $h_c$ coincides with the centre
of gravity of the triplet, and there is presumably a cancellation of several
small terms. 

The analogues in the $(b\bar b)$ family were also hard to find. We have now
$\eta_b$,
$h_b(1P)$ and $h_b(2P)$ in the particle listings \cite{Nakamura:2010zzi}. Very
recently, CLEO \cite{Dobbs:2012zn}
reported a measurement of $\eta_b(2S)$ and a confirmation of $\eta_b(1S)$. The
results are 
\begin{equation}
 \label{mes:eq:etab-hb}
\delta(1S)=0.067~, \quad
\delta(2S)=0.049\,\quad
h_b(1P)=9.898~,\quad
h_b(2P)=10.260\,\text{GeV}~.
\end{equation}

Two remarks are in order: 
\begin{enumerate}
 \item The value of $\delta(1S)$ for  $(b\bar b)$ was predicted by HPQCD
\cite{Dowdall:2011wh} to  be
$70\pm9\,$MeV. If you cannot afford lattice QCD, you can attempt a poor-man
estimate from $(c\bar c)$. Assuming a logarithmic potential,
with the distances scaling as $m^{-1/2}$,
and a hyperfine splitting proportional to $|\psi(0)^2|^2/m^2$, one gets
\footnote{I remember doing this computation using the cell phone of Kamal Seth
in a fancy restaurant of Munich}
\begin{equation}\label{mes:eq:etab-log}
 \delta(1S)_{b\bar b}=\delta(1S)_{c\bar
c}\,\genfrac{(}{)}{}{}{m_b}{m_c}^{-1/2}\sim
65\,\mathrm{MeV}~.
\end{equation}
\item
The ratio of $\delta(1S)$ to $\delta(2S)$ drops from about 2.3 to about 1.4 in
$(c\bar c)$ to $(b\bar b)$. This reflects how anomalously high is the
$\eta_c(2S)$ due to its vicinity of the OZI threshold.
\end{enumerate}
\subsubsection{The origin of the spin-dependent forces}
The phenomenology of the spin-dependent forces in charmonium was greatly
inspired by QED and by nuclear forces.

It was early stressed that the one-gluon-exchange contribution to the $c\bar c$
interaction is very similar to the Coulomb interaction in QED, and thus is
associated to the same spin corrections. The name ``chromomagnetism''
was given to the analogue of the magnetic interaction.
We shall come back on the important
role of chromomagnetism in the systematics of hyperfine splittings in ordinary
hadrons \cite{DeRujula:1975ge} and in speculations about possible multiquarks
\cite{Jaffe:1999ze}.

The one-meson-exchange picture of the
nuclear forces, which was very popular in the 60s and early 70s. The
nucleon--nucleon potential was written $V=\sum V_i$, where each $V_i$
corresponds to a single partial wave in the $t$-channel, and thus gives a very
specific spin dependence in the $s$-channel.

Schnitzer \cite{Schnitzer:1975ux,Schnitzer:1976td}, for instance, analysed the
splittings assuming a vector meson
exchange, and a scalar confinement. Thus the $1/r$ terms gives spin-spin,
tensor and spin-orbit terms similar to the terms describing the fine and
hyperfine structure of atoms, or similar to the terms associated with $\omega$
exchange in nuclear forces. The term proportional to $r$ gives a negative
spin-orbit force. Several variants were discussed by various authors
\cite{Khare:1978hh,Pumplin:1975cr,Pignon:1978ck}.

The method to reach full consistency of this approach was explained by
Gromes (see, e.g.,\cite{Lucha:1991vn} for references) and Eichten and Feinberg
\cite{Eichten:1980mw}. The problem is that the effective potential in the
Schr\"odinger equation contains two types of contributions: an intrinsic spin
dependence due to the nature of the exchanged object, and a relativistic effect
known as Thomas precession. Even a scalar interaction, when reduced to an
effective non-relativistic potential, contains a central term and a related
spin-orbit one. 
\subsubsection{Orbital mixing}
The states with natural parity, except $\slj3P0$, contains two partial waves,
which can mix. Let us consider $J^{PC}=1^{--}$. The wave--function reads
\begin{equation}\label{mes:eq:orb-m1}
 \psi=\frac{u(r)}{r} |\slj3S1\rangle+\frac{w(r)}{r}|\slj3D1\rangle~,
\end{equation}
and after some algebra, it can be shown that the coupled radial equations read
\begin{equation}\label{mes:eq:orb-m2}
\begin{aligned}
  -\frac{u''(r)}{m}+V_c(r)\,u(r)+\sqrt8\,V_t(r)\,w(r)&=E\,u(r)~,\\
-\frac{w''(r)}{m}+\left[\frac{6}{m\,r^2}+V_c(r)-3\,V_{ls}(r)-2\,
V_t(r)\right]w(r)+\sqrt8\,V_t(r)\,u(r) &=E\,u(r)~                  
\end{aligned}
\end{equation}
first derived many years ago for the deuteron. Here $V_C=V+V_{ss}$ is the
spin-triplet central potential.  Solving  \eqref{mes:eq:orb-m2} requires a
regularisation of $V_t(r)$ which behaves as $r^{-3}$ at short distances in most
models. Often the result, for e.g., $\psi''$ is written as
\cite{Rosner:2001nm}
\begin{equation}\label{mes:eq:orb-m3}
\psi(3770)=a|\slj3D1,n=1\rangle+b_1|\slj3S1,n=1\rangle+b_2|\slj3S1,
n=2\rangle+\cdots
\end{equation}
where the states in the r.h.s.\ are obtained by neglecting the coupling of the
two equations. One could debate whether $b_2$ is larger than $b_1$ because
$\psi'$ lies very close to $\psi''$, or $b_1$ is larger due to the
favourable node  structure of $J/\psi$ and $\psi''$. This is
not obvious: in atomic physics, the analogue of \eqref{mes:eq:orb-m3} requires a
summation over all
$b_i$ (in the discrete and continuous spectrum). See, e.g.,
\cite{Richard:1979zn,Richard:1979fc}.

The orbital mixing is responsible for the coupling of $\psi(3770)$ to $e^+e^-$
that made its observation possible. It might also influence the pattern of its
hadronic decays. See, e.g., the discussion by Rosner \cite{Rosner:2001nm}.

Very likely, a good fraction of the orbital mixing is due to the coupling to
decay channels. For instance, the coefficients in \eqref{mes:eq:orb-m3} are
estimated in \cite{Eichten:2004uh}, which uses an improved version of the
Cornell model. 
\subsection{Summary for heavy quarkonia}
\begin{figure}[htp]
 \begin{center}
 \includegraphics[width=.75\textwidth]{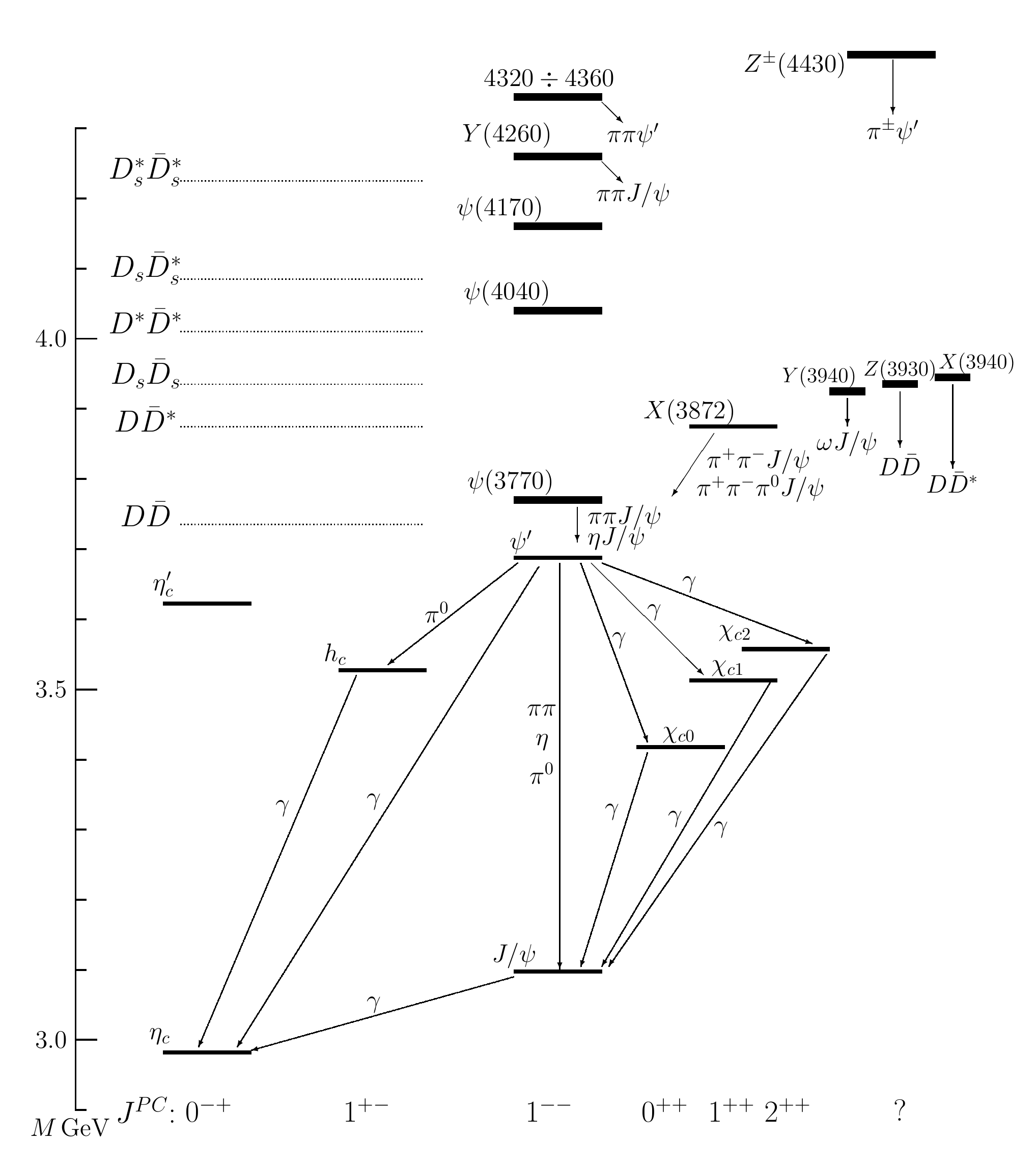}
%./SourcesArticles/VoloshinProgress/spectrum.pdf}
 % spectrum.pdf: 586x662 pixel, 72dpi, 20.67x23.35 cm, bb=0 0 586 662
\end{center}
 \caption{Levels of charmonium, and transitions among them (from
\cite{Voloshin:2007dx}).}
 \label{mes:fig:charmonium}
\end{figure}

In Fig.\ \ref{mes:fig:charmonium}, reproduced from \cite{Voloshin:2007dx} with
the kind permission of the author, are shown the levels of charmonium. The
states labelled $X$, $Y$, $Z$ will be discussed in another section.
It is remarkable to see exactly the levels predicted by the quark model. On the
left, the spin singlet states, with relative angular momentum $\ell=0$ for the
two $\eta_c$, and $\ell=1$ for $h_c$. In the middle, the various states with
spin triplet, and $\ell=0,1$, and even one example with $\ell=2$, and the
various recoupling of spin and orbital momentum. The hierarchy of orbital vs.\
radial excitations, and the patterns of fine and hyperfine splittings are now
well described with simple potentials that are now better understood from the
underlying QCD.
\begin{landscape}
\begin{figure}[htp]
\begin{center}
\scalebox{.6}[.8]{\includegraphics{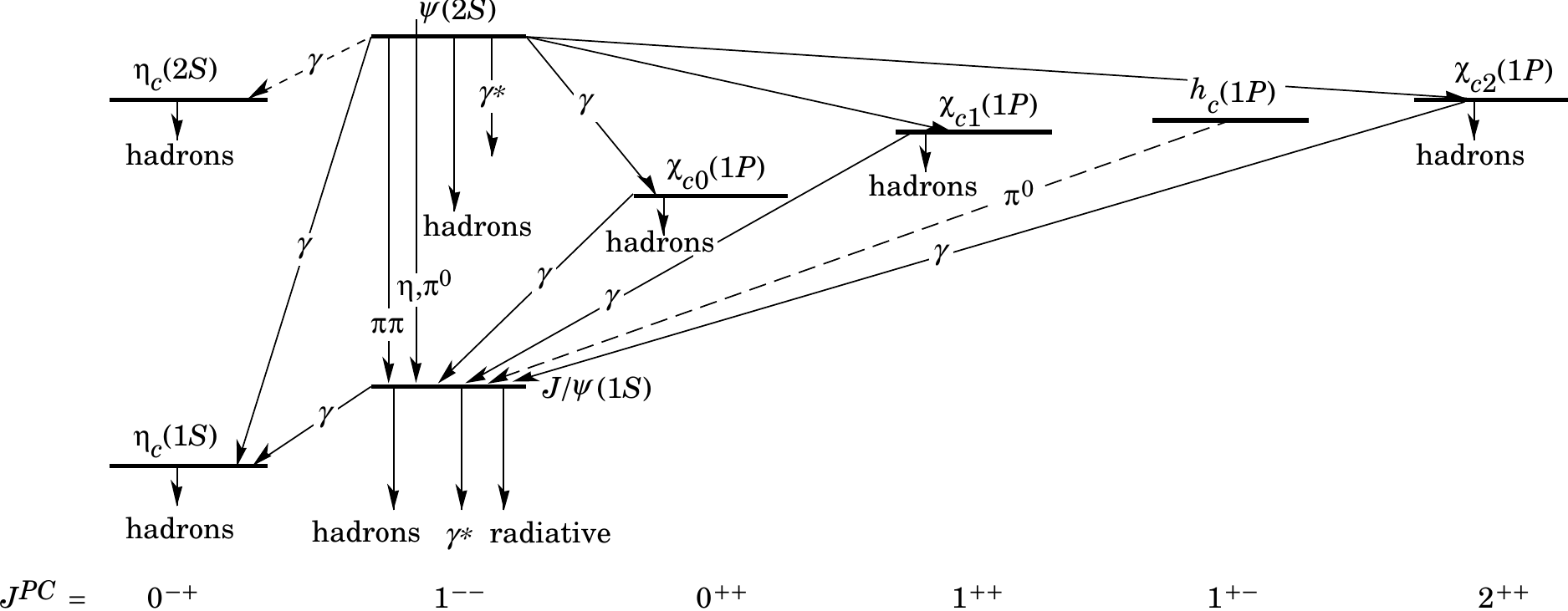} }
\scalebox{.6}[.8]{\includegraphics{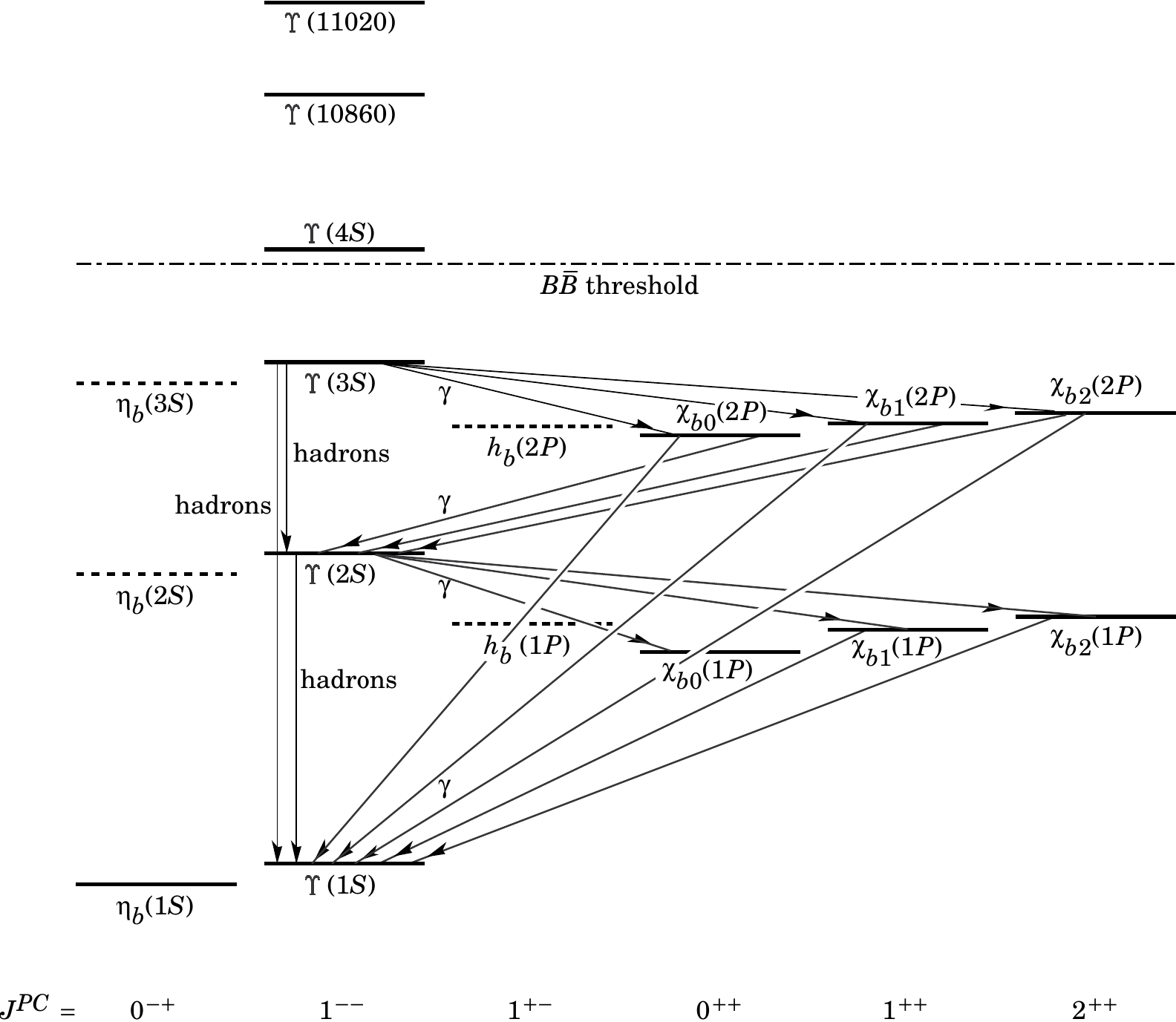} }
\end{center}
 \caption{Charmonium and bottomonium spectra. Borrowed from PDG
\cite{Nakamura:2010zzi}. Some states indicated by a dotted line have been
identified recently.}
 \label{mes:fig:heavyQQ}
\end{figure}
\end{landscape}
In Fig.\ \ref{mes:fig:heavyQQ}, we reproduce the lower part of the charmonium
spectrum, and the bottomonium spectrum, borrowed from the Review of Particle
Properties \cite{Nakamura:2010zzi}. 

The spectra are very similar. This is due to the property of flavour
independence. The differences are also understood, due to explicit
mass-dependence of the fine and hyperfine terms.

\subsection{Light mesons}\label{mes:sub:light}
The simple quark model is in principle not applicable to light quarks, but some
reckless attempts were encouraging. For instance, Martin \cite{Martin:1980rm}
managed to fit $(b\bar b)$, $(c\bar  c)$, $(s \bar s)$ and $(c\bar s)$ with a
single potential. The game was extended further by Bhaduri et al.\
\cite{Bhaduri:1981pn}, however their uniform regularisation of the hyperfine
interaction gave poor predictions for $J/\psi-\eta_c$ and similar splittings. 
Their model was refined by several authors, see, e.g.,~\cite{Brau:2004ie}, with
a regularisation of spin forces that depends on the system under consideration. 

Isgur et al.\ \cite{Capstick:1986kw} have explained why  there are some
good reasons to link potential models, with a minimal amount of relativity, to
the underlying QCD.

Let us give two examples. The first positive-parity excitations of $(q\bar q)$
with $I=1$ are $a_0(980)$, $b_1(1235)$, $a_1(1260)$ and $a_2(1320)$. In the
quark model, they correspond to the partial wave $\slj3P0$, $\slj1P1$,
$\slj3P1$ and $\slj3P2$. The level order of this 1P multiplet, and even the
pattern of spacings is very similar to what is observed in the charmonium 1P
states. 

If one looks at meson with high spin $J$ and plot the square mass $M^2$ against
$J$, one finds an almost perfect linear behaviour for this ``Regge
trajectory''. If a linear potential is plugged into a Schr\"odinger equatiion
with relativistic kinematics, this linear behaviour is reproduced. 

\subsection{Heavy-light mesons}\label{mes:sub:h-l}
This is an even more dangerous playing ground. Remember, for instance, that an
electron is more relativistic in the hydrogen atom, $(p,e^-)$, than in the
positronium atom, $(e^+,e^-)$. There is nevertheless a very interesting
spectroscopy of $D$ and $B$ mesons. 

In first approximation, the reduced mass governing the internal dynamics is the
same for $(c\bar q)$ and $(b\bar q)$ mesons, so the excitation spectra should
be very similar. This is rather obvious in potential models, but it remains
valid much beyond this framework. This is one of aspects of ``heavy quark
symmetry''~\cite{Isgur:1989vq,Isgur:1989ed,Neubert:1993mb}. 
\subsection{Mathematical developments}\label{mes:sub:math}
The quark model motivated some studies about the properties of the Schr\"odinger
operator with confining interactions. See, e.g.,
\cite{Quigg:1979vr,Grosse:847188} for
a review and references.

Among the subjects, let us mention the following.
\subsubsection{Level order}
A sufficient condition has been derived that transforms the Coulomb degeneracy
\begin{equation}\label{mes:eq:Coul-deg}
E(1,\ell)=E(2,\ell-1)=\cdots=E(\ell+1, 0)~,
\end{equation}
into a series of inequalities: the condition is related to the sign of the
Laplacian of the
potential. As $\Delta V>0$ for the funnel potential \eqref{mes:eq:cb+linear} or
any plausible interquark potential, this theorem explains the observed
${\rm 1P<2S}$
ordering in quarkonium, and ${\rm 2P<3S}$ in bottomonium. Note that $\Delta
V\le0$
for the last electron of an alkalin atom and
$\Delta V\ge0$ for a muon inside a nucleus: this explains the observed breaking
of the Coulomb degeneracy for alkalin atoms and muonic atoms.

Similarly, the sign of the second derivative of the potential as a function of
$r^2$ governs the breaking of the HO degeneracy,
\begin{equation}\label{mes:eq:HO-deg}
E(1,\ell)=E(2,\ell-2)=E(2,\ell-4)=\cdots ~.
\end{equation}
This explains why ${\rm 2S<1D}$, i.e., $m(\psi')<m(\psi'')$.

There are also results on the spacing of levels as a function of the radial
number. For the harmonic oscillator, there is a remarkable equal spacing,
$E(n+1,\ell)-E(n,\ell)=E(n,\ell)-E(n-1,\ell)$. For a linear potential, the
spacing decreases. 
\subsubsection{The wave function at the origin}
Many decay widths are proportional to the square wave function at the origin,
which for S-states, reads 
\begin{equation}\label{mes:eq:wforigin}
p_n=|\Phi_n(0)|^2=\frac{1}{4\,\pi} u'_n(0)^2~.
\end{equation}
if the reduced radial wave function is arranged to be real. The studies on
quarkonium gave the opportunity to remember the rule by Schwinger (see, e;g.,
\cite{Quigg:1979vr})
\begin{equation}\label{mes:eq:Scwhinger}
u'(0)^2=2\,\mu\,\int_0^\infty \frac{\dd V}{\dd r} u^2(r)\,\dd r~.
\end{equation}
This explains why the $p_n$ are independent of $n$ for a purely linear
interaction, and, conversely, why the normalisation factor in
\eqref{mes:eq:linear-s} is
given by the derivative.

There are sufficient conditions ensuring that $p_n$ increases or
decreases as a function of $n$ \cite{Grosse:847188}.
\subsubsection{The dependence upon the masses}
If flavour independence is accepted as a strict rule, one has to study how the
energy levels and the wave functions evolve when the reduced mass is modified.

As the kinetic energy term is $\vec p^2/(2\,\mu)$, obviously $E\searrow$ as
$\mu\nearrow$. The effect is pronounced when going from $(c\bar c)$ to $(b\bar
b)$. On the other hand, the reduced mass governing $(Q\bar q)$ does not change
much from $Q=c$ to $Q=b$. Hence the OZI threshold $(Q\bar q)+(\bar Q q)$ has
almost the same energy for $Q=b$ and $Q=c$. This explains why the $\Upsilon(1S)$
is more deeply bound than
$J/\psi$ with respect to this threshold, and why there are more narrow states in
the bottomonium spectrum than in the charmonium one. 

One can go further and notice that in a strictly flavour-independent
interaction, the inverse reduced mass enters linearly. There is a known result
that for an Hamiltonian depending linearly on a parameter, say
$H=A+\lambda\,B$, the ground state energy $\epsilon_0$, or the sum of first
energies $\sum_{i=0}^N \epsilon_i$ is a concave (or convex upward) function of
$\lambda$. Once the constituent masses are added, this implies for the
ground state of each angular-momentum sector, 
\begin{equation}\label{mes:eq:BM}
 (Q\bar Q)+(q\bar q)\le 2(Q\bar q)~,
\end{equation}
an inequality independently derived by Martin and Bertlmann
\cite{Bertlmann:1979zs,Grosse:847188} and Nussinov
\cite{Nussinov:1983hb,Nussinov:1999sx} for potential models, and extended much
beyond this framework \cite{Witten:1983ut,Nussinov:1999sx}. Early applications
introduced dangerously light quarks, but still, the
inequality is satisfied for the spin-triplet states of $(c\bar c)$, $(s\bar s)$
and $(c\bar s)$. Safer is the prediction that the (spin averaged) ground
state of $(b\bar c)$ is above the middle of $(c\bar c)$ and $(b\bar b)$. 

Note that in atomic physics, it is reasonable to believe that the
Born--Oppenheimer potential that binds di-atomic molecules remains unchanged if
the nuclei are replaced by isotopes. Thus \eqref{mes:eq:BM} holds for H$_2$,
D$_2$ and HD, and this governs their abundance at thermodynamic equilibrium
\cite{reif1965fundamentals}.
\subsubsection{Concavity with respect to the spin
coefficients}\label{mes:subsusb:spin-conc}
In the same vein, one can explain the sign of the error when treating the spin
corrections to first order. 
For the hyperfine effect of S-state alone the singlet and triplet Hamiltonians
are respectively, in an obvious notation,
\begin{equation}\label{mes:eq:Hs-Ht}
H_s=H_0-3\,V_{ss}~,\quad H_t=H_0+ V_{ss}~,
\end{equation}
so that the fictitious eigenstate of $H_0$ alone lies at
\begin{equation}\label{mes:eq:Es-Et}
E_0\ge\frac14 E_s+\frac34 E_t~.
\end{equation}
Note that the inclusion of the tensor force would slightly lower $E_t$, with 
$\slj3S1$ becoming $\slj3S1-\slj3D1$.

It is straightforward to extend this to the case of two parameters. Consider the
generic spin-triplet Hamiltonian 
\begin{equation}\label{mes:eq:Htriplet}
H_J=H_t+\alpha_J\,V_{ls}+\beta_J\,V_t~,
\end{equation}
so that $H_0+3\,H_1+5\,H_2=9\,H_t$, then the fictitious spin-triplet state free
of spin-orbit and tensor corrections lies at
\begin{equation}\label{mes:eq:Ecdm}
E_t\ge\left[E_{J=0}+3\,E_{J=1}+5\,E_{J=2}\right]/9~,
\end{equation}
i.e., above the naive centre of gravity, as discussed earlier in this section. 
Thus if the $\slj1P1$ is found at
the location of the naive of gravity, it has received some \emph{attractive}
spin--spin downward shift.\footnote{I remember discussions on  this point at a
Workshop
organised in Genoa, to celebrate years of charm physics with antiprotons, and
later  with Kamal Seth.}

%
%\clearpage
\markboth{\sc An introduction to the quark model\hspace*{1cm}
}{\sc Baryons}
\section{Baryons}\label{se:bar}
\vspace*{-.2cm}
\begin{flushright}
 \sl Tres faciunt collegium\footnote{Three makes a company}\\
\rm Latin sentence
\end{flushright}
\vspace*{-1.2cm}
\subsection{Introduction}\label{bar:sub:intro}
The quark model of baryons was developed in the 60s, in particular by
Dalitz and his collaborators, and several other groups, at a time when 
many nucleon and hyperon resonances were already known.

These early studies, and the more recent work by Isgur and Karl, Gromes et al.,
Stancu et al., Cutosky et al., cannot be separated from the harmonic oscillator 
(HO) model, which provides a powerful classification scheme, and an efficient
tool to implement the symmetry constraints. This is reviewed, e.g., by Hey and
Kelly~\cite{Hey:1982aj}.

The HO model has reached a high degree of refinement, with all the
possible corrections carefully listed, and their tentative interpretation in
QCD (one-gluon-exchange).

There has also been attempts to solve the three-body problem, using known
techniques developed in atomic or nuclear physics, such as Faddeev equations, 
hyperspherical harmonics or variational methods. Then any
potential can be used, for instance a funnel interaction with a superposition
of Coulomb and linear contributions.

Note that in most early studies, it was implicitly assumed that the potential
is pairwise, and the question has been addressed of the link between the
interquark potential in baryons and the quark--antiquark potential in mesons.
This will be discussed in Sec.~\ref{bar:sub:mes-bar}. In fact, it has been
anticipated for many
years, that the confining interaction is of three-body nature, i.e., depends
simultaneously upon all relative distances. This makes the calculations
technically more involved, but makes a link with the lattice QCD which favours
such potentials \cite{Takahashi:2000te}.

The role of relativistic effects is of course important.  In the latest works by
Isgur et al., a relativistic kinematics was adopted \cite{Capstick:1986bm}.
There are even models based on the Bethe--Salpeter equation
\cite{Metsch:2003ix}.

\subsection{Jacobi coordinates}\label{bar:sub:jaco}
\begin{figure}[htp]
 \centering
\includegraphics{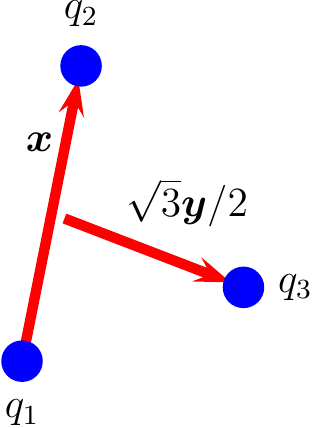}
 \caption{Jacobi coordinates for the relative motion of the quarks inside a
baryon}
 \label{bar:fig:jac}
\end{figure}

In the case of equal masses, a set of Jacobi coordinates that diagonalise the
kinetic energy is (see Fig.~\ref{bar:fig:jac})
\begin{equation}\label{bar:eq:jac1}
 \vec R=\frac{\vec r_1+\vec r_2+\vec r_3}{3}~,\quad
\vec x=\vec r_2-\vec r_1~,\quad
\vec y=\frac{2\,\vec r_3-\vec r_1-\vec r_2}{\sqrt3}~,
\end{equation}
where the factor in $\vec y$ makes it easier to implement the symmetry
constraints, as explained shortly. Then, once the centre of mass motion is
removed, the intrinsic Hamiltonian reads
\begin{equation}\label{bar:eq:h3}
 h=\frac{\vec p_x^2}{m}+\frac{\vec p_y^2}{m}+V(\vec x,\vec y)~,
\end{equation}
where the potential energy, already written as translation-invariant, has to be
scalar.

In the case of two different masses, say $(m,m,M)$, $\vec R$ is modified, but
one can keep $\vec x$ and $\vec y$ as above. Then the intrinsic Hamiltonian
becomes
\begin{equation}\label{bar:eq:h3p}
 h=\frac{\vec p_x^2}{\mu_x}+\frac{\vec p_y^2}{\mu_y}+V(\vec x,\vec y)~,
\end{equation}
with the reduced masses now given by
\begin{equation}\label{bar:eq:redmass}
 \mu_x=m~,\quad \mu_y^{-1}=(m^{-1}+2\,M^{-1})/3~.
\end{equation}
If the three masses are unequal, the Jacobi coordinate $\vec y$ becomes
$\vec y\propto (m_1+m_2)\vec r_3-m_1\,\vec r_1-m_2\,\vec r_2$, and it is
straightforward to derive the reduced masses $\mu_x$ and $\mu_y$ as functions
of the individual masses $m_i$.

Later in this section, we shall elaborate more on the potential energy $V$. Let
us just mention here
two popular choices. The harmonic oscillator corresponds to
$V=K(r_{12}^2+r_{23}^2+r_{31}^2)$, where $r_{ij}=|\vec r_{ij}|$ and $\vec
r_{ij}=\vec r_j-\vec r_i$. It becomes $V=3\,K (\vec x^2+\vec y^2)/2$ for
$(m,m,m)$ and $(m,m,M)$, and a more complicated quadratic form of $\vec x$ and
$\vec y$ for $(m_1,m_2,m_3)$. The pairwise models reads
$v(r_{12})+v(r_{23})+v(r_{31})$. 
\subsection{Permutation symmetry}\label{bar:sub:perm}
Schematically, the wave function of a baryon in the quark model is
\begin{equation}\label{bar:eq:fullwf}
 \Psi=\psi(\vec x,\vec y)\,\psi_s\,\psi_i\,\psi_c~,
\end{equation}
with orbital, spin, isospin and colour parts, and this wave function has to be
antisymmetric with respect to the  permutation of the identical quarks. 

If there are two identical quarks, as in $\Xi^-=(ssd)$ or in $\Lambda=(uds)$ in
the limit where isospin symmetry is exact, then each term in
\eqref{bar:eq:fullwf} is either symmetric or antisymmetric. For the ground
state of $\Lambda$, with isospin $I=0$, the colour, spin and isospin parts are
antisymmetric and the orbital part symmetric. This means spin 0 for the light
quarks, and thus $s=1/2$ for the three quarks, isospin $I=0$, and $\psi(\vec
x,\vec y)$ even
in $\vec x$. For instance, in the harmonic oscillator, $\psi(\vec x,\vec
y)\propto\exp[-a\,\vec x^2-b\,\vec y^2]$. 

There are two type of P-states: one
where the degree of freedom associated to $\vec\lambda$ is excited, with the
same symmetry pattern as the ground state. Another wave where the light quarks
have their spin coupled to $s_{qq}=1$, and the orbital wave function $\psi(\vec
x,\vec y)$ is odd in $\vec x$. In the harmonic oscillator, the orbital wave
functions are of the type
\begin{equation}\label{bar:eq:Lambda1P}
\vec y\,\exp[-a\,\vec x^2-b\,\vec y^2]~,\qquad
\vec x\,\exp[-a\,\vec x^2-b\,\vec y^2]~.
\end{equation}
For the former, the quark spin $\mathsf{s}=1/2$ and the angular momentum
$\vec\ell_y$
 gives a baryon spin $J=1/2$ or $J=3/2$. For the latter, both $\mathsf{s}=1/2$
and $\mathsf{s}=3/2$ are possible, and thus a variety of values for the spin
$J$.
Usually, the spin orbit splittings are small in baryons (see below), and the
states with various combinations of spins and angular momenta are nearly
degenerate. One exception is the $\Lambda(1405)-\Lambda(1520)$ pair, which has
motivated an abundant literature. The most plausible explanation is the
coupling of one of these states to a nearby threshold.

For three identical quarks, imposing the constraints of antisymmetrisation is
more delicate. The colour coupling $3\times3\times3\to1$ is antisymmetric, thus
the rest of the wave function should be symmetric.  This was, indeed, one of
the motivations for introducing colour. For the $\Omega^-$, this is rather
simple: the spin wave-function is symmetric, isospin does not exists, and the
orbital wave function is symmetric. For instance, $\psi(\vec x, \vec
y)=\exp[-a(\vec x^2+\vec y^2)]$ in the HO case. This is similar for the $\Delta
(1232)$ multiplet, as the isospin coupling $1/2\times1/2\times1/2\to3/2$ is
symmetric. 

For the first parity excitations of $\Omega^-$, with $J^P=(1/2)^-$ or $(3/2)^-$,
the spin and the
orbital wave functions are neither symmetric nor antisymmetric, but their
combination is symmetric. One cannot escape here to introduce the concept of
\emph{mixed symmetry}. The prototype is given by the Jacobi coordinates $\vec
x$ and $\vec y$ of \eqref{bar:eq:jac1}. The permutation $P_{12}$ reveals $\vec
x$ odd and $\vec y$ even, but a circular permutation $P_\to$ gives
\begin{equation}\label{bar:eq:circperm}
P_\to[\vec y+i\,\vec x]= j\, [\vec y+i\,\vec x]~,
\end{equation}
where $j=\exp(2\,i\,\pi/3)$, as usual. Obviously, if two pairs, say $z=v+i\,u$
and $Z=V+i\,U$, have the same permutation properties as $\vec z=\vec y+i\,\vec
x$, then their coupling give a symmetric term, and antisymmetric one, and a new
pair of mixed symmetry (again in the complex notation), respectively
\begin{equation}\label{bar:eq:coupl-mix-s}
\begin{aligned}
\RE[z\,Z^*]&=u\,U+v\,V~,\\
\IM[z\,Z^*]&=v\,U-u\,V~,\\
[z\,Z]^*&=(u\,U-v\,V)- i\,(u\,V+v\,U)~.
\end{aligned}
\end{equation}
In the case of the P-states of $\Omega^-$, we have a pair of mixed-symmetry
orbital wave functions, $\{\psi_x,\psi_y\}$, with either $\vec x$ or $\vec y$
excited, which for the harmonic oscillator read $\{\psi_x,\psi_y\}=\{\vec x,\vec
y\}\,\exp[-a(\vec x^2+\vec y^2)]$, and a pair of spin 1/2 wave functions
\begin{equation}\label{bar:eq:spin12}
 S_x=\frac{1}{\sqrt2}\left[
\uparrow\downarrow\uparrow-\downarrow\uparrow\uparrow\right]~,\quad
 S_y=\frac{1}{\sqrt6}\left[
2\uparrow\uparrow\downarrow-\uparrow\downarrow\uparrow-
\downarrow\uparrow\uparrow \right]~,
\end{equation}
which bears an obvious analogy with the Jacobi variables \eqref{bar:eq:jac1}.
With a spin-independent interaction, the P-states of $\Omega^-$ have a wave
function of the type $[\psi_x\,S_x+ \psi_y\,S_y]/\sqrt2$.

For the nucleon ground state, the orbital part is symmetric, but the spin and
isospin parts are both of mixed symmetry, with \eqref{bar:eq:spin12} for the
spin, its analogues $\{I_x,I_y\}$ for isospin, and the combination
$[S_x\,I_x+S_y\,I_y]/\sqrt2$ being symmetric. 

The most interesting example of baryon in  this respect consists of the
antisymmetric spin--isospin wave function $[S_x\,I_y-S_y\,I_x]/\sqrt2$ which
requires a fully antisymmetric orbital wave function. This wave function is
excited in any interquark separation. For instance, in the specific case of
the
harmonic oscillator, it reads $\vec x\times \vec y\,\exp[-a(\vec x^2+\vec
y^2)]$, i.e., $\ell^P=1^+$. This is one of the states predicted in the
three-quark model and absent in the quark-diquark model. See below the section
on diquarks.
\subsection{Solving the three-body problem for baryons}\label{bar:sub:modHO}
\subsubsection{The harmonic oscillator}
After rescaling, the linear oscillator reads $h=p^2+x^2$ with energies
$\epsilon_n=1+2\,n$ and wave functions $\varphi_0(x)\propto\exp(-x^2/2)$ and 
$\varphi_n(x)\propto[-\dd /\dd x+ x]^n\,\varphi_0(x)$.

For equal masses $m$, the three-body problem with  $V=2\,K/3
(r_{12}^2+r_{23}^2+r_{31}^2)$ is governed by the Hamiltonian
\begin{equation}\label{bar:eq:HOeqm}
H(m,m,m)=\frac{\vec p_x^2}{m}+\frac{\vec p_y^2}{m}+K(\vec x^2+\vec y^2)~,
\end{equation}
which corresponds to a sum of two independent spatial oscillator, or six linear
oscillators. The energies 
$E=(6+4\,n_x+2\ell_x+4\,n_y+2\,\ell_y)\sqrt{K/m}$ have a cumulated   number of
excitations $N=2\,n_x+\ell_x+2\,n_y+\ell_y$.
Here, the radial and orbital excitations in each variable are counted
separately. 

There is a considerable degeneracy of the levels of given $N\ge2$, which
contains contains several $J^P$ states. The literature on baryons within HO has
developed a systematic notation which is used beyond this model. In particular,
the multiplets are denoted $[d,\ell^P]^{(n)}$, where $d$ is the spin--flavour
multiplicity, $\ell$ the total orbital momentum, and $(n)$ the radial occurrence
($n=0$ is omitted). For instance, the ground state is labelled as $[56,0^+]$,
and the Roper resonance and its partners $[56,0]'$. The ground state contains
the octet of Fig.~\ref{hi:fig:octet} and the decuplet of
Fig.~\ref{hi:fig:decuplet}, i.e., $2\times 8 + 4\times 10=56$ states. The
first orbital excitation is $[70,1^-]$, and the first state with a full
antisymmetric orbital wave function is $[20,1^+]$.

The case of two different masses reads
\begin{equation}\label{bar:eq:HOmmM}
H(m,m,M)=\frac{\vec p_x^2}{m}+\frac{\vec p_y^2}{\mu}+K(\vec x^2+\vec y^2)~,
\end{equation}
with $\mu$ given by \eqref{bar:eq:redmass}. There is still an exact decoupling,
and the energies are 
\begin{equation}\label{bar:eq:EmmM}
E(m,m,M)=\sqrt\frac{K}{m}(3+4\,n_x+2\,\ell_x)+
\sqrt\frac{K}{\mu}(3+4\,n_y+2\,\ell_y)~.
\end{equation}
Here  the lowest excitation is explicitly
associated with the heavier quark(s),i.e., this is an excitation in the $_vec
x$ variable. This point will be further discussed in the section on
double-charm baryons. 

The case of three different masses is also solvable exactly, though this is
slightly less known.\footnote{I remember discussions on this point with J.-L.
Basdevant, and with D.O. Riska}\@ Once the kinetic energy is written as
\eqref{bar:eq:h3p}, one can use a rescaling $\vec x\to \vec x/\sqrt{\mu_x}$ and 
$\vec y\to \vec y/\sqrt{\mu_y}$. The Hamiltonian becomes
\begin{equation}\label{bar:eq:HOm1m2m3}
H(m_1,m_2,m_3)=\vec p_x^2+\vec p_y^2+A\,\vec x^2+B\,\vec
y^2+2\,C\,\vec x\,\vec y~,
\end{equation}
and one is left with the diagonalisation of a $2\times 2$ matrix.
\subsubsection{First-order perturbation around the harmonic oscillator}
We now turn to more general models for the interaction. The most spread method
consists of writing the potential as
\begin{equation}\label{bar:eq:HOpert}
V=K(\vec x^2+\vec y^2)+\delta V~, 
\end{equation}
and treat the second term as a perturbation. If $K$ is optimised, this is the
first step of a variational expansion, to be discussed below. This is of course
not very accurate, especially at short distances. If $\delta V$ tentatively
pushes down the Roper resonance (to be discussed shortly) from the $N=2$ level
of
the harmonic oscillator to the vicinity of the $N=1$ level, then perturbation
theory is in principle not applicable, as it requires that the shifts are small
compared to the initial spacings. 

Isgur and Karl \cite{Isgur:1978wd} have noticed 
the nice patterns of the $N=2$ multiplet splitting when perturbed by a
pairwise $\delta V$. It is illustrated in Fig.~\ref{bar:fig:N2pert}. 
Gromes and Stamatescu \cite{Gromes:1979xn} pointed out that the upper part
survives if the
HO is perturbed by a symmetric 3-body potential, while the $[56,0^+]'$
decouples. 
The same conclusion was reached by Taxil et al.~\cite{Richard:1989ra}, in the
more general situation of a nearly hypercentral potential.

The upper
part of Fig.~\ref{bar:fig:N2pert} fits rather well the spin-averaged values of
the observations.
\def\phpr{\phantom{'}}
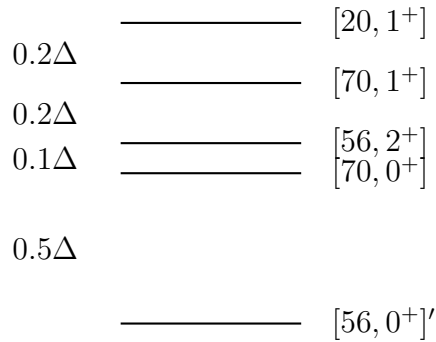
\begin{figure}
\setlength{\unitlength}{4cm}
\begin{center}
\begin{picture}(1.5,1.5)(0,0)
\thicklines
\put(0.3,.1){\line(1,0){.6}}
\put(0.3,.6){\line(1,0){.6}}
\put(0.3,.7){\line(1,0){.6}}
\put(0.3,.9){\line(1,0){.6}}
\put(0.3,1.1){\line(1,0){.6}}
\put(0,.35){\makebox(.1,0){$0.5\Delta$}}
\put(0,.65){\makebox(.1,0){$0.1\Delta$}}
\put(0,.80){\makebox(.1,0){$0.2\Delta$}}
\put(0,1.00){\makebox(.1,0){$0.2\Delta$}}
\put(1.1,.1){\makebox(.15,0){$[56,0^+]'$}}
\put(1.1,.6){\makebox(.15,0){$[70,0^+]\phpr$}}
\put(1.1,.7){\makebox(.15,0){$[56,2^+]\phpr$}}
\put(1.1,.9){\makebox(.15,0){$[70,1^+]\phpr$}}
\put(1.1,1.1){\makebox(.15,0){$[20,1^+]\phpr$}}
\end{picture}
\end{center}
\caption{\label{bar:fig:N2pert}
Splitting pattern of the $N=2$ multiplet for a two-body perturbation of the
harmonic oscillator. [Check whether the 2-body character is required.] There is
an overall shift (not shown), and, in addition, a shift $0.1\Delta$,
$0.2\Delta$, $0.3\Delta$ or $\pm 0.5\Delta$ of the sublevels.}
\end{figure}

This pattern and the one for the $N=3$ multiplet have been further discussed by
Taxil et al.\ \cite{Richard:1989ra}, Stancu et al.\ \cite{Stancu:1991cz}, etc. 
\subsubsection{Converged variational methods}
Let us consider the ground state with equal masses, but the generalisation to
unequal masses is straightforward.  The above method can be read as the first
term in a systematic
expansion
\begin{equation}\label{bar:eq:HOexp}
\Psi(\vec x,\vec y)=\sum_n c_n\,\phi_n(\vec x,\vec y)~, 
\end{equation}
where $n$ denotes the number of excitations and other quantum numbers
associated to the solutions $\phi_n(\vec x,\vec y)$ of the harmonic
oscillator. The matrix elements can be calculated with subtle recursion
relations.  The convergence as a function of the maximal $n$ is excellent for
the energy but elusive for the short-range correlations.

A variant is the so-called Gaussian expansion, widely used in theoretical
chemistry and other quantum problems. It reads
\begin{equation}\label{bar:eq:Gauss-exp}
\Psi(\vec x,\vec y)=\sum_i \gamma_i\,\exp[-(a_i\,\vec x^2+b_i\vec
y^2+2\,c_i\,\vec x.\vec y)]~.
\end{equation}
Note that the permutation symmetry is violated in an individual term  if
$a_i\neq b_i$ and $c_i\neq 0$, but
is restored in the summation either explicitly (correlated Gaussians) or by the
numerical adjustment. Different techniques are used
\cite{Suzuki:1998bn,Hiyama:2003cu} to avoid redundancies in the expansion and
guide towards an efficient optimisation. 
\subsubsection{Faddeev equations}
The Faddeev equations for the three-body problem have been written for
short-range potentials in momentum space. Later, they have been translated into
position space and adapted to the confining interactions. They imply a
summation
over the internal angular momentum, but even with the crudest truncation, they
provide an excellent wave function. The first paper (to my knowledge)  where
the Faddeev equations are applied to baryons is \cite{SilvestreBrac:1985ic}.
See, also, \cite{Richard:1992uk}.
\subsubsection{Hyperspherical expansion}
The 6-dimensional vector $\{\vec x,\vec y\}$, car be written
$(\varrho,\Omega_5)$ in spherical coordinates. The five angles $\Omega_5$
include the physical angles $\hat x$ and $\hat y$ and $\arctan(y/x)$. The
method consists of expanding $\Psi(\vec x,\vec y)$ into partial waves in this
space. For a harmonic oscillator, $V\propto \varrho^2$, and one recovers the
familiar results. Otherwise, one gets an infinite set of coupled equations
\begin{equation}\label{bar:eq:Hyper-exp}
 -u_{[L]}''(\varrho)+ \frac{(L+3/2)(L+5/2)}{\varrho^2}\,u_{[L]}+\sum_{[L]'}
V_{[L],[L]'}(\varrho)\,u_{[L]'}(\varrho)=E\,u_{[L]}(\varrho)~,
\end{equation}
where $[L]$ denotes the ``grand'' angular momentum and its associated magnetic
numbers.

Keeping only one term in the expansion, the lowest hyperspherical harmonics,
gives the hyperscalar approximation, which is already very good, as often
stressed by the Genoa group \cite{Giannini:2003xx}. The first application of
this method to
baryons is by Guimaraes et al.\ \cite{Giumaraes:1981gn}, and Hasenfratz et
al.~\cite{Hasenfratz:1980ka}. The convergence was studied
in the thesis work by Taxil, and is summarised in
\cite{Richard:1983mu,Richard:1992uk}.

The splitting pattern of Fig.~\ref{bar:fig:N2pert} has been re-analysed, with
the starting
point being now the hyperscalar approximation.  The upper part is not modified,
but the radial excitation now decouples. This gives more freedom to push this
state down, but this is not sufficient.  H\o gaasen and I have noticed that the
same hyperspherical potential $V_{00}$ governs the ground state, its radial
excitations and the first negative parity
excitation. The effective angular momentum in the radial equation is $\ell=3/2$
for the former, and $\ell=5/2$ for the latter. Then for any reasonable model
for the interquark interaction, the radial excitation is \emph{above} the
orbital one \cite{Hogaasen:1982rb}.  The result relies on the local character of
the
interaction and
remains valid for local three-body terms. It
does not hold if the potential is velocity or spin dependent.\footnote{I thank
W.~Plessas for an enlightening  discussion on this point, many years ago.}

\subsection{Light baryons}
This is the first sector to which the quark model was applied. One cannot
compare the refinement of the model and the quality of the fit with the
analogues for heavy quarkonia. Nevertheless, the quark model explains the rough
features of the observed spectrum, in the sense that any low-lying resonance
corresponds to a level predicted in the quark model.

Among the problem is that of the Roper resonance. This denotes the state
$N^*(1440)$ with the same $I=1/2$ and $J^P=(1/2)^+$ as the nucleon.
Experimentally it lies at about the same mass as the orbital excitations of
negative parity, perhaps slightly below. In the simplest quark model, it is a
radial excitation. In the harmonic oscillator, it belongs to the $N=2$
multiplet and the first orbital excitation to $N=1$. With a standard Coulomb +
linear interaction, with an accurate treatment of the three body problem, there
is no way to make these states degenerate. This is a serious limitation for the
naive quark model, which misses the relativistic effects and chiral dynamics.
For instance a model with Goldstone-boson exchange, proposed by Glozman and
Riska \cite{Glozman:1995fu}, and further developed by the Graz group
\cite{Glozman:1997ag,Glozman:1998xs}, gets the Roper
resonance right. See, also, \cite{Cano:1998wz}, and the comments by 
Isgur on this approach \cite{Isgur:1999jv}.
In recent lattice simulations with large pion mass (this facilitates the
computation, see next lectures), the same ordering as potential models is
obtained. When the pion mass decreases (and the cost of the computations
increases), there is a dramatic change in the level ordering, and the
experimental one is  recovered~\cite{Lin:2011da}.
\subsection{Missing states and the diquark alternative}
Conversely, there are several instances where a state predicted by the
three-quark model is missing in the experimental spectrum. As often stressed by
Isgur and Karl (their papers in the bibliography and private communications),
most experimental data deal with pion or photon scattering off a nucleon. This
favours states with one quark excited and the two remaining quarks nearly
untouched, and suppresses states with double internal excitation. However,
experiments with higher statistics should be able to
see the missing states.

A drastic alternative is the diquark models. I should better say the diquark
model\emph{s}, as there exists different points of view, and also, the model is
regularly reinvented by authors who do not quote the pioneering papers. For a
rather comprehensive review, see \cite{Anselmino:1992vg}.

Diquarks
have been used for many purposes, in particular
multiparticle production. In spectroscopy, the quark--antiquark model retains
only those baryons consisting of a compact diquark and a quark. The question is
whether the diquark is postulated as an elementary constituent or somehow
generated dynamically. 

Martin has given a partial answer \cite{Martin:1985hw}: for three-quark systems
with high
orbital momentum $\ell$, the ground state consists of two quarks mainly in
S-wave with the third quark carrying most of the orbital momentum, while for
low $\ell$, the three quarks share $\ell$ symmetrically. This explained, at
last, a long standing issue: the Regge trajectories of mesons and baryons are
linear and have the same slope. 

A  digression here about the Regge trajectories. They were invented to explains
some properties of potential scattering, and later, in a different form, for
the high-energy scattering of hadrons. In this later context, a Regge
trajectory is the squared mass of hadrons, as a function of the spin $J$. The
observed trajectories are beautifully linear. For mesons, this was interpreted
a flux of constant section linking the quark to the antiquark.

An intriguing observation is that the baryons also belong to linear
trajectories, with the same slope. This is of course natural in the
quark--diquark model, as
the energy is that of colour 3 linked to a colour $\bar 3$. We now understand
that this occurs spontaneously.

A question that is sometimes eluded is whether diquark are just an effective
degree of freedom that simplifies the dynamics of baryons, or a constituent to
be used on the same footing as quarks in the ``lego'' puzzle of hadrons.
Recently, Maiani and his collaborators described the new $X$, $Y$, $Z$
resonances as diquark--antidiquark states \cite{Maiani:2005qv}, as done earlier
for baryonium.
Frederikkson et al.\ \cite{Fredriksson:1981mh} raised the important question for
the three-diquark systems
(though there is a slight confusion with the quantum numbers, the paper remains
relevant): do we have a ``demon'' deuteron made of three diquark, in addition
to the ordinary deuteron made of two baryons?
\subsection{Baryons with a single heavy flavour}
Some years ago, this sector was marginal for quark dynamics, with only the
ground state of $(Quu), (Qud), \ldots, (Qss)$. Of course, the measurement of
the lifetimes and semi-leptonic branching ratios were very exciting, to compare
the role of $W$-emission, $W$-exchange, $W$-formation in various flavour
configurations. But this is out of the scope of this review. 

In recent years, a lot of  data came from CLEO, B-factories, Tevatron, etc., and
we expect many results from LHCb. It has even been suggested that we can learn
more on
light quark dynamics for the $(Qqq)$ systems than from $(qqq)$.  This is a
domain where the data on charm are or will be confirmed by the data on beauty. 

Just to give an example, take $\Omega_b$. When its mass was
published by D0 \cite{Abazov:2008qm}, it was a surprise that
$\Omega_b-\Lambda_b$ was found 
significantly higher than $\Omega_c-\Lambda_c$. See, e.g.,
\cite{Klempt:2009pi,Dorigo:2009iq}. The CDF collaboration got a
lower mass \cite{Aaltonen:2009ny}, as this is now confirmed by LHCb
\cite{LHCb-CONF-2011-060}. The mass of this state
was predicted by Lipkin et al.\ \cite{Karliner:2008sv}\footnote{I thank Marek
Karliner for a
correspondence on this point.}, on the basis of the chromomagnetic interaction
\cite{De Rujula:1975ge,Jaffe:1999ze}. See, also,
\cite{Jenkins:2007dm,PhysRevD.79.014502}. The lesson is that the spectrum has
regularities, due in particular to flavour independence, and any departure
should receive an explanation or disappear.

\begin{figure}[htp]
\begin{center}
\includegraphics{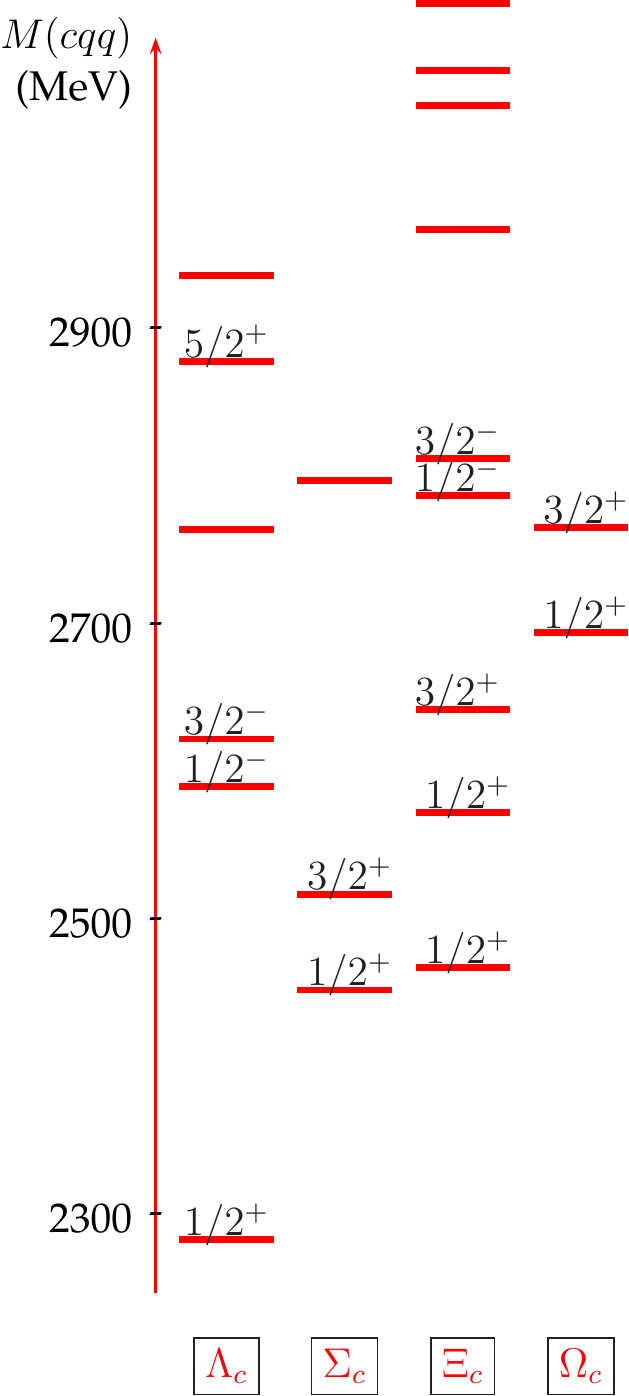}
 \hspace{.5cm}
\includegraphics{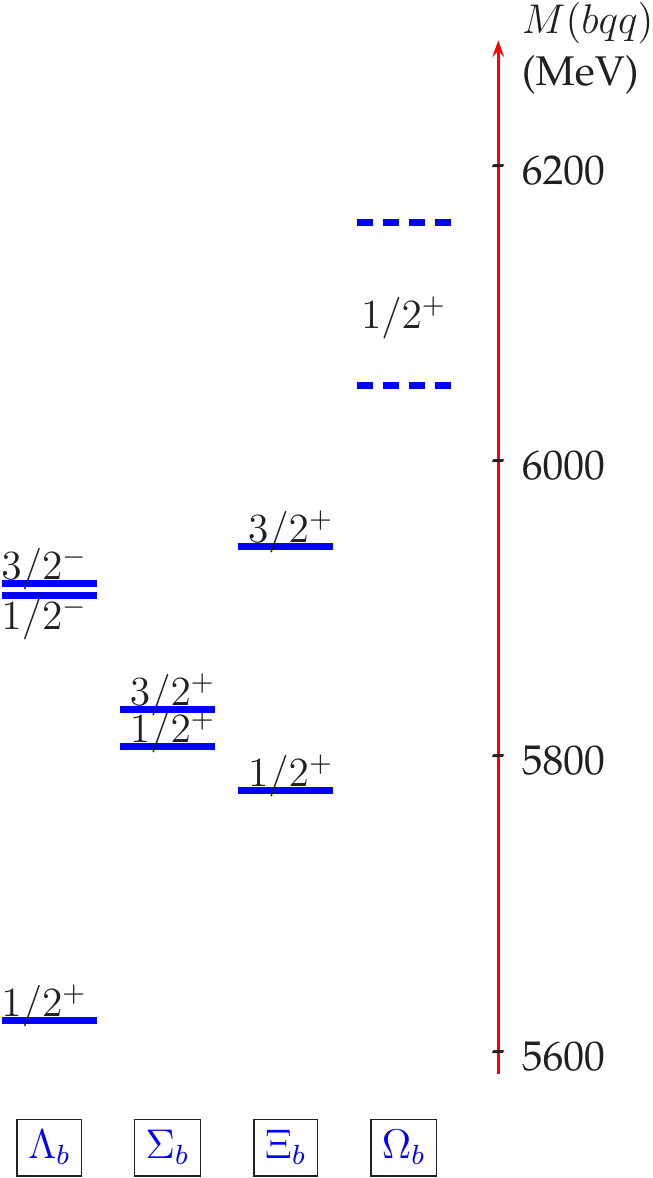}
\end{center}
 \caption{Comparison of $(cqq)$ and $(bqq)$ experimental spectra. Note that some
of the quantum numbers have not been established, and are the quark model
predictions}
 \label{bar:fig:singleQ}
\end{figure}

The spectrum of single-charm and single-beauty baryons are shown in
Fig.~\ref{bar:fig:singleQ}. It includes the very recent discoveries at LHC
\cite{Aaij:2012aa,Chatrchyan:2012ni}. The spectra are arranged so that
$\Lambda_c$ and $\Lambda_b$ coincide. Clearly the excitation energies are very
similar.

\subsection{Baryons with two heavy quarks}\label{bar:sub:double}
It becomes intriguing or even annoying that no $(QQq)$ has been
firmly identified yet. We have just the results by the SELEX collaboration at
Fermilab \cite{Mattson:2002vu,Ocherashvili:2004hi}, and negative results in a
few other attempts. As Brodsky
often insists in Conference talsk or private communications, see, e.g.,
\cite{Brodsky:2011zs}, it is important to detect
double-charm baryons. We also advocate that one should look for  double charm
baryons, or even, in the same experiments, at ``ordinary'' double-charm baryons
and possible double-charm mesons to be discussed in the next chapter. The
production of two $c\bar c$ pairs in $e^+e^-$ collisions occurs at a rate that
is somewhat larger than expected. It has been used to analyse the mass spectrum
of $e^+e^-\to J/\psi+X$, where $X$ is mainly $c\bar c$, thanks to the OZI rule.
This led to the discovery of $\eta_c(2S)$. One should reasonably expect that
sometimes the two pairs hadronise as $(ccq)+(\bar c \bar q\bar q)+ (\bar c q)+
\cdots$.

The dynamics of $(QQq)$ baryons is rather rich, with, in the same hadron, the
relativistic motion of a light quark around a coloured source, as in $D$
mesons, and the slow motion of two heavy quarks, as in charmonium. The $(QQq)$
baryons were listed in the pioneer paper by Gaillard et al.\ on charm physics
\cite{Gaillard:1974mw}. The first detailed study in potential models was given
in \cite{Fleck:1989mb} and has been improved by several authors, see, e.g.,
\cite{Gershtein:2000nx}

At first sight, a quark--diquark approximation seems tempting. Indeed, the
average distance between the two heavy quarks, $r_{QQ}$, is smaller than the
distance $r_{Qq}$ between one of the heavy and the light one. However, if the
first excitations occurs within the $QQ$ subsystem, and in the quark--diquark
picture, one should reconstruct a new diquark for each of them.

An alternative  approach is the Born--Oppenheimer approximation, as done for
H$_2{}^+$
in atomic physics. For a given $\vec x$, one estimates the light-quark
energy and deduce the effective $QQ$ interaction%
\footnote{One actually works with the Jacobi variables and thus the recoil
effects are properly included}. 
If this procedure is applied to a simple quark model, one
can compare the results to the exact numerical solution of the three body
problem \cite{Fleck:1989mb}, and it turns out that the Born--Oppenheimer
approximation is extremely accurate%
\footnote{With an energy slightly below the exact one, unlike a
variational approximation}. Now, the effective $QQ$ potential of $(QQq)$ can be
calculated on the lattice.

The problem is therefore where are the first excitations of this system. The
question has been raised in several recent papers, for instance Cohen, Roberts,
etc.\ \cite{Cohen:2006jg,Eakins:2012jk}. The answer governs the strategy to be
adopted for these systems. 

If the $(QQ)$ subsystem enters the regime where the Coulomb interaction
$\propto \alpha_s(M)/r$ dominates, then the spacing will behave as
$\alpha_s^2\,M$ (not $\alpha_s\,M$ as suggested by a misprint in
\cite{Eakins:2012jk}). Hence the diquark is frozen in its ground state, and the
dynamics is that of a light antiquark around a colour $\bar 3$ source. This is
very similar to the physics of flavoured mesons $(\bar Q q)$, and this symmetry
is promising \cite{Cohen:2006jg,Eakins:2012jk}.

However, the Coulombic limit of heavy-quark spectroscopy is rather elusive, as
noted years ago by \cite{Billoire:1977mp}. For instance, if one uses the simple
quarkonium potential \eqref{mes:eq:funnel} and study the level spacing
(2S)--(1S) of $(Q\bar Q)$ as a function of the quark mass $M$, one finds that it
decreases for $M\lesssim 6$ and increases for $M\gtrsim6\,$GeV. So, for the
double
charm baryons, the first levels are probably excitations in the diquarks, and
can be described in the Born--Oppenheimer limit with an effective $QQ$
interaction.

\subsection{Triple heavy flavour}\label{bar:sub:ccc}
A classic introduction is a paper by Bjorken \cite{Bjorken:1985ei} who named
this sector ``the
ultimate deal of baryon spectroscopy''. This is the analogue of charmonium for
baryons, i.e., a sector where the heavy-quark dynamics is probed
without any complication due to light quarks. There are already several
predictions for the lowest states. Of course, $(bbb)$, $(bbc)$, etc., states
are expected as well.

One of the issues is whether $(ccc)$ has s normal hierarchy of excitations,
with first the orbital excitation with negative parity, and next the radial
excitation which is the analogue of the Roper resonance. For light baryons, an
anomalous ordering is observed, probably due to the chiral dynamics of light
quarks, which cannot be mimicked by a static, spin-independent potential
\subsection{Spin splittings}\label{bar:sub:spin}
As for mesons, there are different possible approaches, and some compromises.

Thanks to the charismatic role of Isgur and Karl, the chromomagnetic
interaction is the best known mechanism to explain the spin-splittings of
ground state baryons. It even works in sector where it should be prohibited.
Take $\rho-\pi$: it gets a colour factor $-16/3$, a spin factor $4$
($\vec\sigma_1.\vec\sigma_2=+1$ for triplet and $-3$ for singlet), and
$|\phi(0)|^2=\langle \delta^{(3)}(\vec r)\rangle$ and results in about 0.6\,GeV.
For $\Delta-N$, the colour factor is $-8/3$, the spin factor $6$, and $\langle
\delta^{(3)}(x)\rangle$ is probably reduced, as the wave function is more
extended. Hence a value $\Delta-N$ of about 0.3\,GeV is roughly compatible.

The systematics of hyperfine splittings in a model based on 
\begin{equation}\label{bar:eq:hyper-ss}
 V_{ss}=\sum_{i<j}\frac{2}{3}\,\frac{\alpha_s}{m_i\,m_j}\,\frac{2\,\pi}{3}\,
\delta^{(3)}(\vec r_{ij})\,\vec \sigma_i.\vec\sigma_j~,
\end{equation}
was remarkable. See, e.g., 
\cite{DeRujula:1975ge,Isgur:1977ef,Isgur:1978wd,Isgur:1979be}.
In particular, the
chromomagnetic interaction, with its very specific $(m_i\,m_j)^{-1}$ dependence
explains why the $\Lambda$ is ligther than $\Sigma$, a long standing puzzle.
See the paper by the Orsay group (Le Yaouanc et al.). This approach has been
regularly updated
\cite{LeYaouanc:1976kp,LeYaouanc:1976ne,Cohen:1981ut,Richard:1983mu,
Anselmino:1989ji,Karliner:2008sv}
 See, also, the interesting historical remarks by
Lipkin \cite{Lipkin:1990vj}, in particular on the pioneer contribution by A.
Zakharov.

The chromomagnetic model of hyperfine splittings has been used to make
predictions for heavy baryons. It was for instance predicted that the
$\Sigma_Q-\Lambda_Q$ splitting will remain large as the mass of the heavy quark
increases, while $\Sigma_Q^*-\Sigma_Q$ will decreases. See, e.g.,
\cite{Richard:1983mu}, and for a recent upgrade in the beauty sector,
\cite{Karliner:2008sv}.

For the fine structure, the situation is less convincing. We have seen that the
fine structure of mesons requires a sizeable spin-orbit interaction, which
explains the ordering $\slj3P0<\slj3P1<\slj3P2$. This spin-orbit potential is
explained 
as to be resulting from a partial cancellation between the positive spin-orbit
of the
vector Coulomb potential and the negative spin-orbit of the scalar confinement.
It has been argued \cite{Isgur:1978wd,Reinders:1980af} that the cancellation
might be more
effective for baryons, but still the complete vanishing of the spin-orbit force
is not understood. 

There is definitely a problem with the spin-orbit splittings in baryons, which
are usually very tiny, at least smaller than expected by extrapolation of meson
spectroscopy, but sometimes surprisingly large. For instance
the $N(1520)-N(1535)$ difference (with $J^P=(1/2)^-$ and $J^P=(3/2)^-$,
respectively) is small, but $(\Lambda(1520)-\Lambda(1405)$ is large. So, a
strategy initiated by Isgur and Karl consists of setting to zero the spin-orbit
forces in baryon, and to call for contributions beyond the simple quark model
for the exceptions.

Another approach, also present in the phenomenology of mesons, relies on the
role of hadron loops. 
See,  for instance, \cite{Cottingham:1980cd,Tornqvist:1985fi}. For sure, the
$\Delta\leftrightarrow \pi N$ coupling contributes to the $\Delta-N$ spacing.
But the best advertised splitting is between $\Lambda(1405)$ and
$\Lambda(1520)$. Both are negative-parity excitations of the ground-state
$\Lambda$, with isospin $I=0$. The $\Lambda(1405)$ has $J^P=(1/2)^-$, and
$\Lambda(1520)$ has $J^P=(3/2)^-$. Hence, in the quark model, they
correspond to the same spin structure as the ground state $\Lambda$, a light
diquark with spin $s_q=0$ giving a total quark spin $s=1/2$. The motion
of the $s$ quark with respect to the $(ud)$ diquark has $\ell_y=1$. Thus there
are two possibilities, $J=1/2$ and $J=3/2$. 
The vicinity the $\bar K N$ threshold is presumably responsible
for this unusually large separation $\Lambda(1405)$ and $\Lambda(1520)$. The
pioneer here is Dalitz \cite{Dalitz:1960du}. 

A third possibility consists of modifying the part of the quark dynamics
dealing with the spin effects. For instance an instanton-induced interaction,
inspired but the work of t'Hooft, was adapted by the Bonn group
\cite{Metsch:2003ix}. This was
rather successful. Even more advertised is the class of models with a kind of
pion-exchange between quarks inside the same hadron. This is named
``Goldstone-boson exchange''.

Interesting models have been built were both chromomagnetism and
Goldstone-boson exchange contribute to the spin splittings.%
\footnote{I remember a very interesting discussion with Glozman on this issue.
Once a pion is emitted, the quark spin is flipped and thus the intrinsic
spin--spin interaction at the quark level is modified. So it is perhaps not so
easy to superpose within the same model Goldstone exchange and gluon exchange.}

\subsection{Convexity properties}\label{bar:sub:convex}
There are not many exact results on the three-body problem. Everything becomes
rather involved when going from two to three constituents. For instance, if you
consider three distinguishable particles in a local potential, it is not
obvious that the lowest energy for each spin $J$ is an increasing function of
$J$.\footnote{I thank Andr\'e Martin and Victor Mandelzweig for several
discussions on this point.}\@ 

Among the simple results, one may mention: the energies decrease if any mass
$m_i$ is increased; \dots

One point that was rather debated is the convexity as a function of the inverse
masses. We have seen in the two-body problem that (the symbol $(..)$  is meant
for the ground state energy) that $(m,m)+(M,M)\le 2(m,M)$. Is is thus tempting
to generalise as 
\begin{equation}\label{bar:eq:concave3}
 (m,m,m')+(M,M,m')\le 2\,(m,M,m')~,
\end{equation}
for any value of the ``spectator'' quark of mass $m'$. This was conjectured
\cite{Richard:1983mu,Nussinov:1983vh}. After some investigation, it turns out
that the rule remains true for reasonably-shaped
potentials, but that there are  counterexamples involving very sharp confinement
and large mass ratios \cite{Lieb:1985aw,Martin:1986da}.

The lesson of \eqref{bar:eq:concave3} is that heavy quarks like to cluster
together.
\subsection{Link between meson and baryon spectroscopy}\label{bar:sub:mes-bar}
\subsubsection{The baryon potential}
In the late 70s, there were clearly experts on baryon spectroscopy, obviously
not restricted to heavy quarks, and on the other side, experts on quarkonia. 
But the question was already addressed of the link between the quark--antiquark
interaction in mesons and the three-quark potential in baryons. 

One-gluon-exchange, or more generally, any colour-octet exchange gives a factor
1/2 for the interaction of a pair of quarks in baryons as compared to
the quark--antiquark interaction in mesons.  This is the analogue
of the $(-1)$ factor for one-photon exchange in $e^+e^+$ as compared to
$e^+e^-$, or the Fermi--Yang rule  $G$ for the exchange of mesons, between $NN$
and $N\bar N$, where $G$ is the $G$-parity of the meson or set of
mesons.\footnote{The Fermi--Yang rule relates the potentials with the same
isospin,
for instance, $pp$ and $p\bar n$.}\@

It is thus reasonable to try, as an educated guess, to apply the same ``1/2''
rule to the whole potential. After all, when a quark in a baryon, say
quark $\#1$,  is well separated from the two others, $\#2$ and $\#3$, it feels a
colour $\bar3$
at separation $\vec r\simeq\vec r_{12}\simeq\vec r_{23}$, exactly as a quark
feeling an antiquark at separation $\vec r$, and the ``1/2'' rule reproduce
perfectly this equivalence.

This leads Stanley and Robson, Greenberg, Cohen, Lipkin, etc.
\cite{Cohen:1980su,Stanley:1980fe,Greenberg:1981xn}, to discuss the merits of
this rule, and to propose a combined phenomenology of both meson and baryon
sectors. I remember once Martin making an adventurous extrapolation of his
$(c\bar c)$ and $(b\bar b)$ potential down to $(s\bar s)$ \cite{Martin:1980rm}.
In the good old days
of Inquisition, you would be burned at the stake for much less, because $(s\bar
s)$ is anything but non-relativistic. Nevertheless, the agreement was good. I
tried
the ``1/2'' rule, and got the mass of $\Omega^-(sss)$ within just a few MeV
\cite{Richard:1980tg}. There are
certainly many regularities in the hadron spectrum that are mimicked by
potential models, but the reason of not fully understood. 

The main issue is whether the interaction in baryons is pairwise. In
nuclear physics, the interaction among nucleons is not pairwise: current
potentials fitting the nucleon--nucleon scattering data and the properties of
the deuteron underestimates the binding of few-nucleon systems. Three- and even
four-nucleon forces have to be introduced. In atomic physics, there are
empirical potentials between for instance two He atoms (Asiz, etc.). They can
be reproduced by a (complicated but feasible) Born--Oppenheimer calculation,
where the energy of four electrons around two fixed $\alpha$ particles is
properly minimised. Then the interaction energy of three He atoms can be
estimated, by minimising the energy of six electrons around three fixed
$\alpha$'s. The interaction contains three-body terms, which are not very
important in practice, but are there.

Similarly, any physical interpretation of the linear confinement in mesons
leads to a generalisation for baryon which is of the three-body type. See, for
instance,
\cite{Artru:1974zn,Dosch:1975gf,Hasenfratz:1980ka,Carlson:1982xi,Bagan:1985zn,
Fabre:1997gf}, but many important papers have been probably omitted. In early
days, some approaches were a little empirical. For instance, in
\cite{Hasenfratz:1980ka}, the same variant of the bag model used for quarkonia
and hybrid mesons was applied to a bag containing three heavy quarks surrounded
by their gluon field. At very large separations, the bag that minimises the
gluon energy is Y-shaped, with three tubes joining at the centre of the baryon.
Today, the potential of baryons is derived in lattice QCD and even in AdS/QCD,
as will be explained by the other lecturers. 

\begin{figure}[htp]
 \centering
\includegraphics{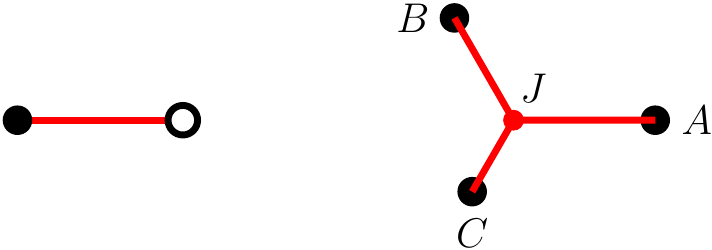}
\caption{The string linking the quark to an antiquark in a meson (left) or the
three quarks of a baryons has a minimal length.}
 \label{bar:fig:Fermat-T}
\end{figure}

The main result is that if you consider that $V(r)\sim b\,r$ in
\eqref{mes:eq:funnel} corresponds to a gluon tube of constant section whose
linear behaviour $\propto r$  minimizes the energy, the analogue for a baryon
is a $Y$-shape set of three tubes whose cumulated length minimises the energy,
i.e.,
\begin{equation}\label{bar:eq:Y}
V(\vec r_1,\vec r_2,\vec r_3)=b\,\min_{J} (r_{1J}+r_{2J}+r_{3J})~,
\end{equation}
where the junction $J$ is adjusted to minimise the sum. The solution is well
known. For a flat triangle, with, say, $\angle213>120^\circ$, then $J$
coincides with the quark \#1, other wise, $J$ sees each side at $120^\circ$, as
shown in Fig.~\ref{bar:fig:Fermat-T}.

As a French lecturing in Italy, I cannot resist mentioning here the work of
Fermat and Torricelli, and of Napoleon. The two first names will not be too much
a surprise, and, indeed, Torricelli had an interesting correspondence
with several scientists, including Fermat. Napoleon Bonaparte invaded Italy,
 among other countries, and his troops behaved wildly, in particular in Naples,
and the following sentence is reported ``Non tutti i francesi sono ladri, ma
buona parte s\`i''. Napoleon was also a fine mathematician, and his theorem on
triangles is relevant for the confinement of baryons.

\begin{figure}[htp]
 \centering
 \includegraphics{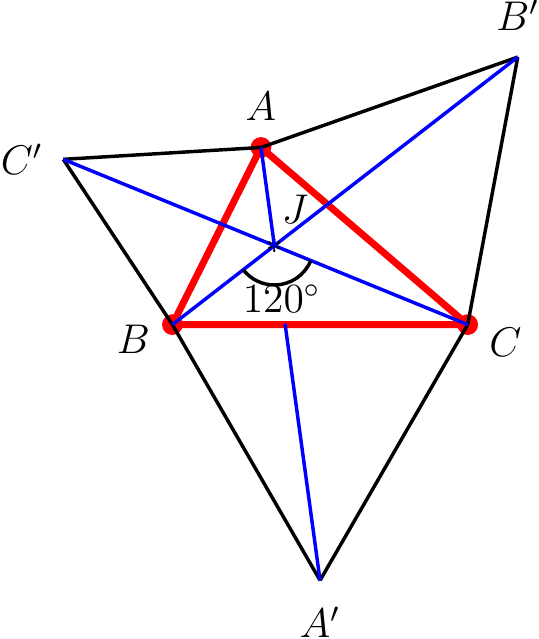}
 \caption{If the quarks are located at $A$, $B$ and $C$, and if $A'$ is the
third vertex of an external equilateral triangle built on the side $BC$, and
similarly for $B'$ and $C'$, then the straight line $AA'$, $BB'$ and $CC'$
intersect at the Fermat--Torricelli junction.  Moreover, the minimal sum of
distances, $Y=JA+JB+JC$ is equal to $Y=AA'=BB'=CC'$}
 \label{bar:fig:napoleon1}
\end{figure}

The problem of the minimal cumulated distance to three given points was solved
by Fermat and Torricelli. A possible construction is shown in
Fig.~\ref{bar:fig:napoleon1}.

\begin{figure}[htp]
 \centering
\includegraphics{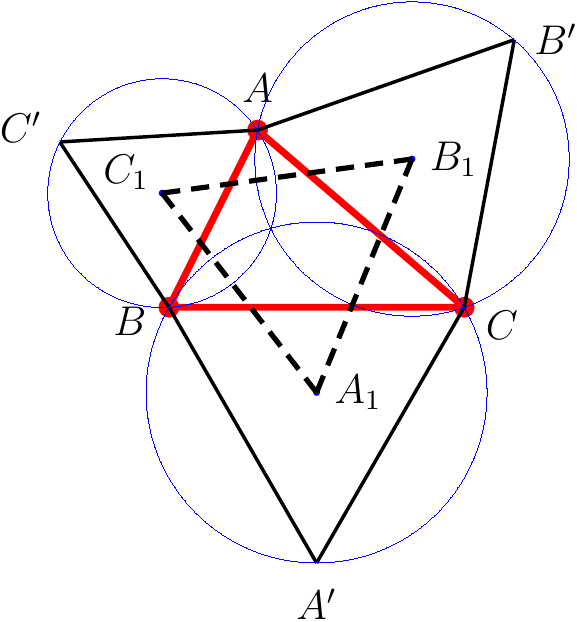}
 \caption{The Fermat--Torricelli junction is at the intersection of the
circumcircles of the external equilateral triangles which are at the basis of
Napoleon's theorem.}
 \label{bar:fig:napoleon2}
\end{figure}

One may also remark that the circumcircles of the external triangle such as
$BCA'$ intersect in $J$. The construction is also that of Napoleon's theorem,
which states that the centres of these  external triangles form an equilateral
triangle. See Fig.~\ref{bar:fig:napoleon2}. This is an  interesting example of
symmetry restoration starting from an asymmetric triangle. 
\subsubsection{Inequalities between meson and baryon masses}
\label{bar:subsub:mesbarineg}
Presumably, the short range contributions are pairwise and fulfil the 1/2 rule,
and the long range linear confinement is given by the Fermat--Torricelli
minimal length $Y$ (times the string tension). It is easy to check that the
latter is bound by ($p$ is the perimeter)
\begin{equation}\label{bar:eq:Y-ineg}
 \frac{p}{2}\le Y\le \frac{p}{\sqrt3}~,
\end{equation}
(the lower bound is saturated for a flat triangle, the upper one for an
equilateral triangle), 
so that for the whole baryon potential,
\begin{equation}\label{bar:eq:V-ineg}
 V\ge \frac12[v(r_{12})+v(r_{23})+v(r_{31})]~.
\end{equation}

In the 60s, physicists working on few-nucleon systems, or on quantum systems
under gravitational interaction, have derived inequalities between three-body
and two-body energies. In our case,
\begin{equation}\label{bar:eq:H3vsH2}
 V\ge \frac12[v(r_{12})+v(r_{23})+v(r_{31})]~.
\end{equation}
and hence for the Hamiltonians
\begin{equation}\label{bar:eq:H3vsH2a}
H_3=\frac{\vec p_1^2}{2\,m}+ \cdots +V\ge \frac12
\left\{\left[\frac{\vec p_1^2}{2\,m}+
\frac{\vec p_2^2}{2\,m}+v(r_{12})\right]+\cdots\right\}~.
\end{equation}
Since the minimum of the sum is larger than the sum of minima, the above
inequality immediately implies $E_3\ge (3/2)\,E_2$, or after adding the
constituent masses \cite{Ader:1981db},
\begin{equation}\label{bar:eq:H3vsH2b}
 2 \,M(qqq)\ge 3 \,M(q\bar q)~.
\end{equation}
In other words, the matter is heavier per quark in the baryon form than in
the meson form. The inequality \eqref{bar:eq:H3vsH2b} has been generalised and
improved in several ways. See, e.g.,
\cite{Basdevant:1989pt,Basdevant:1989pv,Basdevant:1992cm} for some details and
references to the pioneer  works, in particular by Hall, Post, Fisher, Ruelle,
L\'evy-Leblond, etc.

In the form $(qqq)+(\bar q\bar q\bar q)\ge 3\,(q\bar q)$, the inequality
\eqref{bar:eq:H3vsH2b} implies that, at least within a simple quark model,
baryon--antibaryon ``annihilation'' (more precisely rearrangement) into three
mesons is energetically allowed, even at rest. 

On the other hand, if the mass ratio becomes large enough, the inequality is
inverted, and becomes
\begin{equation}\label{bar:eq:QQQqqq}
(\bar Q\bar Q\bar Q)+(qqq)\le 3\,(\bar Q q)~,
\end{equation}
meaning that a triply-flavoured antibaryon would not annihilate into ordinary
matter. The critical value of $M/m$ is model dependent. There is in
\eqref{bar:eq:QQQqqq} another manifestation of the tendency of heavy
constituents to cluster together, as seen already in \eqref{mes:eq:BM}.

``Why do it the easy way when you can do it the hard way?'' was the favourite
watchword of a popular TV series in the late 60s.  An alternative (and not so
rigorous) way of understanding $(qqq)+(\bar q\bar q\bar q)\ge 3\,(q\bar q)$ is
as follows. The reasoning will be useful in the section about multiquarks.

Consider the case where the interaction is pairwise and the 1/2 rule is exact.
Otherwise replace some equalities
by inequalities.  Both $(qqq)+(\bar q\bar q\bar q)$ and $(q\bar q)+(q\bar
q)+(q\bar q)$ systems can be artificially considered as governed by a 6-body
Hamiltonian (for more mathematical rigour, one can add a very weak overall
harmonic confinement that makes all levels discrete and normalisable)
\begin{equation}\label{bar:eq:H6}
 H_6[\{g_ij\}]=\sum_i\frac{\vec p_i^2}{2\,m}+\sum_{i<j}
g_{ij}\,v(r_{ij})~,\qquad \sum_{i<j} g_{ij}=3~.
\end{equation}
Let $E_6[\{g_ij\}]$ denote the ground state. From the variational principle, it
is maximal for the fully symmetric case where $g_{ij}=1/5\ \forall i,j$. 
The baryon-antibaryon case corresponds to a set of coefficient (``*'' means
``times'') $\{g_{ij}\}=\{6*(1/2),9*0\}$, and the three-meson configuration to 
$\{g_ij\}=\{3*1,12*0\}$. The latter has the same mean value 1/5 as the
former, but a much larger variance, about 0.17 vs.\ about 0.06, and this makes
its energy lower.

%
%\clearpage
\markboth{\sc An introduction to the quark model\hspace*{1cm}
}{\sc Multiquarks}
\section{Multiquarks}\label{se:mul}
\vspace*{-.2cm}
\begin{flushright}
\sl
Le plus grand d\'er\`eglement de l'esprit,\\ c'est de croire les choses parce
qu'on veut qu'elles soient,\\ et non parce qu'on a vu qu'elles sont en effet.
\footnote{The biggest disorder of the spirit, it is to believe things because we
want that they are, and not because we saw that they are indeed.}\\ 
\rm Bossuet
\end{flushright}
\vspace*{-1.2cm}
\subsection{Introduction}\label{mul:sub:intro}
We are now entering a delicate chapter. An atomic physicist could not feel
satisfied with only the hydrogen atom and its variants, and the H$^-$ and
H$_2{}^+$
ions, or a nuclear physicist with the deuteron, tritium and ${}^3$He: they have
to understand why some small systems are less easily bound than others, and to
explore more complicated structures.
Similarly, one has to know whether there exist hadrons beyond ordinary
mesons and baryons. Besides multiquarks, several systems have been
considered, either exotics, i.e.,  with quantum numbers that cannot be matched
by $(q\bar q)$ nor $(qqq)$, or crypto-exotics, with the same quantum numbers
but different internal structure.  The latter are more difficult to identify:
one should first detect the exhaustive spectrum of ordinary hadrons, and look
at super-numerous peaks, but the signature can be hidden by mixing of ordinary
and extra-ordinary hadrons. 
\subsubsection{Glueballs} They used to be fashionable, and motivated several
experiments. In a glueball, a few constituent gluons experience a confining
interaction. There is a considerable phenomenology of glueballs, and many
predictions based on lattice QCD, with noticeable fluctuations along the years
\cite{Mathieu:2008me}.
\subsubsection{Hybrid mesons}
At first sight, this is just another piece of the folklore, combining mesons
and glueballs. However, it acquired some ground. 

Consider H$_2{}^+$ in the Born--Oppenheimer limit, which is a very good
approximation. At given $pp$ separation $r$, one estimates the minimum of the
electronic energy, which, when supplemented by the direct Coulomb repulsion,
built the effective $pp$ potential. In the Schr\"odinger equation, this
potential
gives the ground-state and the first orbital and vibrational excitations. If
one takes the second electronic energy, one gets another effective potential
generating a second set of levels. 

\begin{figure}[htp]
 \centering
 \includegraphics[width=.4\textwidth]{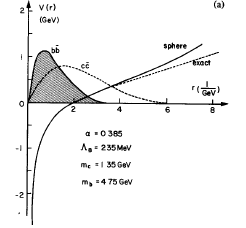}
% quarkoniumpot.png: 246x228 pixel, 96dpi, 6.51x6.03 cm, bb=0 0 184 171
\hspace*{1cm}
\includegraphics[width=.5\textwidth]{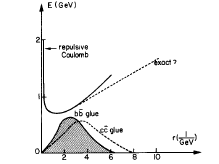}
% hybride.png: 219x164 pixel, 96dpi, 5.79x4.34 cm, bb=0 0 164 123
\caption{Left: ordinary $Q\bar Q$ potential, right: effective $Q\bar Q$
potential for hybrids. }
 \label{mul:fig:hyb}
\end{figure}

Now, replace the protons by a quark and an antiquark, and the electron by the
gluon field, you first get the ordinary quarkonium and next the hybrids. 
Pioneer papers are \cite{Giles:1977mp,Horn:1977rq}.
The quark--antiquark potential within an excited gluon field was
explicitly estimated in a variant of the bag model \cite{Hasenfratz:1980jv},
where gluons are treated classically and confined in volume whose shape and size
is optimised: a $Q\bar Q$ potential is derived, very similar to the
Coulomb-plus-linear one, and a second potential is obtained, with a repulsive 
Coulomb interaction and a confinement smoother than for quarkonia.
See Fig.~\ref{mul:fig:hyb}.
\subsubsection{Hadron molecules}
It is regularly rediscovered that the Yukawa mechanism leading to the nuclear
forces is by no means restricted to the nucleon--nucleon interaction. 
Any particle containing light quarks or antiquarks can emit or absorb light
mesons and thus enter a nuclear-type of interaction. 

The hidden-charm sector is no exception. Already in 1975, Iwazaki suggested that
$\psi'$ could be
$(c\bar c q\bar q)$~\cite{Iwasaki:1975pv}, where $q$ denotes $u$ or $d$. In
1976, Voloshin and Okun \cite{Voloshin:1976ap} suggested the existence of
molecules made of a charmed meson and an anticharmed meson. De Ruj\`ula, Georgi
and Glashow \cite{DeRujula:1976qd}  were intrigued by one of the recurrence of
$J/\psi$, with the same quantum numbers $J^{PC}=1^{--}$, whose  branching ratios
into $D\bar D$, $D^*\bar D+ \mathrm{c.c.}$ and $D^*\bar D{}^*$ were anamalous
as compared to naive spin counting. They suggested a $D^*\bar D{}^*$
structure for the $\psi(4.03)$. In fact, the anomalies in the branching ratios
were explained by the node structure of the state: roughly speaking, a two-body
decay calls for a range of relative momentum, and the decay is either
suppressed or enhanced if this momentum is near a zero or a maximum of the
probability in momentum space. 

More recently, T\"ornqvist \cite{Tornqvist:1991ks,Tornqvist:1993ng}, and also,
Ericson and Karl \cite{Ericson:1993wy}, Manohar and
Wise \cite{Manohar:1992nd}, etc., 
stressed that the one-pion exchange between two charmed mesons or between a
charmed and an anticharmed meson, can be attractive. This potential is somewhat
weaker than the potential between a proton and a neutron, but of the same order
of magnitude. It is experienced by heavy particles, and thus could lead to
binding. 
Remember that for an Hamiltonian $H=\vec p^2/m+ g\,V(r)$, where $V(r)$ is
attractive or contains an attractive part, the existence of a  discrete
spectrum depends on $m\,g$. 

The discovery of the $X(3872)$ just at the $D^*\bar D$ threshold was
considered as the success of the prediction by T\"ornqvist and others. Some
further measurements have confirmed the existence of this state. However, some
decay properties, such as $X\to\psi(2S)+\gamma$ larger than
$X\to\psi(1S)+\gamma$, suggest the structure of a radially excited P-state.
Presumably both charmonium and molecular components exist in the wave function.
The question now is whether this is just a charmonium state modified by the
coupling to its nearest threshold, or whether there are really two separate
states in the spectrum, one mainly $2\,\slj3P2$ state of charmonium, another
mainly a $D^*\bar D+ \mathrm{c.c.}$ molecule, both with $J^{PC}=1^{++}$.

I cannot review the literature on molecules, especially the papers based on
effective theories. A superficial reading gives the feeling that for any pair
of hadrons, one can write up an effective Lagrangian that will lead to a
molecule. The approach based on conventional nuclear forces is more predictive.
Remember the Yukawa interaction in a S-wave contains a spin--isospin factor
$\vec\sigma_1.\vec\sigma_2\,\vec\tau_1.\vec\tau_2$, and thus is either
attractive or repulsive and varies by a factor 1:9 in magnitude. 
After the discovery of the $X(3872)$, the literature exploded. See, e.g., the
reviews by Swanson \cite{Swanson:2006st} and by Nielsen et
al.~\cite{Nielsen:2009uh}. Recent contributions include
\cite{Ohkoda:2011vj,Ohkoda:2012hv}
\subsection{Baryonium}\label{mul:sub:baryonium}
In the 70s, bumps were tentatively seen in antiproton cross sections, and peaks
in inclusive $\gamma$-ray spectra of $\bar p p\to\gamma X$. The generic name
``baryonium'' was invented for new mesons preferentially coupled to the
baryon--antibaryon channels. For a review of the early results, see
\cite{Montanet:1980te}.

A molecular picture of baryonium was proposed, e.g., by Shapiro and his
collaborators \cite{Shapiro:1978wi}, Dover et al.\
\cite{Buck:1977rt,Dover:1979zj,Dover:1980bs}, etc. The
interaction is given by meson-exchange (Fermi and Yang \cite{Fermi:1959sa} have
given a simple rule
to transform meson exchange into a particle--particle system to the
corresponding
particle--antiparticle system) at large distances, and exhibit a noticeable
attraction in some partial wave. This suggests the possibility of $N\bar N$
bound states, in the same way that the exchange of mesons leads to a bound state
of two
nucleons. The role of annihilation remains however uncertain in this approach.

A four-quark picture of baryonium was also elaborated, and in fact was even
anticipated by arguments based on the so-called ``duality'', whose explanation
is beyond the scope of this review. See, e.g., Veneziano and Rossi
\cite{Montanet:1980te} and Jaffe \cite{Jaffe:1977cv,Jaffe:2004ph}. 
Unfortunately, if the authors of the above references were relatively careful in
their predictions, this was not always the case with the followers, who produced
dozens of baryonia and generalisations. 

Chan H.M., H\o gaasen and their collaborators speculated about the role
of the internal colour degrees of freedom \cite{Chan:1978nk}.\footnote{I thank
Chan H.M., H.\ H\o gaasen, P.\ Sorba and B.\ Nicolescu for numerous discussions}
\@
 For $(q^2\bar q^2)$, they named
``true'' baryonium a state where the quarks are in colour $\bar 3$ and the
antiquarks in colour $3$. Such a state is thus easily coupled to the
baryon--antibaryon channels but can be prevented of annihilating too fast if
the quarks are separated from the antiquarks by some angular-momentum barrier.%
\footnote{Note that the structure $(qq-\bar q\bar q)$ for angular momentum
$\ell>0$ is not demonstrated, it is assumed}\@ States with colour structure
$6\bar 6$ were also envisaged and named ``mock'' baryonia. It was suggested
they could be rather narrow. 
Of course, what was not clear in this approach is why four-quark states should
cluster in the form of a diquark and an antidiquark with such colour structure. 
See, e.g., \cite{Gromes:1988ue}.

No baryonium survived better experimental scrutiny, in particular with the
intense and cooled antiproton beam of the LEAR facility at CERN. More recently,
hadrons containing heavy quarks have been observed decaying into $\text{baryon}
+\text{antibaryon} +
\cdots$, and enhancements are seen in the spectrum of the invariant mass $p\bar
p$, probably due to a strong final-state interaction. So, the saga of baryonium
is perhaps not fully over. 

Note also that the mechanisms invented to explain the tentative baryonium states
were applied to other configurations. For instance, Strottmann
\cite{Strottman:1979qu}, Sorba and H\o gaasen  \cite{Hogaasen:1978jw}, and De
Swart et al.\ \cite{deSwart:1980db} studied baryon states with four quarks and
one antiquark (the name ``pentaquark'' was used later). Other studies dealt with
$(q\bar q-q\bar q)$ \cite{Tow:1978pq} or with dibaryons~\cite{Aerts:1977rw}.
\subsection{Chromomagnetic binding}\label{mul:sub:chromo}
\subsubsection{The chromomagnetic interaction}
As mentioned earlier, an appealing aspect of the quark model for mesons and
baryons is the systematics of hyperfine effects such as $J/\psi-\eta_c$ or
$\Delta-N$. Lipkin \cite{Lipkin:1990vj,Lipkin:1992xq}  likes to stress that
Sakharov, though isolated, had
interesting remarks on the subject. A popular reference is the paper of DGG
\cite{DeRujula:1975ge}, where hyperfine forces among quarks are described in
analogy with the Breit--Fermi interaction in atomic physics and attributed to
one-gluon-exchange. See also \cite{Jaffe:1999ze}.

This mechanism was used in the bag model \cite{DeGrand:1975cf}, and in several
variants of
potential models, with a component 
\begin{equation}\label{mul:eq:ss}
 V_{ss}=\sum
_{i<j}\frac{K}{m_i\,m_j}\,\vec\sigma_i.\vec\sigma_j\,
\tilde\lambda_i.\tilde\lambda_j
\,v_{ss}(r_{ij})~,
\end{equation}
where $v_{ss}$ is either a contact term, to be treated at first order, or a
regularised contact term, to be inserted in the wave equation. We already
stressed the importance of the strength being proportional to 
$(m_i\,m_j)^{-1}$, for instance for the $\Sigma-\Lambda$ difference, but now we
will first consider the case of equal masses.  For angular momentum
$\ell>0$, there are additional terms, in particular a tensor component.
\subsubsection{The dibaryon \boldmath$H$\unboldmath}
At the end of 1976, Jaffe \cite{Jaffe:1976yi} studied the $(ssuudd)$ system
in the SU(3)$_{\rm F}$ limit, assuming that $\bar v_{ss}=\langle v_{ss}(r_{ij})
\rangle$ has the same value for all the pairs, thus concentrating on the
properties
of the operator
$\mathcal{O}=\sum\vec\sigma_i.\vec\sigma_j\,\tilde\lambda_i.\tilde\lambda_j$ in
this dibaryon system, as compared to the baryons. He discovered a very
intriguing
coherence,  namely that for some configurations $\langle \mathcal{O}\rangle$
can be attractive and \emph{larger} (in absolute value) than its cumulated value
for two baryons constituting the threshold!
This is really remarkable. For the positronium molecule and for two positronium
atoms constituting its threshold, both described by a potential
$\sum\,g_{ij}\,r_{ij}^{-1}$, we have the \emph{same} cumulated strength $\sum
g_{ij}=-2$. 

Using for $\bar v_{ss}$ the same value as in the light baryons,
Jaffe predicted a ``di-lambda'', named $H$, bound by about 150\,MeV below the
baryon--baryon threshold. In the SU(3)$_F$ limit considered by Jaffe, the
$N$, $\Lambda$, $\Sigma$ or $\Xi$ receives from \eqref{mul:eq:ss} a downward
shift of 150\,MeV (this is half of the $\Delta-N$ splitting), and thus the
$\Lambda\Lambda=N\Xi$ thresholds benefit from about 300\,MeV. Thus the H
receives an additional 150\,MeV as compared to this threshold.  This was a
dramatic change in the physics of multiquarks. Earlier speculations, such as
baryonium, proposed metasable states, lying above their lowest threshold,
but perhaps not too broad. The $H$, as initially proposed, was stable against
any dissociation, and thus decaying weakly. 

As the MIT physicists are very convincing and
powerful, the $H$ was searched for in about 20 experiments.\footnote{For
comparison, the pentaquark analogue discussed in the next section was
searched in a single experiment, and is vaguely in the agenda of the COMPASS
experiment at CERN. \textsl{"Selon que vous serez puissant ou mis\'erable, les
jugements de cour vous rendront blanc ou noir."} (According as you're feeble, or
have might, High courts condemn you to be black or white)
Les animaux malades de la peste, Jean de la Fontaine}\@

Unfortunately, it was later realized that the first calculation by Jaffe needed
some improvements,
namely \cite{Oka:1983ku,Rosner:1985yh,Karl:1987cg,Fleck:1989ff}: \textsl{i)}
treat the
spin-independent interaction (instead
of assuming that $(q^6)$ and $(q^3+q^3)$ starts from the same line before
switching on the hyperfine effects, \textsl{ii)} consider SU(3)$_{\rm
F}$ breaking that spoils the coherence of the configurations $\Lambda\Lambda$,
$N\Xi$, etc., , and \textsl{ii)} use for $\bar v_{ss}$ values consistently
computed from 6-body wave functions. Each effect dramatically reduces the
binding, and the $H$ becomes eventually unbound, though there is some model
dependence. Recent lattice simulations \cite{Beane:2010hg,Inoue:2010es} leaves
the door open.
\subsubsection{The heavy pentaquark \boldmath$P$\unboldmath}
In 1987, Gignoux et al.\ \cite{Gignoux:1987cn} and Lipkin
\cite{Lipkin:1987sk} realised  independently that the scenario
for the $H$ is repeated in a baryon configuration involving a heavy antiquark
and four light quarks.\footnote{Lipkin, in fact, had some precursor idea in
this direction at Baryon 80 in Toronto. Anyhow, I remember Lipkin arriving at
Les Houches for the Workshop on the ``Elementary Structure of Matter'', handing
me a freshly issued preprint,  myself welcoming him with a copy of our latest
manuscript,  and our astonishment when we realized that both papers were dealing
with the same exotic baryon!}\@

Let us consider a configuration $(\bar Q qqqq)$ in the limit $m_Q$ for the
antiquark and exact SU(3)$_F$ for the quarks. If the quarks have spin 0 and
form a triplet of flavour. This means $(uuds)$, $(udds)$ or $(udss)$. Then the
chromomagnetic operator
$\mathcal{O}=\sum\vec\sigma_i.
comp\vec\sigma_j\,\tilde\lambda_i.\tilde\lambda_j$ has an expectation value
which is negative and twice larger than for $\Lambda$. This means that if you
compare the pentaquark\footnote{The word was adopted at this occasion} $P(\bar
Q uudqs)$ to its thresholds $(\bar Q q)+ (qqq)$, and adopt the same short-range
correlation factor $\bar v_{ss}$, you find the pentaquark bound by aboundkarl
150\,MeV. This pentaquark was searched for in an experement at Fermilab
\cite{Aitala:1997ja,Aitala:1999ij}, but the results were not conclusive.

This pentaquark was further analysed, and its was found, as for $H$, that the
corrections tend to spoil the binding. They are: breaking of SU(3)$_F$,
finite mass for the heavy antiquark, self-consistent estimate of $\langle
v_{ss}(r_{ij})\rangle$, or even better, inclusion of $v_{ss}$ in an accurate
five-body calculation, etc. See, for instance,
\cite{Fleck:1989ff,Lichtenberg:1998dm,Lipkin:1998pb}
\subsubsection{Other configurations}
The chromagnetic mechanism remains fascinating, even so it does not bind as
easily as first thought. It contributes to the structure of the $X$, $Y$,
$Z$ states \cite{Hogaasen:2005jv}, and to the short-range hadron--hadron
interaction. Leandri and Silvestre-Brac have scanned the configurations where
interesting coherences occur \cite{Leandri:1989su,Leandri:1997ge}.
Then they investigated whether the stability suggested by the chromomagnetic
operator survived a more detailed calculation, with mostly negative results.
\subsection{Binding by central forces}\label{mul:sub:central}
In a novel by Alphonse Daudet, one finds this sentence: `` \dots il ferma la
porte \`a double tour. Malheureusement, il avait oubli\'e la fen\^etre \dots''
(He double locked the door, but forgot the windows). Similarly, if the
chromomagnetism cannot bind multiquarks, why not try chromoelectricity?

Believe it or not, it works, and there is an amusing side effect, or I should
perhaps say, a miracle: while the literature on multiquarks is usually a
jungle,  of controversed, here a consensus has been
reached that some configurations should support stable tetraquarks, i.e., states
below the threshold for their dissociation into two mesons. 

I will first summarise the results obtained from an empirical pair-wise ansatz
for the multiquark interaction, and in the next subsection, explain the changes
brought by an improved modelling of the dynamics. 
\subsubsection{The additive model}
As explained in Sec.~\ref{bar:sub:mes-bar}, the potential for mesons can be
extrapolated to
baryons by the ``1/2'' rule, i.e., $V=[v(r_{12})+v(r_{23})+v(r_{31})]/2$, where
$v$ is the quarkonium potential. It can be read as the exchange of a colour
octet in the $t$-channel. Similarly, in nuclear physics, there are isoscalar
exchanges, leading to an isospin-independent interaction, and isovector
exchanges, which get a $\vec\tau_1.\vec\tau_2$ factor, such a pion-exchange.
The analogy is of course delicate, as local and global symmetries are not on
the same footing. Nevertheless, the multiquark phenomenology in simple
constituent models has been very often based on the assumption that 
\begin{equation}\label{mul:eq:pot-add}
 V=-\frac{16}{3}\sum_{i<j} \tilde\lambda_i.\tilde\lambda_j\,v(r_{ij})~,
\end{equation}
which reduces to the 1/2 rule in the case of baryons. Here is the version for
quarks. For an antiquark, $\tilde\lambda_i \to -\tilde\lambda_i^*$

This additive model leads to coupled equations, using either a $(q\bar
q)-(q\bar q)$ basis with colour 1 or 8 for the pairs, or a $(qq)-(\bar q\bar
q)$ basis with internal colour $\bar3-3$ or $6-\bar 6$. 

Solving the four-body problem is a considerable task, and requires a lot of
care. It can be done by variational methods based, for instance on a
superposition of Gaussian wave functions, or by hyperspherical expansion. 

The hyperspherical expansion consists of considering the $n$ Jacobi variables
\footnote{if necessary rescaled to be associated with the same reduced mass} of
a $(n+1)$-body problem, as a unique $3\,n$-dimensional vector, and to use
spherical coordinates $(\varrho,\Omega)$ to describe it. In the particular case
of the harmonic oscillator with equal masses, the potential is isotropic,
$V\propto \sum r_{ij}^2\propto \varrho^2$, and a single radial equation allows
one to recover the standard results for this interaction. In general, one deals
with an anisotropic potential in the $3\,n$-dimensional space, and thus an
infinite set of coupled equations (the same situation would occur for a
single paticle in a potential that is not rotation-invariant). One obtains a
rather satisfactory
convergence as a function of the number of equations actually included.
\subsubsection{Results for equal masses}
If this colour-additive potential, without spin-dependent terms, is applied to
$(q,q\,\bar q,\bar q)$, no bound state is found below the $(q\bar q)+(q\bar
q)$ threshold. Even though  such a model is rather primitive, it already
indicates that
there is no proliferation of narrow multiquarks. 

One may wonder why the positronium molecule is stable whereas the $(q,q,\bar
q,\bar q)$ is not in such a simple quark model. The generic Hamiltonian for
these systems is
\begin{equation}\label{mul:eq:H4}
 H_4=\sum_i\frac{\vec p_i^2}{2\,m}-\frac{[\sum \vec p_i]^2}{8\,m}+\sum_{i<j}
g_{ij}\,v(r_{ij})~,\quad \sum_{i<j}g_{ij}=2~,
\end{equation}
where $v(r)=-1/r$ for the atomic problem, and $v(r)=\lambda\,r + \cdots$ in the
quark problem. For the thresholds (atom--atom or meson--meson), the molecule,
and the ``true'' $T=(\bar 33)$ and ``mock'' $M=(6\bar6)$ colour wave-functions
of the tetraquark system, the coefficients are given in Table~\ref{mul:tab:gij}.
\begin{table}[htp]
 \centering
 \caption{Coefficients $g_{ij}$ of the potential energy for the threshold, the
positronium molecule and the $T$ and $M$ type of tetraquark, average and
variance}
 \label{mul:tab:gij}
\vskip .1cm
\begin{tabular}{l|rrrrrr|rr}
\hline
% \backslashbox{State}{{}\\[8pt]Pair} & 12 & 34 & 13 & 24 & 14 & 23 & $\bar g$
% & $\Delta g$\\
\backslashbox{State}{{}Pair\\[-12pt]} & 12 & 34 & 13 & 24 & 14 & 23 & $\bar g$
& $\Delta g$\\
\hline\\[-8pt]
Threshold & 0 & 0 & 1 & 1 & 0 & 0 &1/3& 0.22\\[2pt]
Ps$_2$ & $-1$ & $-1$ & 1 & 1 & $-1$ & $-1$&1/3&0.89\\[2pt]
$T$ &1/2 & 1/2 & 1/4 & 1/4 & 1/4 & 1/4&1/3&0.01\\[2pt]
$M$ &$-1/4$& $-1/4$ & 5/8 & 5/8 & 5/8 & 5/8 &1/3&0.17\\
\hline
\end{tabular} 
\end{table}

We repeat here in some more detail a reasoning sketched at the end of 
Sec.\ \ref{bar:subsub:mesbarineg} dealing with  the comparison of mesons and
baryons. Symmetry breaking is known to lower the energy of the ground state. For
instance, the linear oscillator $h(0)=p^2+x^2$ has a ground state at
$\epsilon(0)=1$ which becomes $\epsilon(\lambda)=1-\lambda^2/4$ for
$h(\lambda)=h(0)+\lambda\,x$. More generally, for $H(\lambda)=H_0+\lambda\,H_1$,
where $H_0$ is even and $H_1$ odd, one gets the same lowering
$\epsilon(\lambda)\le
\epsilon(0)$ of the ground state, and this is true for symmetries other than
parity. This is easily seen by using the variational principle, with the
symmetric ground state of $H_0$ as a trial wave function for $H(\lambda)$.

Thus, if one considers the family of Hamiltonians \eqref{mul:eq:H4}, the ground
state energy is maximal if all couplings are equal, i.e., $g_{ij}=\bar g$ \,
$\forall i,j$ and  $\Delta g=0$.

If $H_4=H_0+\lambda\,H_1$ with $H_0$ symmetric and $H_1$ a given asymmetric
potential energy, then the ground state $\epsilon(\lambda)$ is maximum at
$\lambda=0$ and concave. Thus the larger $\lambda$ for $\lambda>0$, the lower
the energy. And the larger $|\lambda|$ for $\lambda<0$, the lower the energy.

This is rigorous, so far. But the energy, as a function of $\lambda$ is nearly
symmetric, and for $H_4=H_0+\lambda\,H_1$, the energy $\epsilon(\lambda)$
decreases approximately as $\epsilon_0-\lambda^2\, \tilde\epsilon_0$.
Furthermore, in the case of \eqref{mul:eq:H4}, $H_0$ corresponds to the
symmetric case $g_{ij}=\bar g$, and one can probe different $H_1$, i.e.,
different ways of breaking the symmetry among the pairs in the potential:
the energy decreases approximately according to the variance $\Delta g$ of the
distribution of the coefficients~$\{g_{ij}\}$. 

From the values listed in Table \ref{mul:tab:gij}, one sees that $\Delta g$
is larger for $Ps_2$ than for two non interacting atoms.  This explains (again,
not very rigorously) why Ps$_2$ is stable in atomic physics. On the other hand, 
for the $T$ or $M$ tetraquarks, $\Delta g$ is smaller than for two
non-interacting mesons. This suggests that in the simple additive model
without spin-dependent terms, there is no stable tetraquark, and the  mixing of
colour configurations does not improve much the situation. In other word, the
tetraquark is penalised as compared to the positronium molecule by the
\emph{non-Abelian} character of the theory. 
\subsubsection{Binding for unequal masses}
The lesson of the above subsection is that the tetraquark potential is too
symmetric with respect to permutations, as compared to the potential governing
the two-meson threshold. So to get binding, another effect is required. We have
mentioned chromomagnetism, we shall see below the possibility of improving the
modelling of spin-independent forces, to better conform with QCD. There is
another effect, which is conceptually simpler: introduce symmetry breaking in
the kinetic energy, i.e., mix different flavours in the multiquarks.

As stated earlier, any symmetry breaking lowers the ground-state energy, but
does not necessarily improve binding. Usually the threshold benefits more from
symmetry breaking, thus the stability deteriorates. 

This is the case, for instance, in atomic physics, when one breaks permutation
symmetry for $(\mu^+,\mu^+,\mu^-,\mu^-)$. For simplicity, let us do it in the
same way in the two charge sectors, i.e., keeping an eigenstate of charge
conjugation. This corresponds to $(M^+,m^+,\linebreak[2]{M⁻,m^-)}$, and one can
use the scaling properties of the Coulomb problem to impose
$M^{-1}+m^{-1}=2\,\mu^{-1}$ without any loss of generality. This corresponds to
maintaining the threshold energy constant. What is observed is that the
positronium-line molecule is stable for $M=m$, but becomes unstable for
$M\gtrsim 2.2\,m$ or $m\gtrsim 2.2\,M$. For a large value of the mass ratio, the
ground-state consists of a protonium atom which cannot polarise a positronium
atom to attach it. (Of course, some metastability of an hydrogen atom and an
antihydrogen atom cannot be exclude, but this is a much more delicate issue.)

The same mechanism renders $(Q,\bar Q, q,\bar q)$ less bound than the
equal-mass tetraquark in any simple, flavour-independent, central potential.
This is of course rather unfortunate, since the best candidates today, to be
discussed in Patricia's lectures following mine, are just in the hidden charm
or hidden beauty sector! The mechanism here is presumably long-range forces,
such as pion exchange, and spin-dependent forces. 

Another possibility if symmetry breaking with two masses consists to give up
charge conjugation and keep particle identity in the two sectors. In atomic
physics, the configuration reads $(M^+,M^+,m^-,m^-)$. It is well known that the
 binding energy of the hydrogen molecule (in units of the threshold energy to
have a scale-independent statement) is much larger than that of the positronium
molecule. Historically, the hydrogen molecule was studied in the
Born--Oppenheimer--Heitler--London limit, starting from $M\to\infty$, while the
question of the stability of the positronium was addressed later by Whyleer,
Ore, Hylleraas, etc. 

The Hamiltonian can be written as follows
\begin{equation}\label{mul:eq:MMmm}
 H(M^+,M^+,m^-,m^-)=H(\mu^+,\mu^+,\mu^-,
\mu^-)+\frac14\left(\frac1m-\frac1M\right)\left[\vec p_1^2+\vec p_2^2-\vec
p_3^2-\vec p_4^2\right]~,
\end{equation}
 In
\eqref{mul:eq:MMmm}, with the choice of the inverse of $\mu$ averaging the
inverse masses, the two Hamiltonians have the same threshold, but the hydrogen
molecule benefits from the term breaking charge-conjugation, and thus has a
lower ground state! In other words, the stability of the hydrogen molecule is a
consequence of the stability of the positronium molecule, \textit{via} the
variational principle.

However, the Coulomb character of the potential hardly matters, what is 
crucial is property that the same potential governs both systems. Thus it is
not a surprise that in a simple quark model with flavour independence, one 
finds
that  $(Q,Q,\bar q,\bar q)$ becomes stable if the mass ratio $M/m$ is large
enough. This was found by Ader et al.\ \cite{Ader:1981db} and in many subsequent
papers, see, e.g.,
\cite{Heller:1985cb,Zouzou:1986qh,Carlson:1987hh,SilvestreBrac:1993ss,%
Brink:1998as,Janc:2004qn ,Yang:2009zzp,Vijande:2007rf,Ohkoda:2012hv}, and
references there. 

The question is now how large $M/m$ should be. It depends on the details of the
model, it depends how good is the four-body calculation, and whether or not
spin-dependent corrections are also added. For long, it was believed that with
current pairwise models, $(b,b,\bar q,\bar q)$ was required, where $q=u,\,d$.
Two $b$ or not two $b$? The question was answered by Jang and Rosina
\cite{Janc:2004qn}, who
got binding with $Q=c$, using an additive  potential model fitting mesons and
baryons.
This was confirmed by Barnea et al.\ \cite{Vijande:2007rf}.
\subsection{Multiquarks in a Steiner-tree potential}\label{mul:sub:steiner}
\subsubsection{A new confinement for tetraquarks}
As already discussed, the pairwise interaction with colour factors is not
dictated by QCD. It is a convenient tool for preliminary investigations. For
sure, the ansatz looks reasonable for a $N$-body colour-singlet, when a quark is
well separated from the $(N-1)$ remaining constituents: the quark feels the
same interaction as in a quarkonium. But there is no fundamental reason why
the interaction should be pairwise. 

Years ago, Artru \cite{Artru:1974zn}, followed by many others
\cite{Dosch:1975gf,Hasenfratz:1980ka,Carlson:1982xi,Bagan:1985zn,Fabre:1997gf,%
Dmitrasinovic:2009dy}, pictured the mesons as a
simple string linking the quark to the antiquark, and baryons as three strings,
each linking a quark to an intersection named  \emph{junction}. In the strong
coupling regime of QCD,
this corresponds to a flux of gluons linking the quarks. The flux having  a
constant cross section, the potential grows linearly. 

\begin{figure}[!htb]
\begin{center}
\includegraphics{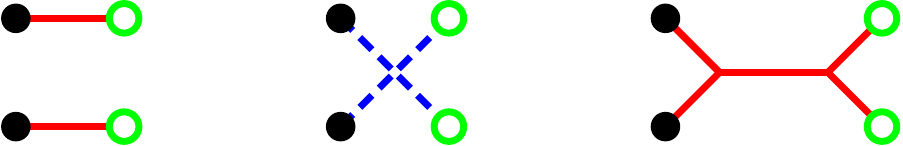}
\end{center}
 \caption{Simple string model for the tetraquark: the potential is the minimum
of two meson--meson links (flip--flop) and a connected contribution.}
\label{mul:fig:string-tetra}
\end{figure}

The analogue for tetraquarks of the $Y$-shape interaction of baryons
consists of two terms, 
\begin{enumerate}
 \item A flip-flop interaction, $V_f$, which is  the minimum of
$v(r_{13})+v(r_{24})$ and $v(r_{14})+v(r_{23})$,
\item A connected Steiner-tree, $V_s$, which links the quarks to the antiquarks
with a minimal length.
\end{enumerate}
The combination of flip-flop and Steiner tree, with a linear behaviour, 
namely,
\begin{equation}\label{mul:eq:V4}
 V_4=\min(V_f,V_s)~,
\end{equation}
thus involving a discrete two discrete minimisations and a minimisation over
the continuous variables describing the location of the junction of the Steiner
tree. This interaction was  proposed by Carlson et al.\ \cite{Carlson:1991zt},
Vijande et al.~\cite{Vijande:2007ix}, and Ay et al.\ \cite{Ay:2009zp}.
The model \eqref{mul:eq:V4} is schematically pictured in
Fig.~\ref{mul:fig:string-tetra}

\subsubsection{The flip-flop interaction}
The flip-flop interaction was proposed in a celebrated paper by Lenz et al.\
\cite{Lenz:1985jk}, with, however, a quadratic behaviour for $v(r)$. In this
paper is
discussed the possibility of extending the quark model of ordinary mesons and
baryons to the study of multiquark states and of the hadron--hadron
interaction. 
\subsubsection{The connected string}
\begin{figure}[htp]
 \centering
\includegraphics[width=.7\textwidth]{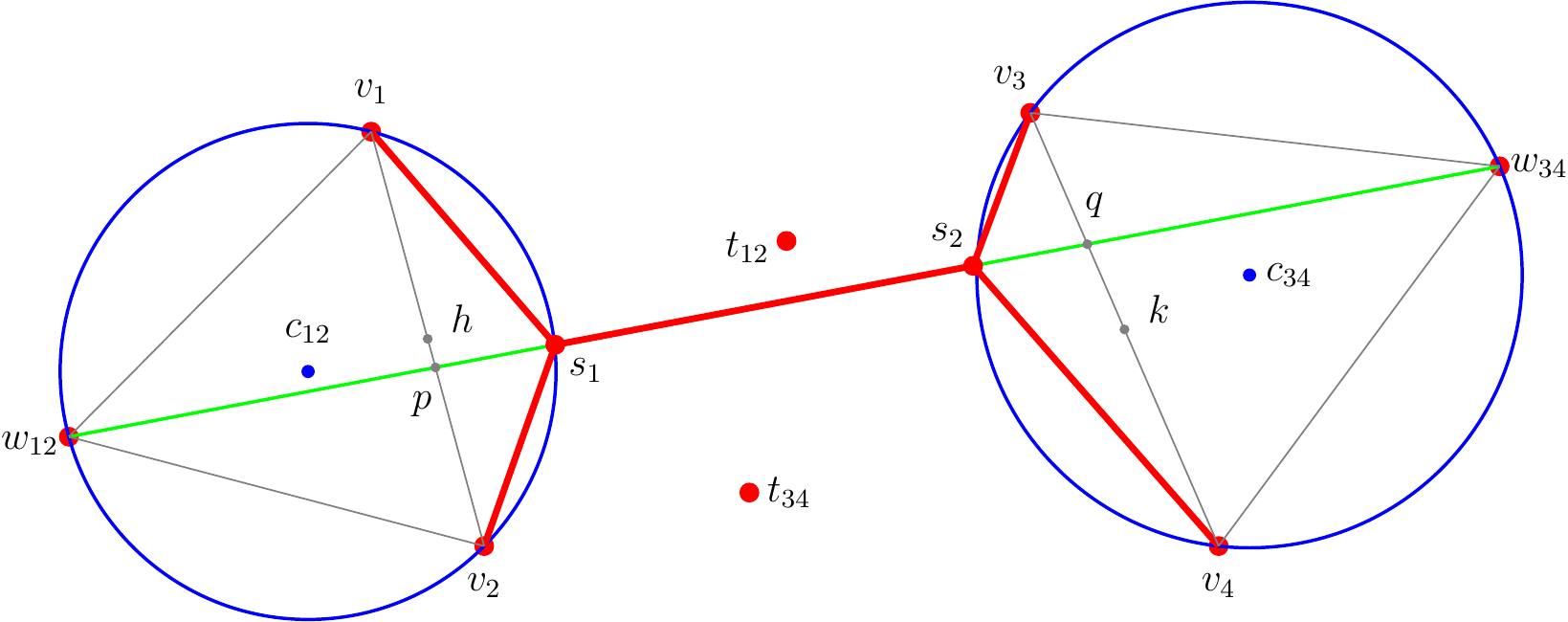}
 \caption{In a plane, the Steiner-tree potential can be constructed as the
Fermat--Torricelli minimal sum of distances for baryons. The length of
interest is equal to distance between the two auxiliary points $w_{12}$ and
$w_{34}$. More precisely, to the quarks $v_1$ and $v_2$ are associated the
Melznak set $s=\{w_{12},t_{12}\}$ and to the antiquarks $v_3$ and $v_4$ are
associated $\bar s=\{w_{34},t_{34}\}$, so that the triangles $(w_{12},v_1,v_2
)$ and $(w_{34},v_3,v_4)$ are equilateral. The length of the minimal Steiner
tree linking $v_1,v_2$ to $v_3,v_4$, which involves the optimisation of several
continuous variables is equal to the maximal distance from $s$ to $\bar s$,
which is just a discrete optimisation ,and thus very much faster.}
\label{mul:fig:Steiner-p}\end{figure}

It has interesting properties. In the planar case, it can be
constructed by duplicating Napoleon's construction for baryons, see
Fig.~\ref{mul:fig:Steiner-p}.

\begin{figure}[!thb] 
 \centering
\scalebox{.8}{\includegraphics{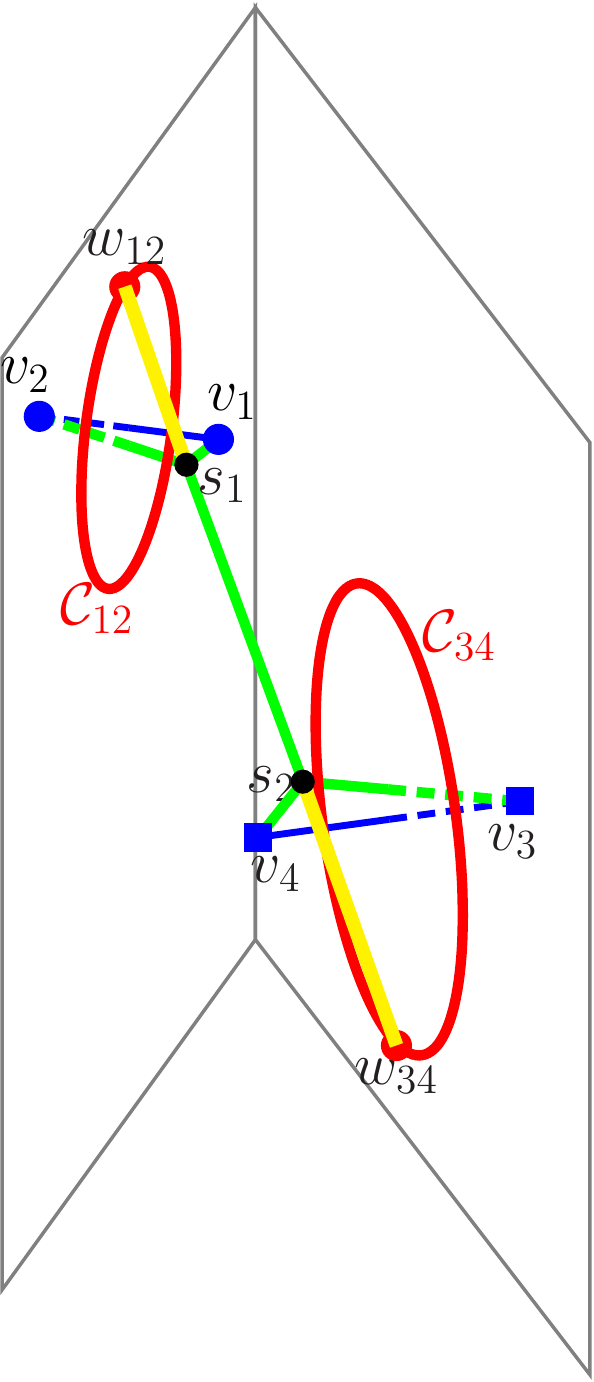}}
 \caption{The minimal length of the Steiner tree
$v_1s_1+v_2s_1+s_1s_2+s_2v_3+s_2v_4$ linking the quarks to the
antiquarks is the maximal distance $w_{12}w_{34}$ between their Melznak's
circles. Any point of
the circle $\mathcal{C}_{12}$ makes an equilateral triangle with the quarks
$v_1$ and $v_2$. The same holds in the antiquark sector.}
 \label{mul:fig:Steiner-s}
\end{figure}

In space, the length of the minimal Steiner tree is the maximal distances
between two circles. The first one includes all the points making an equilateral
triangle with the quarks, the second one for the antiquarks. See
Fig.~\ref{mul:fig:Steiner-s}. Now if you wish for the best algorithm to estimate
the
maximal distance between two circles in space, you have to  read the reviews on
computer-assisted cartoons.
 The software instructions in this branch of art are such that once a few basic
elements are drawn, you are guided to deduce auxiliary pieces of the picture. So
experts on multiquark physics can seek positions at Hollywood!

\begin{figure}[!bch]
\begin{center}
\includegraphics[width=.2\textwidth]{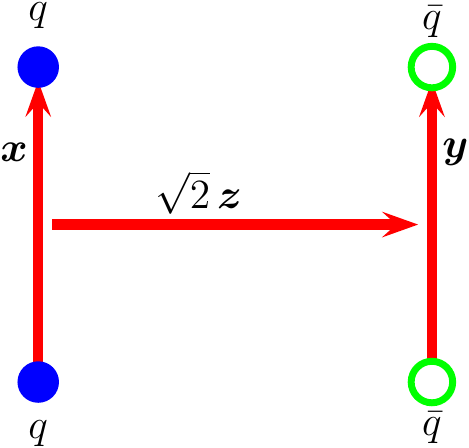}
\end{center}
\caption{Confining interaction for a tetraquark in the model by Barbour and
Ponting : $V=a\,(x+y+z)$}
 \label{mul:fig:barbour}
\end{figure}

A simplified version
of the connected interaction $V_s$ can be found in the work by Barbour and
Ponting \cite{Barbour:1978yz,Barbour:1979qi}. A set of Jacobi coordinates is
shown in Fig.~\ref{mul:fig:barbour}.
It reads (besides the centre of mass)
\begin{equation}\label{mul:eq:jacobi4}
\vec x=\vec r_2-\vec r_1~,\quad
\vec y=\vec r_4-\vec r_3~,\quad
\vec z=\frac{\vec r_4+\vec r_3-\vec r_2-\vec r_1}{\sqrt2}~, 
\end{equation}
so that the intrinsic part of the four-body Hamiltonian reads
\begin{equation}\label{mul:eq:H4a}
 H_4=\frac{\vec p_x^2}{m}+\frac{\vec p_y^2}{m}+\frac{\vec p_z^2}{m}+V~.
\end{equation}
In the case of a
T-baryonium, with colour $\bar 3$ coupling of the two quarks, and $3$ for
the antiquarks, Barbour and Ponting adopted a confining interaction
\begin{equation}\label{mul:eq:BP}
 V=a(x+y+z/\sqrt{2})~.
\end{equation}
Then $H_4$ exactly splits into three independent pieces, and the ground state
is at $e_0\,a^{2/3}\linebreak[3]{m^{-1/3} [2+2^{-1/3}]}$ above the threshold
at 
$2\,e_0\,a^{2/3}\,m^{-1/3}$. Here $e_0$ is the negative of the first zero of
the Airy function, see Sec.~\ref{mes:sub:central}.

Now, if one gives up the square geometry of Barbour and Ponting, and minimises
the string potential by varying the location of the junctions, one gets a lower
energy, but still above the threshold, for equal masses.
\subsubsection{Rigorous results}
We now discuss the four-body energy with the full linear potential $V_4$ of
Eq.~\eqref{mul:eq:V4}, starting with the exact result, and then, in the next
subsection the numerical results.

There are not many exact results for the four-body problem, and with the
cumbersome potential \eqref{mul:eq:V4}, it looks difficult to derive a simple
upper bound. In particular, $V_4$ is smaller than the flip-flop $V_f$ alone,
which in turn is smaller than the potential $a\,r_{13}+a\,r_{24}$ which governs
the threshold. This means that the effective meson--meson interaction is
attractive. In a world with on or two dimensions, this would imply binding. But
in our valley of tears in three dimensions, this is not the case.

Nevertheless, it is easily found from the construction of
Fig.~\ref{mul:fig:Steiner-s} that 
\begin{equation}\label{mul:eq:Ay}
V_s\le a\left[(x+y)\frac{\sqrt3}{2}+z\sqrt2\right]~,
\end{equation}
and it has been shown \cite{Ay:2009zp} that this bound also holds for $V_4$. 
Thus one gets a upper bound
\begin{equation}\label{mul:eq:H4up}
 H_4\le \frac{\vec p_x^2}{m}+\frac{\vec p_y^2}{m}+\frac{\vec
p_z^2}{m}+a\left[(x+y)\frac{\sqrt3}{2}+z\sqrt2\right]~,
\end{equation}
which is exactly a sum of three independent terms. 
Is is easily found that in this model $(QQ\bar q\bar q)$ becomes stable for a
mass
ratio $M/m\gtrsim 6402$. This is of course very crude, as the inequality in the
potential is saturated only when the circles in Fig.~\ref{mul:fig:Steiner-s}
are in the same plane, and do not overlap. 
\subsubsection{Numerical results}
The four-body problem with the interaction $V_4$ can be solved using standard
techniques. Here, the difficult
y comes from the numerical estimate of the
matrix elements of the potential. 
It is found that binding is obtained even for equal masses
\cite{Vijande:2007ix}, and, of course, it becomes better for $(QQ\bar q\bar q)$.
However, the minimisation in \eqref{mul:eq:V4} and in the flip-flop itself
implies a continuous rotation of the internal colour degree of freedom. Hence
colour cannot contribute to the antisymmetrisation. In other words, the
stability if for distinguishable quarks and antiquarks.
The connected term, or Steiner tree, contributes marginally, as compared to the
flip-flop. 
\subsubsection{String model of pentaquark}
\begin{figure}[htp]
\begin{center}
 % Pentaquark not connected and connected 
\includegraphics{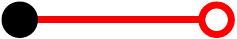}
\hskip .4cm
\includegraphics{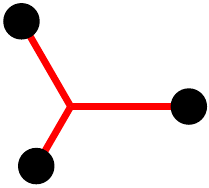}
\hskip 1cm
%Sch\'ema connexe pour le pentaquark %%%%%%%%
\includegraphics{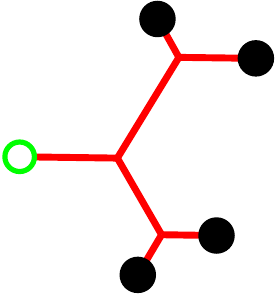}
\end{center}
 \caption{Simple string model for the pentaquark configurations}
 \label{mul:fig:pentastring}
\end{figure}

The model can be adapted to the case of pentaquarks, i.e., four quarks and one
antiquark \cite{Richard:2009rp}. The potential is pictured in
Fig.~\ref{mul:fig:pentastring}. Again, the interaction contains flip-flop
terms, i.e., the lowest of the meson + baryon string potential, and some
connected diagrams, the rule being to take the minimum.

Again, the calculation corresponds to distinguishable quarks, even in the case
where for simplicity the same mass is adopted. Then it is found that the
pentaquark is stable.
\subsubsection{String model of hexaquarks}
In a  recent study, six-quark configurations were studied, both $(q^3\bar q^3)$
and $(q^6)$ \cite{Vijande:2011im}.

\begin{figure}[htp]
\begin{center}
\scalebox{.7}{
% q^3 qbar^3 not connected and connected
\hskip 1cm
\includegraphics{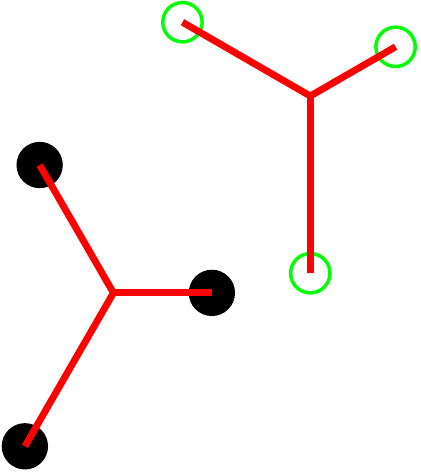}
% q^3 qbar^3 not connected and connected 
\includegraphics{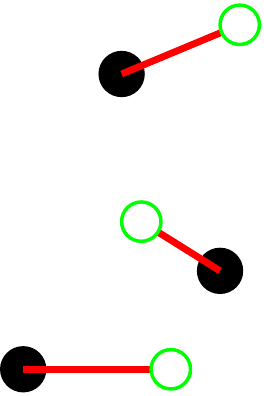}
\hskip 1cm
\includegraphics{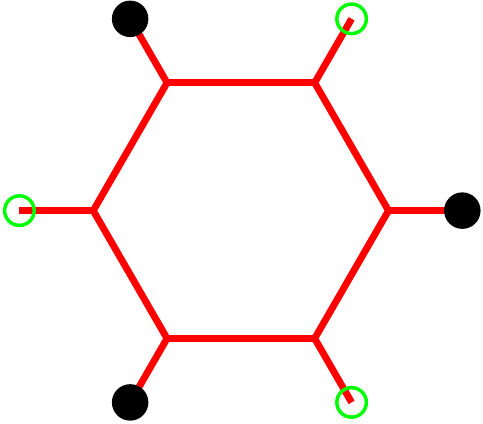}
\hskip 1cm
\includegraphics{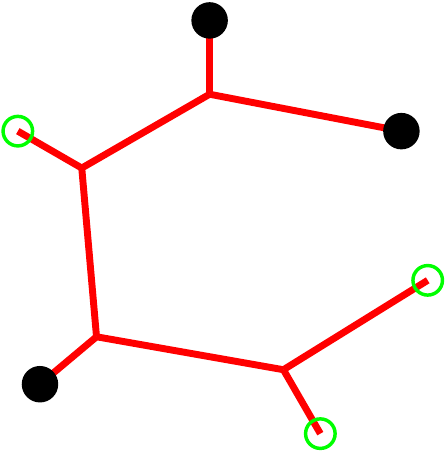}
}
\end{center}
 \caption{Potential for $(q^3\bar q^3)$: disconnected terms of
baryon--antibaryon type of three-meson type (with suitable minimisation over the
permutations) and some connected terms.}
 \label{mul:fig:q3qbar3}
\end{figure}

The string potential for $(q^3\bar q^3)$ is shown in Fig.~\ref{mul:fig:q3qbar3}.
Missing is a term consisting of a meson and a separated tetraquark. For
quarks and antiquarks with equal masses, but distinguishable, a weak binding is
obtained. The lowest threshold is $(q\bar q)^3$. If one introduces two
different masses and look at $(Q^3\bar q{}^3)$, the binding first is improved
when $M/m$ increases from $M/m=1$. But for very large $M/m$, the lowest
threshold becomes $(QQQ)+(\bar q\bar q\bar q)$ and $(Q^3\bar q{}^3)$ becomes
unstable.

\begin{figure}[htp]
\begin{center}
\scalebox{.75}{
 % q^6 not connected 
\includegraphics{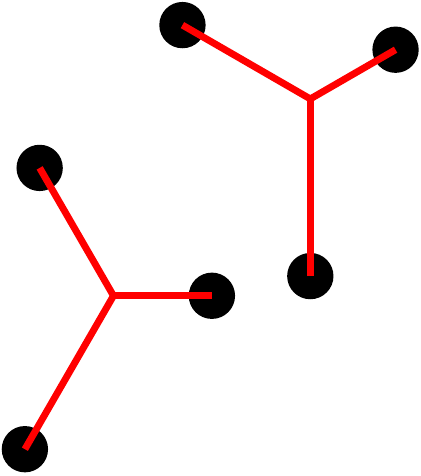}
\hspace{2cm}
% q^6 Connected
\includegraphics{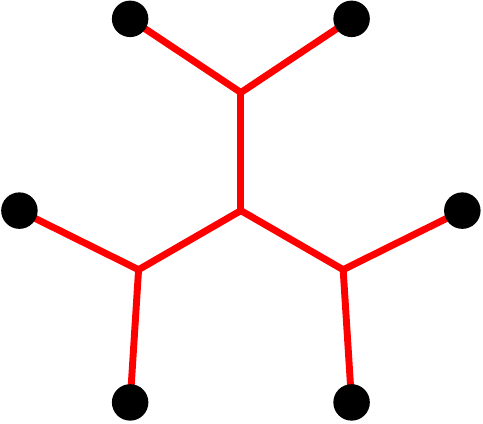}
}
\end{center}
 \caption{Contributions to the linear part of the $(q^6)$ potential,
disconnected and conected terms. A minimisation over the permutations is
implied. }
 \label{mul:fig:q6}
\end{figure}

In the case of $(q^6)$ the potential is either of baryon--baryon type, with all
possible permutations, or given by a connected string. See
Fig.~\ref{mul:fig:q6}. For equal masses, a comfortable binding is obtained. For
two different masses, i.e., $(Q^3q^3)$, the binding deteriorates. As the mass
ratio $M/m$ further increases, the 6-quark system becomes unbound and the
ground state consists of two isolated baryons, $(QQQ)+(qqq)$. 
\subsubsection{The next challenge: combining string and full antisymmetrisation}
The above string model, with mainly a flip-flop interaction and some connected
contributions, has many appealing features. However, it is not suited for
implementing the constraints of antisymmetrisation, and it tends to produce
several stable multiquarks that are not physical. Consider for instance the
tetraquark case. The potential is taken as the minimum (in units of the
string tension) of  $r_{13}+r_{24}$ which corresponds to a singlet--singlet
state of colour, or $r_{14}+r_{23}$ which corresponds to another
singlet--singlet, and the double-$Y$ string which corresponds to $(\bar3-3)_1$.
For distinguishable quarks and antiquarks, this is plausible, and the string
potential is just a Born--Oppenheimer potential, once the coloured gluon field
is minimised. Some change is required to implement this dynamics with full
account of the symmetry constraints.
\subsection{The light pentaquark}
The claim by the LEPS collaboration \cite{Nakano:2003qx} that there
exists a rather narrow baryon $\theta^+$ with mass about 1.54\,GeV, charge
$Q=+1$ and strangeness $S=+1$ stimulated much activity in the physics of hadron.

First, other experimental groups looked back at their old data already on
tapes, and also found some evidence for this state or for its SU(3)$_F$
partners (In the paper by Diakonov et al.\ \cite{Diakonov:1997mm} having
motivated the experimental search, the $\theta^+$ supposedly belongs to an
antidecuplet $\bar{10}$). Later, the experimental evidence became much weaker.
For a summary, see, e.g., the review by PDG \cite{}.

Second, the original calculation within chiral dynamics was revisited
\cite{Jaffe:2004qj,Walliser:2005pi} and led to different conclusions. The
pentaquark was also studied with QCD sum rules or Lattice QCD, with a variety
of conclusions. There were also studies within potential models, with detailed
five-body calculation. See, for instance, \cite{Hiyama:2007ur}. The light
pentaquark can not be reproduced in such calculations, unless one introduces
ad-hoc extra assumptions.

\clearpage
\markboth{\sc An introduction to the quark model\hspace*{1cm}
}{\sc Outlook}
\section{Outlook}\label{se:out}
\begin{flushright}
 \sl Important problems never receive a definite solution\\
\rm Confucius
\end{flushright}
About 50 years after the first speculations on quarks, what does remain?
Let us just list of few items for the discussions after the lectures.
\paragraph{\boldmath SU(3)$_{\rm F}$ symmetry} is now becoming somewhat
controversial. Sometimes, the SU(3)$_{\rm F}$ symmetry is almost exact.
This is the case for the rates of $J/\psi\to p\bar p, \ \Lambda\bar\Lambda,
\ldots$. In other cases, SU(3)$_{\rm F}$ looks badly broken. For instance, if
one looks at $K$ to $\pi$ production in high-energy collisions, one finds a
large ``strangeness suppression''. Perhaps the elementary couplings for $u\bar
u$ or $s\bar s$ production are very similar, but the small differences in
masses are amplified by a tunnelling factor which is exponential, like in the
Gamow theory of $\alpha$ decay.

\paragraph{From flavour symmetry to flavour independence}
In hadron spectroscopy, it is delicate to attempt a blind extension of
SU(3)$_{\rm F}$ to SU(4) or SU(5)  to include charm and beauty. The breaking
is too large to be treated as a perturbation. A new understanding of the
symmetry has been provided by the quark model. The basic interaction is flavour
independent, or has a well-defined flavour dependence in the spin-spin and
spin-orbit terms. Then the same basic interaction is used to build the various
$(q\bar q{}')$ mesons and $(qq'q'')$ baryons.

\paragraph{The bootstrap idea} was conceptually superb, but operationally
somewhat inefficient. Still, a kind of duality exists\footnote{G.F.~Chew,
private communication, years ago, that is very thankfully remembered}: for
instance, it
is natural to consider the $\Delta$ baryon in some context the $(qqq)$ partner
of $N=(qqq)$ with all
quark spins aligned, but to describe the  pion-nucleus scattering, there is
nothing better than $\Delta$ as a $\pi-N$ resonance. When the quark model was
invented, it was
considered as a relief, with at last a systematic and coherent scheme to
describe the hadrons. Nowadays, for any hadron $a$ of any mass $m$, one finds a
pair of existing hadrons $b$ and $c$ and an effective Lagrangian coupling $a$,
$b$ and $c$, such that $a$ appears as a kind of molecule. The overall picture
is seemingly lost. Is that a big step backwards?

\paragraph{From quark models  to QCD} This will be covered in some of the next
lectures.
It is fascinating how a somewhat empirical model, but based on the acute
observation of facts, became a beautiful theory.

\paragraph{A premium to simplicity} The quark model, even in its simplest
variant,  works rather well. Presumably, the quark model incorporates a lot of
subtle dynamical effects in an
effective way, and enables one to make successful predictions.

\paragraph{Hadron dynamics beyond hadron decay} 
An ambitious application of the quark model deals with the hadron--hadron
interaction. This is the continuum part of the sections devoted to loop
corrections and to multiquark spectroscopy.

Let us mention the calculation of the $KN$ interaction in the exotic $S=+1$
sector of strangeness, and the extensive work done to understand the short-range
part of the nucleon--nucleon interaction in terms of quarks.
%
%\newpage
\subsection*{Acknowledgements}
I would like to thank the organisers of the Ferrara school Niccol\`o Cabeo for
the opportunity to present this survey of the quark model. A slightly shortened
version will be included in the Proceedings.
It is a pleasure to thanks M.\ Asghar for several constructive comments on an
early version of this review.
			
\begin{small}
%  \bibliographystyle{unsrt}
%  \bibliography{biblio}

\end{small}
\end{document}